\documentclass[a4paper,11pt]{article}

\usepackage[T1]{fontenc}

\DeclareUnicodeCharacter{202F}{\,}
\usepackage{caption}
\usepackage{jheppub} 

\usepackage[mathlines]{lineno}

\usepackage{comment}
\usepackage{graphicx}
\usepackage{dcolumn}
\usepackage{bm}
\usepackage{hyperref}
\usepackage{amsmath} 
\usepackage{upgreek}
\usepackage{mathtools}
\usepackage{multirow}

\usepackage{tikz}
 \usepackage[compat=1.1.0]{tikz-feynman}
\usepackage{subcaption} 

\title{\boldmath Study of Central Exclusive Production of $\pi^+\pi^-$, $K^+K^-$ and $p \bar{p}$ Pairs in Proton-Proton Collisions at $\sqrt{s} = 510$ GeV with the STAR Detector at RHIC}

\collaboration{\includegraphics[height=17mm]{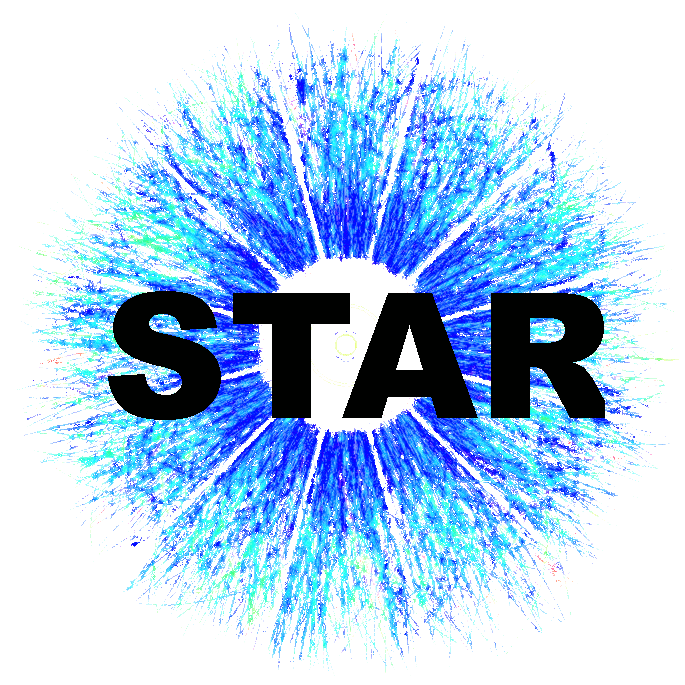} \\[6pt] STAR Collaboration}

\emailAdd{Tomas.Truhlar@fjfi.cvut.cz}

\abstract{We report on the first measurement of the Central Exclusive Production process in proton-proton collisions: $pp \ \rightarrow \ p h^+ h^- p$ (where $h = \pi, K, p$) at the center-of-mass energy $\sqrt{s} = 510$ GeV with the STAR experiment at RHIC. At this energy, the process is dominated by a double Pomeron exchange mechanism. Hence, it provides a clean environment for investigating Pomeron interactions by measuring fully reconstructed final states involving only two hadrons and two forward scattered protons. The oppositely charged hadron pairs are measured within the central detector of STAR. The forward scattered protons are measured in the Roman Pot system allowing the verification of the event's exclusivity. Differential fiducial cross sections within the STAR acceptance are presented as a function of the difference in the azimuthal angle between the outgoing protons. The invariant masses of the charged hadron pairs are measured up to approximately 3 GeV and the square of the four-momentum transfer ($t_1$ and $t_2$) of the two forward-scattered protons in the range $0.3 \text{ GeV}^2 < -t_1 , -t_2 < 1.6 \text{ GeV}^2$. The differential fiducial cross sections of the forward protons as a function of the $|t_1 + t_2|$ are also presented. All results for the $\pi^+\pi^-$ pair are presented in three mass ranges. A comparison with GRANIITTI Monte Carlo predictions are also presented, where the spectra include continuum and resonant contributions. The observed spectra are consistent with double Pomeron exchange, including resonances seen in previous studies.}

\begin{document}
\maketitle
\flushbottom

\newcommand{\Pom}{P} 
\newcommand{\DPE}{DPE} 
\newcommand{\pT}{p_{\textsf{T}}} 
\newcommand{\pTmiss}{p_{\textsf{T}}^{\text{miss}}} 
\newcommand{\nFit}{N_{\text{hits}}^{\text{fit}}} 
\newcommand{\nDedx}{N_{\text{hits}}^{\langle dE/dx\rangle}} 
\newcommand{\zvtx}{z_{\text{vtx}}}
\newcommand{\dpt}{dP_\mathrm{T}}

\section{Introduction}\label{sec:Introduction}
Central Exclusive Production (CEP) in proton-proton collisions is a diffractive process in which the two incoming protons remain intact after the interaction~\ref{eq:CEP}. The protons are measured in the forward direction and the recoil system $X$ is produced in the central region of rapidity and is separated from forward protons by large rapidity gaps, ${\Delta \eta_1},{\Delta \eta_2}$, with no additional particle production. By definition, ``exclusive'' means that all particles in the final state are measured, which includes both the forward protons and the central system. 
\begin{equation}\label{eq:CEP}
p + p \rightarrow  { p} \ensuremath{\stackrel{\Delta \eta_1}{\oplus}} {X} \ensuremath{\stackrel{\Delta \eta_2}{\oplus}} {p}
\end{equation}
where $\oplus$ denotes a rapidity gap.

Measurement of CEP of charged hadron pairs in the Double \Pom omeron Exchange (DPE) process~\cite{Barone:2002cv} provides a unique opportunity to study quantum chromodynamics (QCD), the theory of strong interactions. 
The validity of QCD has been demonstrated in many experiments since the 1970s, in particular at hadron colliders, like the measurements of CEP at the Intersecting Storage Rings (ISR) by the AFS~\cite{Akesson1986} experiment, at RHIC by the STAR~\cite{Rafal20} experiment, and at the Large Hadron Collider (LHC) by the CMS and TOTEM~\cite{CMS} and ATLAS~\cite{Atlas} experiments, and also by the measurement of central production (CP), where the forward protons are not measured, at the Tevatron by the CDF~\cite{PhysRevD.91.091101} experiment and the LHCb experiment~\cite{LHCb:2014,LHCb:2025}, where a glueball interpretation is possible. The CEP was also studied in fixed target experiments. A review of those experiments can be found in~\cite{Albrow:2014}. While these results confirm the validity of QCD, they also make clear that the non-perturbative regime remains insufficiently understood.

The forward protons are measured using detectors placed in special vessels called Roman Pots (RP)~\cite{Bltmann2004}. The centrally produced hadrons are measured in the Time Projection Chamber (TPC)~\cite{Anderson2003} and in the Time of Flight (TOF) systems~\cite{Llope:2009zz}. The rapidity gaps $2.1 < | \eta | < 5.0$ are ensured by vetoes using the Beam-Beam Counters (BBC)~\cite{Whitten2008} acceptance. This experimental setup provides full control over the interaction’s kinematics and allows for a direct verification of its exclusivity.


There are three possible mechanisms of CEP: double photon exchange, photoproduction, and DPE. The last one is expected to be the dominant CEP mechanism~\cite{Ganguli1980, Forshaw1997, Kirk2014} at RHIC energies and within STAR acceptance of the forward protons. In photoproduction and double photon exchange, the transverse momentum of one or both scattered protons is typically too small to be detected in the RP detectors. Consequently, the requirement of observing both forward protons strongly suppresses these mechanisms, making DPE the dominant CEP mechanism within STAR acceptance.

\begin{figure}[ht]
\centering
    \begin{subfigure}[b]{0.32\textwidth}
\begin{center}
\begin{tikzpicture}
    \begin{feynman}
        \vertex (p1) at (-2.0, 1.5) {\( p \)};
        \vertex (p2) at (-2.0, -1.5) {\( p \)};
        
        \vertex (p1p) at (2.0, 1.5) {\( p \)};
        \vertex (p2p) at (2.0, -1.5) {\( p \)};
        
        \vertex (g2) at (0, 1.5) [dot];
        \vertex (g4) at (0, -1.5) [dot];
        
        \vertex (g5) at (0.0, 0) [dot];

        \vertex (g6) at (0.8, 0) [dot];

        \vertex (h1) at (2.0, 0.5) {\( h^+ \)};
        \vertex (h2) at (2.0, -0.5) {\( h^- \)};
        
        \diagram* {
            (p1) -- [fermion] (g2) -- [fermion] (p1p),
            (p2) -- [fermion] (g4) -- [fermion] (p2p),
            
            (g5) -- [scalar] (g6),
            (g6) -- [fermion] (h1),
            (h2) -- [fermion] (g6)
        };

        \draw[double, decorate, decoration={zigzag, amplitude=0.8mm, segment length=2mm}] (g2) -- (g5);

        \draw[double, decorate, decoration={zigzag, amplitude=0.8mm, segment length=2mm}] (g5) -- (g4);

        \draw[fill=black, draw=black] (0, 1.5) ellipse (0.08cm and 0.08cm); 

        \draw[fill=black, draw=black] (0, -1.5) ellipse (0.08cm and 0.08cm); 
        
        \draw[fill=gray!30, draw=black] (g5) ellipse (0.2cm and 0.2cm);

    \end{feynman}
\end{tikzpicture}
\end{center}    
    \end{subfigure}
    \begin{subfigure}[b]{0.32\textwidth}
\begin{center}
\begin{tikzpicture}
    \begin{feynman}
        \vertex (p1) at (-2.0, 1.5) {\( p \)};
        \vertex (p2) at (-2.0, -1.5) {\( p \)};
        
        \vertex (p1p) at (2.0, 1.5) {\( p \)};
        \vertex (p2p) at (2.0, -1.5) {\( p \)};
        
        \vertex (g2) at (0, 1.5) [dot];
        \vertex (g4) at (0, -1.5) [dot];
        
        \vertex (g5) at (0.0, 0) [dot];
        
        \vertex (h1) at (2.0, 0.5) {\( h^+ \)};
        \vertex (h2) at (2.0, -0.5) {\( h^- \)};
        
        \diagram* {
            (p1) -- [fermion] (g2) -- [fermion] (p1p),
            (p2) -- [fermion] (g4) -- [fermion] (p2p),
            
            (g5) -- [fermion] (h1),
            (h2) -- [fermion] (g5)
        };

        \draw[double, decorate, decoration={zigzag, amplitude=0.8mm, segment length=2mm}] (g2) -- (g5);

        \draw[double, decorate, decoration={zigzag, amplitude=0.8mm, segment length=2mm}] (g5) -- (g4);

        \draw[fill=black, draw=black] (0, 1.5) ellipse (0.08cm and 0.08cm); 

        \draw[fill=black, draw=black] (0, -1.5) ellipse (0.08cm and 0.08cm); 
        
        \draw[fill=gray!30, draw=black] (g5) ellipse (0.2cm and 0.2cm);

    \end{feynman}
\end{tikzpicture}
\end{center}    
    \end{subfigure}
    \begin{subfigure}[b]{0.32\textwidth}
    \begin{center}
\begin{tikzpicture}
    \begin{feynman}
        \vertex (p1) at (-2.0, 1.5) {\( p \)};
        \vertex (p2) at (-2.0, -1.5) {\( p \)};
        
        \vertex (p1p) at (2.0, 1.5) {\( p \)};
        \vertex (p2p) at (2.0, -1.5) {\( p \)};
        
        \vertex (g1) at (0.0, 1.5) [dot];
        \vertex (g2) at (0.0, 1.5) [dot];
        \vertex (g3) at (0.0, -1.5) [dot];
        \vertex (g4) at (0.0, -1.5) [dot];
        
        \vertex (g5) at (0.5, 0) [dot];
        
        \vertex (h1) at (2.0, 0.5) {\( h^+ \)};
        \vertex (h2) at (2.0, -0.5) {\( h^- \)};
        
        \diagram* {
            (p1) -- [fermion] (g1) -- (g2) -- [fermion] (p1p),
            (p2) -- [fermion] (g3) -- (g4) -- [fermion] (p2p),
            
            (g1) -- [gluon] (g3),
            (g5) -- [gluon] (g2),
            (g4) -- [gluon] (g5),
            
            (g5) -- [fermion] (h1),
            (h2) -- [fermion] (g5)
        };

        \draw[fill=black, draw=black] (0, 1.5) ellipse (0.08cm and 0.08cm); 

        \draw[fill=black, draw=black] (0, -1.5) ellipse (0.08cm and 0.08cm); 
        
        \draw[fill=gray!30, draw=black] (g5) ellipse (0.2cm and 0.2cm);

    \end{feynman}
\end{tikzpicture}
\end{center}
    \end{subfigure}

    \addvspace{-17pt}
\caption[Diagrams of CEP of $h^+h^-$ in DPE process]{Diagrams of CEP of $h^+h^-$ in the DPE process. The resonant (left) and continuum  production (middle) are shown. The two-gluon approximation of continuum production in DPE in perturbative QCD (right) is also shown. \Pom omerons are represented by double zigzag lines and gluons by spiral lines.}
\label{fig:Faynman}
\end{figure}
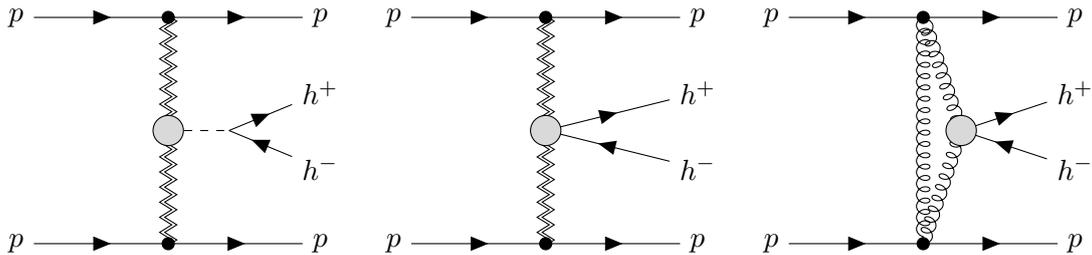

Figure~\ref{fig:Faynman} shows diagrams of $h^+h^-$ pair production, where $h^+h^-$ are $\pi^+\pi^-$, $K^+K^-$, and $p\bar{p}$, in the DPE process. In the non-perturbative QCD regime: central exclusive resonant production (left) and continuum (non-resonant) production (middle) are shown. Representation of continuum production within perturbative QCD is shown in the two-gluon approximation model (right). 

CEP probes the non-perturbative regime of QCD where Regge theory is often used to describe such interactions. In QCD, the \Pom omeron is modeled as a color-singlet object. In the lowest order, it can be represented as a two-gluon exchange. As no color is exchanged in DPE, the color content of each proton is preserved allowing both protons to stay intact after the collision. The presence of large rapidity gaps between the centrally produced system and the outgoing protons is a key signature of such events.

These characteristics make CEP a suitable process for investigating the nature of the \Pom omeron exchange and the non-perturbative regime of QCD. Theoretical non-perturbative approaches in this domain remain challenging~\cite{HarlandLang20140626, Lebiedowicz2016}, as the dynamics involve interference between resonant and continuum contributions and potential re-scattering effects between the outgoing protons.
In DPE, the production of charged hadron pairs often enhances states with quantum numbers $I^GJ^{PC} = 0^+(\mathrm{even})^{++}$, which makes this process particularly sensitive to scalar and tensor mesons. However, this is not an absolute constraint: other states, such as the $f_{1}(1285)$ ($1^{++}$)~\cite{BARBERIS1997225} and the $\eta$ ($0^{-+}$)~\cite{1998398}, have also been observed in DPE. This reflects the more complex nature of the \Pom omeron exchange and the fact that exclusive production does not occur strictly at $\pT=0$, allowing additional quantum numbers to contribute. Consequently, DPE acts as a selective but not exclusive ``quantum number filter," enhancing certain states while still permitting others, which in turn makes it a valuable tool for meson spectroscopy and for probing gluon-rich channels. 
DPE also offers a gluon-rich environment for exploring glueball production~\cite{Ochs_2013}. Experimental confirmation of glueball existence would be yet another strong support for the validity of QCD. However, the exact nature of the \Pom omeron still remains elusive~\cite{EWERZ201431} and the existence of a glueball has not been unambiguously confirmed yet. 


This paper presents the first measurement of $\pi^+\pi^-$, $K^+K^-$, and $p\bar{p}$ production in the CEP processes in proton-proton collisions at $\sqrt{s} = 510$ GeV using the STAR detector at the Relativistic Heavy Ion Collider (RHIC)~\cite{Harrison, HAHN2003245}.
Differential and integrated cross sections are measured within the STAR detector’s acceptance. The results are compared with the previous STAR results on CEP of charged hadron pairs in proton-proton collisions at $\sqrt{s} = 200$ GeV~\cite{Rafal20}. 
Throughout the paper, the natural units $ c = \hbar = 1$ are used. 

\section{Experimental setup}\label{sec:ExpSetup}

\begin{figure}[htbp!]
\centering
\includegraphics[width = 0.8\columnwidth]{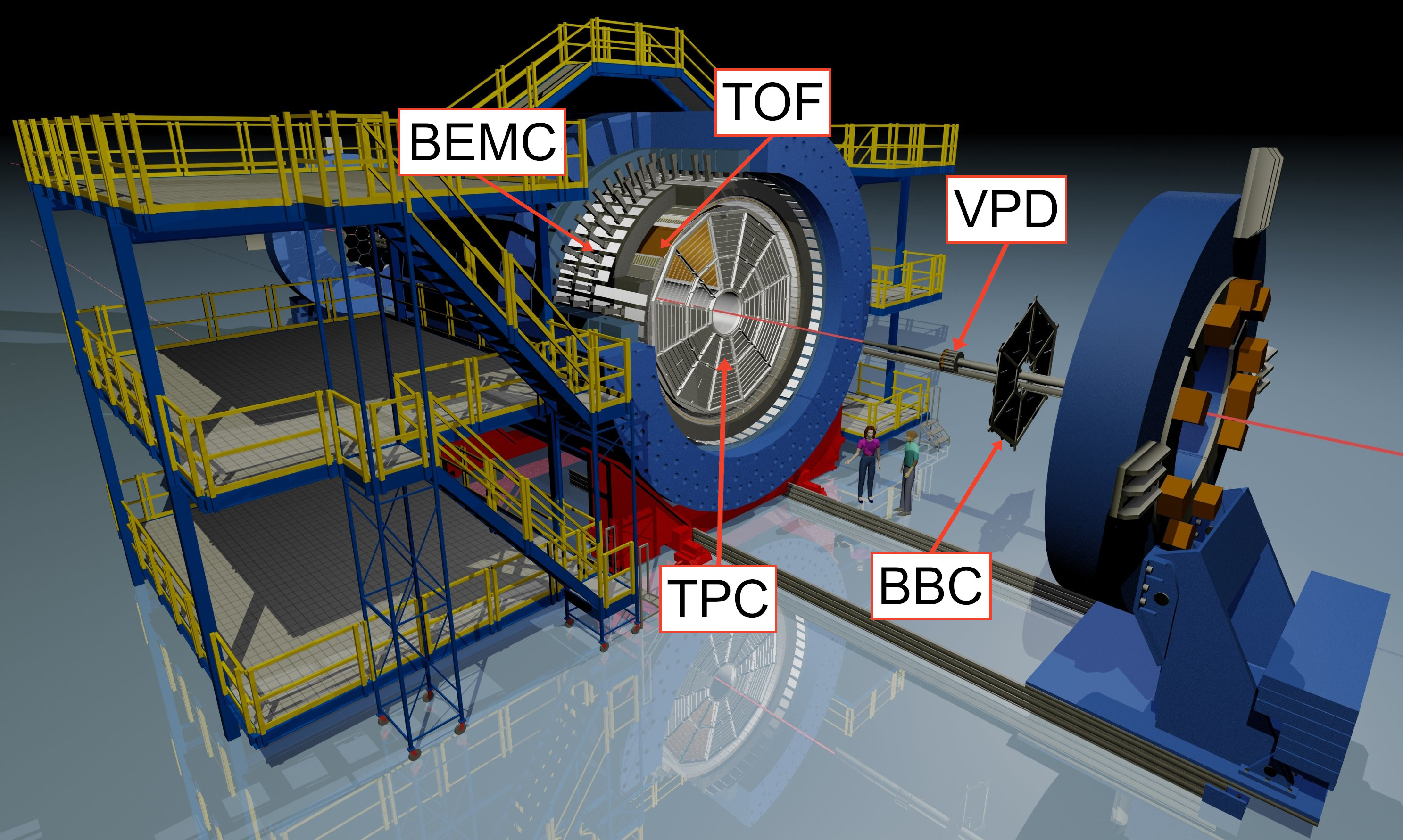}
\caption[Schematic view of STAR experiment]{The schematic view of the STAR experiment. The main sub-detectors, including the TPC, the TOF, the BBC, the VPD and the BEMC are highlighted.}
\label{02star}
\end{figure}

STAR is a general-purpose detector at RHIC, with a 0.5 Tesla~\cite{ACKERMANN2003624} solenoidal magnetic field parallel to the beam axis. This measurement utilizes various subsystems of the STAR detector in a configuration similar to that used for the CEP measurements in proton-proton collisions at $\sqrt{s} = 200$~GeV~\cite{Rafal20}. Figure~\ref{02star} shows the STAR detector with its sub-detectors, namely TPC, TOF, BBC, the Vertex Position Detector (VPD)~\cite{LLOPE201423}, and the Barrel Electromagnetic Calorimeter (BEMC). Furthermore, the STAR experiment includes forward systems: the Zero-Degree Calorimeters (ZDC)~\cite{Adler2001, Adler2003} and the RP system. These sub-detectors are particularly important for the CEP analysis, especially the RP system. In the following, the sub-detectors are briefly described.

The TPC is a gas-filled detector, which provides both tracking information and an average ionization energy loss per unit length ($\langle dE/dx\rangle$) for each track. The momentum reconstruction and the $\langle dE/dx\rangle$ allow identification of charged hadrons: pions, kaons, and protons. It has a cylindrical shape with a length of 4.2 m, an inner radius of 0.5 m and an outer radius of 2 m. The TPC has full azimuthal coverage $0 < \varphi < 2\pi$, and pseudorapidity coverage $|\eta | < 1.0$ for charged particle detection and identification\footnote{The origin is at the interaction point (IP) at the center of the detector. The $z$-axis is aligned with the center of the beam pipe, the $x$-axis points radially outward from the IP toward the outside of the RHIC ring, and the $y$-axis points vertically up. The pseudorapidity $\eta$ is defined in terms of the polar angle $\theta$ as $\eta = -\ln\left( \tan\left( \frac{\theta}{2} \right) \right)$.}.

The TOF system is a cylindrical detector around the TPC extending the particle identification (PID) capabilities of the TPC for particles with momenta up to $\sim$ 3 GeV. The TOF is composed of adjacent Multi-gap Resistive Plate Chambers covering full azimuthal angle and pseudorapidity $| \eta | < 0.9$. It is a fast timing detector with a time resolution between $60-100$ ps. Therefore, it is often used to trigger on charged particle multiplicity in the central rapidity. For the CEP events, it is crucial for triggering on low multiplicity events in the TPC. Moreover, it helps to discriminate in-time TPC tracks from different bunch crossings (out-of-time pile-up).

The VPD is a fast timing detector located on both sides of the IP at forward pseudorapidity ($4.24 < | \eta | < 5.1$). It consists of arrays of scintillator detectors coupled to photomultiplier tubes and provides precise measurement of the event start time and the longitudinal position of the primary vertex using the timing difference between the east and west detectors. Since there are no particles in this rapidity region for the CEP events, the reconstruction of the starting time using the VPD is not possible. 

The BBC is an array of plastic scintillator detectors designed to detect high$-\eta$ particles produced in the forward direction. The BBC consists of two identical detectors placed at $\pm3.74$~m from the IP. Each detector is formed by 36 hexagonal scintillator tiles, covering a pseudorapidity range of $2.1 < | \eta | < 3.3$, and 18 large tiles covering $3.3 < | \eta | < 5.0$, respectively. For the CEP events, it is used as a rapidity gap veto, to ensure the rapidity gaps between the central produced system and forward-scattered protons.  

The ZDC's main function is to measure neutral particles produced in the forward direction, $\theta \lesssim 4$ mrad. They are placed on each side of the IP at a distance of $\pm$18~m, directly in line with the STAR beam pipe behind the DX dipole magnets. These magnets deflect charged particles, including the forward protons. They are used to detect beam energy neutrons produced in the forward direction close to the beam direction. Both BBC and ZDC are used to ensure a rapidity gap within their acceptance, thereby reducing background due to the non-exclusive events at the trigger level.

\begin{figure}[htbp!]
\centering
\includegraphics[width= 0.8\columnwidth]{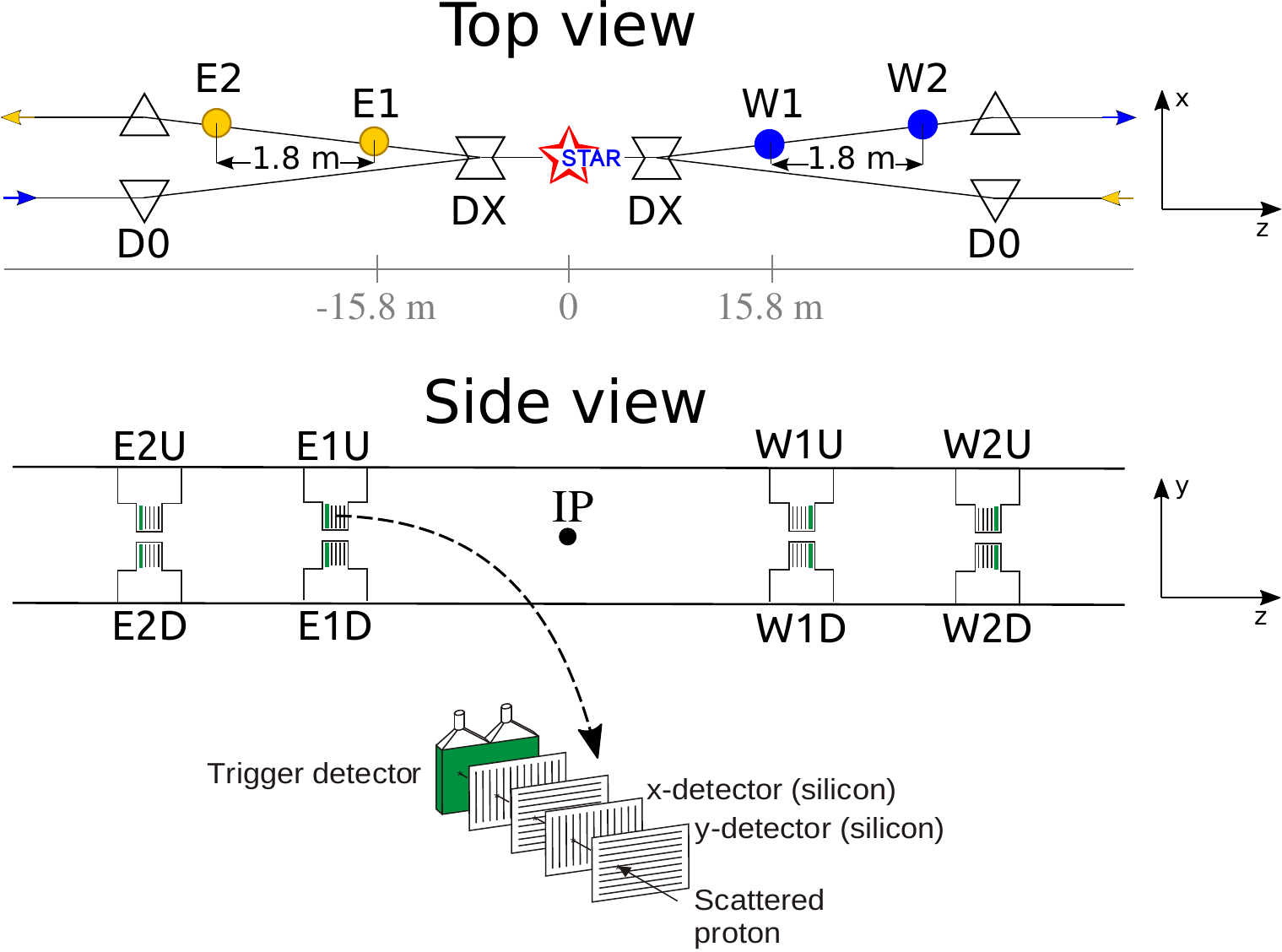}
\caption[Roman Pot Phase II* layout]{
The schematic layout of the STAR Roman Pot system (not to scale), showing top $(x,z)$ and side $(y,z)$ views. The RP stations (E1, E2, W1, W2) are located at $\pm15.8$~m and $\pm17.6$~m from the interaction point, between the DX and D0 dipole magnets. The lower panel shows also the detector package with four silicon strip planes and a trigger scintillation counter. Taken from Ref.~\cite{Wlodek}.}
\label{02rp}
\end{figure}

The RPs are used to detect and measure forward protons scattered at very small angles (few~mrad), whose trajectories are contained in the accelerator beam pipe. For that purpose, the STAR experiment was upgraded with the RP system previously used by the PP2PP experiment~\cite{Bltmann2004}. Two RP stations were installed on each side of the IP at a distance of $\pm$15.8~m and $\pm$17.6~m from the IP. The location of the RPs, top and side views, and the coordinate system are shown schematically in~figure~\ref{02rp}. Each RP station has two movable RP vessels: one above and one below the beam pipe. Each stainless steel RP vessel houses a Silicon Strip Detector (SSD) package and a scintillation counter. The SSD package consists of four SSDs measuring proton position in the $x-y$ plane, two measuring $x-$coordinate and two measuring $y-$coordinate. 

\begin{figure}[htbp!]
        \centering
        \includegraphics[width = .9\linewidth]{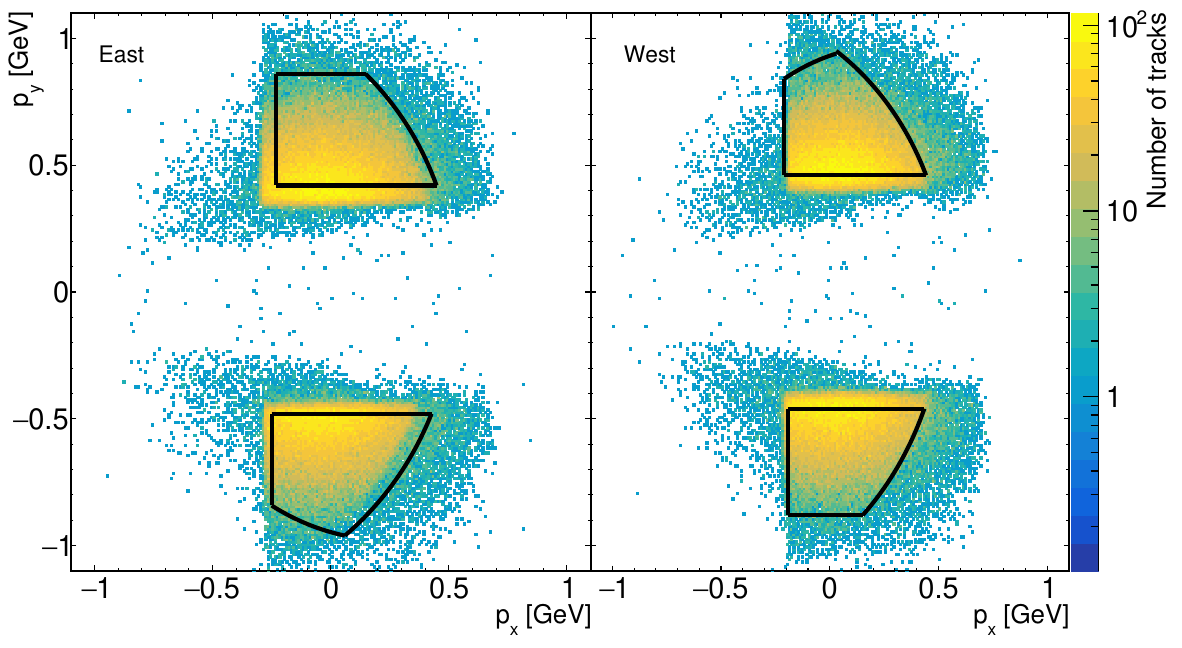}
        \caption[The RP fiducial volume.]{The RP fiducial volume shown with black solid lines on top of the combined distributions of forward protons’ momenta $p_y$ vs. $p_x$ reconstructed with the East and West RP stations.}
        \label{fig:RPFV}
    \end{figure}

The sensitive area of the detector package is  $5\times 8\; \mathrm{cm}^2$ in the transverse size and a depth 3.5~cm. The silicon sensor is $400\;\mu\mathrm{m}$ thick, while the trigger scintillator is 5 mm thick. The strips in the silicon detectors are approximately $100\; \mu$m wide allowing a spatial resolution of $30 \ \mu$m resulting in the transverse momentum resolution at the level of $\sim$10~MeV. The RP sensitive area in the momentum transverse plane can be seen in~figure~\ref{fig:RPFV} with the used fiducial volume illustrated by black solid lines.

The typical distance of the first horizontal strip from the beam is about 2 cm. Having two RP stations on each side allows momentum reconstruction of scattered protons. The scintillation counter was used to trigger on forward protons. Each trigger counter was read out by two photomultipliers. The stainless steel vessel separates the detector package from the machine vacuum allowing operation of the package at normal atmospheric pressure.

There is an RP naming convention that denotes the position of each RP. The RP names are in the following format: side of the IP (E or W), station (1 or 2), and orientation (U or D). For example, the first RP, as measured away from the IP, located on the west side of the IP above the beam line, is called W1U. Moreover, the RP system is divided into four branches of the RP (EU, ED, WU, and WD), in which proton tracks can be reconstructed. A branch of the RP is composed of two RP on the same side of the IP (E or W) and with the same orientation (U or D). For example, the EU branch of the RP is composed of E1U and E2U RPs. 

The RP stations are placed behind the DX magnet, the RHIC-lattice dipole magnet closest to the IP. This positioning allows the measurement of the momentum vector of the scattered protons at the detection point.
The symmetry of the RHIC rings requires that the magnetic fields in the DX magnets are the same, and the relative differences of DX magnet strengths are at the $10^{-3}$ level~\cite{STAR:2023vdw}. The RP layout ensures that no special beam conditions are needed to operate the RP detectors.

After the RP installation, a survey was performed to determine the position in space of the first strip in each detector package in the STAR coordinate system, the so-called baseline survey alignment. 
Corrections to the baseline survey alignment were obtained using elastically scattered protons. The collinearity constraint was applied, which required that the two reconstructed tracks on each side of the IP lie on the same straight line. The alignment method is the same as described in~Ref.~\cite{Wlodek} and used in~Refs.~\cite{Wlodek, Rafal20, STAR:2023vdw}. The obtained corrections are applied to correct the position of reconstructed points. The average final correction over all runs is $20 \pm 10 \ \mu$m, consistent with the one obtained in~Ref.~\cite{STAR:2023vdw}.

\section{Triggers}\label{sec:triggers}

The trigger for CEP events was based on their characteristic event topology. The following trigger conditions were required:
\begin{enumerate}
    \item To ensure at least one proton on each side of the IP, an energy deposit was required in the scintillators, consistent
with a minimum ionizing particle (MIP) in time with the beam crossing in exactly one branch of the RP on each side of the IP. A veto was imposed on the other branches of the RP to reject either proton dissociation or pile-up events. This signal ensures the presence of at least two forward-scattered protons. 
    \item At least two charged particles in central rapidity. This was implemented by requiring the TOF multiplicity $\geq 2$. The upper limit on the TOF multiplicity ($\leq 10$) was applied to reject high multiplicity events. This allowed good acceptance for four-pion events and avoided rejection of CEP events which could overlap with a small level of noise.
    \item A rapidity gap between the central system and forward protons. A veto on a signal consistent with a minimum ionizing particle in small or large BBC tiles, or an energy deposition in the ZDCs above a few tens of GeV on either side of the IP was used to ensure the double-gap topology typical of CEP events. The typical noise in ZDC is at level of few GeV. 
\end{enumerate}

With the above trigger, we acquired the integrated luminosity of $121\pm6$ pb$^{-1}$ with the average instantaneous luminosity equal to $L \simeq 127~\mathrm{\mu b^{-1}\,s^{-1}}$. 
At these luminosities, pileup leads to overlap of exclusive events with additional inelastic interactions in the same bunch-crossing. Such a pile-up is suppressed by the CEP trigger vetoes.
\section{Event selection}\label{chap:EventSelection}

The event selection is designed to select CEP events while maintaining high signal purity and efficiency. 
The applied selection criteria are described in detail in the following subsections and proceed in several steps.

First, events with two forward-scattered protons are selected. To suppress pileup, exactly one primary vertex reconstructed from tracks matched to TOF hits is required. Reconstructed TPC tracks are required to satisfy standard quality and fiducial acceptance criteria.  
The exclusivity of the event is enforced by requiring a balance of the transverse momentum of all measured particles. Finally, PID criteria are applied to classify the selected tracks into the $\pi^+\pi^-$, $K^+K^-$, and $p\bar{p}$ final states.

\subsection{RP tracks}\label{sec:RpTracks}

In order to reconstruct the position and momentum of the forward protons, we follow the approach from~Refs.~\cite{Rafal20,Wlodek,STAR:2023vdw}. First, space points are reconstructed in each RP. Second, events with only one space point per RP are used to reconstruct a track. For these events, the transverse momentum of the protons ($p_x$, $p_y$) is reconstructed based on the positions of space points assuming a constant and uniform magnetic field inside the DX magnet.
To select CEP event candidates, exactly one RP track on each side of the IP is required. Each track must have at least three out of four SSD planes used in its space point reconstruction in each RP.
Furthermore, to ensure high acceptance, RP efficiency, and low systematic uncertainties, the reconstructed protons have to have momenta ($p_x, \ p_y$) inside a fiducial volume, shown in~figure~\ref{fig:RPFV} and defined by an "and" of the following conditions~(\ref{eq:fiducialVolume}):
\begin{equation}
    \begin{aligned}
        &|p_y| > p_y^\text{min} \land  p_x > p_x^\text{min} \\
        &(p_x + p_x^\text{center})^2 + p_y^2 < R^2 \\
        &(p_y > 0 \quad \text{for EU and WU} ) \ \land   (p_y < 0 \quad \text{for ED and WD}) \\
        &|p_y| < p_y^\text{max} \quad \text{for EU and WD} \\
        &(p_x + \bar{p}_x^\text{center})^2 + p_y^2 < \bar{R}^2 \quad \text{for ED and WU},
    \end{aligned}
    \label{eq:fiducialVolume}
\end{equation}
where the parameters ($p_y^\text{min}, \ p_x^\text{min}, \ p_x^\text{center}, \ R^2, \ p_y^\text{max}, \ \bar{p}_x^\text{center}\text{, and } \bar{R}^2$) describe the acceptance of the protons in the RPs and are summarized in~table~\ref{tab:fiducialVol}. The parameters $R$ and $\bar{R}$ define the circles in the acceptance plots in the ($p_x$, $p_y$) momentum space. 

\begin{table}[htbp!]
        \centering
        \begin{tabular}{c|c|c|c|c}
             & EU & ED & WU & WD \\
            \hline
            $p_x^\text{min}$ (GeV) & -0.23 & -0.25 & -0.21 & -0.19 \\
            $p_y^\text{min}$ (GeV) & 0.42  & 0.48  & 0.46  & 0.46  \\
            $p_y^\text{max}$ (GeV) & 0.86  & 0.84  & 0.84  & 0.88  \\
            $p_x^\text{center}$ (GeV) & 0.64 & 0.7  & 0.6   & 0.7   \\
            $R^2$ (GeV$^2$) & 1.36 & 1.5  & 1.3   & 1.5   \\
            $\bar{p}_x^\text{center}$ (GeV) & 0.0  & -0.25 & -0.28 & 0.0   \\
            $p_x^\text{max}$ (GeV) & 0.0  & 0.06 & 0.03 & 0.0   \\
            $\bar{R}^2$ (GeV$^2$) & 0.0  & 0.959 & 0.946 & 0.0  \\
        \end{tabular}
 \caption[Summary of parameters defining the fiducial volume.]{The summary of parameters defining the fiducial volume.} 
  \label{tab:fiducialVol}
\end{table}

Since the cross section drops with increasing scattering angle (i.e. increasing $|p_y|$), most of the scattered protons are at low $|p_y|$. However, the probability of measuring a proton from the beam halo\footnote{The beam halo consists of particles traveling with the main beam, most of which are protons. These particles are typically located at transverse distances greater than $10\sigma$ of the beam size from the beam center.} increases as the RPs get closer to the beam axis (decreasing $|p_y|$). In order to maximize the acceptance, the fiducial volume is chosen to allow the transverse momentum component $|p_y|$ to be as small as possible for each branch of the RP separately, while simultaneously suppressing contamination from beam-halo. The selected fiducial volume therefore represents an optimal compromise between acceptance and background rejection. Hence, a part of yellow region in~figure~\ref{fig:RPFV}, representing the region with the highest statistics, is not used in this analysis.
\subsection{Primary vertex}
\label{sec:zCut}
In this analysis, tracks and primary vertices are reconstructed with the standard STAR procedures prior to this analysis. In order to suppress pileup from multiple inelastic interactions within the same bunch crossing, exactly one primary vertex is required per event. Primary vertices are reconstructed using tracks matched to TOF hits, ensuring that the vertex originates from the triggered bunch crossing. Events with more than one reconstructed primary vertex are therefore rejected.  

The primary vertex position is reconstructed with a resolution of $\sigma_{\mathrm{vtx}} \approx 0.1~\mathrm{cm}$ in each spatial coordinate. The longitudinal distribution of primary vertices is approximately Gaussian, centered at zero with a width of $\sigma_{z_{\mathrm{vtx}}} \approx 60~\mathrm{cm}$. The transverse size of the interaction is $\sigma_{x_{\mathrm{vtx}}} \approx 0.05$~cm and $\sigma_{y_{\mathrm{vtx}}} \approx 0.02$~cm. To ensure uniform detector acceptance, the primary vertex is required to be reconstructed within $\pm100~\mathrm{cm}$ of the nominal interaction point along the $z$ axis.

\subsection{TPC tracks}
\label{sec:tpc_tracks}
In order to select tracks in time with the bunch crossing and to extend the momentum range in which efficient PID can be provided for the central state, only tracks matching valid TOF hits are used. Therefore, tracks are required to have $\pT > 250$~MeV, where the TPC acceptance is high and uniform. This also ensures that the track can reach the TOF detector. Track quality criteria are imposed to achieve good momentum and energy loss resolution: a minimum of 20 hits in the TPC are required for the track reconstruction ($\nFit \geq 20$) out of which a minimum of 15 hits are required to calculate $\langle dE/dx\rangle$ ($\nDedx \geq 15$). In addition to the above, only tracks with a good match to the primary vertex are further analyzed. A track has a good match to a vertex if its distance of closest approach (DCA) to the vertex in the transverse plane ($\mathrm{DCA}_{xy}$) is smaller than 1.5 cm and in the $z$\nobreakdash-direction ($\mathrm{DCA}_z$) is smaller than 1.0 cm. The DCA resolution is $\sigma_{\mathrm{DCA}} \approx 0.2~\mathrm{cm}$ for tracks with $\pT = 250~\mathrm{MeV}$, both in the transverse plane and along the $z$ direction, and it improves with increasing track $\pT$, particularly in the $z$ direction.

To select CEP event candidates, only events with exactly two tracks satisfying the above criteria and with opposite charge are considered. To ensure high geometric acceptance and efficiency, the tracks are further required to have pseudorapidity within the fiducial acceptance defined by: 
    \begin{equation} \label{eq:etarange}
            (-\frac{1}{250}\zvtx - 0.9  < \eta <  -\frac{1}{250}\zvtx + 0.9) \land 
        (|\eta| < 0.9),
    \end{equation}
where $\eta$ is the track's pseudorapidity and $\zvtx$ is the $z$\nobreakdash-position of the primary vertex, in units of centimeters. The equation takes into account the change of the acceptance in pseudorapidity as a function of the $z$\nobreakdash-position of the primary vertex.   

\subsection{Exclusivity cut}
\label{sec:excl_cut}
Finally, an exclusivity requirement is imposed to select CEP event candidates. The detection and reconstruction of forward-scattered protons in the RP detectors, together with the central system, allow the calculation of the missing transverse momentum, defined as the sum of the transverse momenta of all measured particles:
    \begin{equation}\label{eq:pTMiss}
        p_{\textsf{T}}^{\text{miss}} \coloneqq |\left( \vec{p}_{p^{\mathrm{E}}} + \vec{p}_{h^+} + \vec{p}_{h^-} + \vec{p}_{p^{\mathrm{W}}} \right)_{\textsf{T}}|,
    \end{equation} 
where $\vec{p}_{p^{\mathrm{E}}}$ and $\vec{p}_{p^{\mathrm{W}}}$ denote the momenta of the forward protons measured on the east and west sides of the IP, while $\vec{p}_{h^+}$ and $\vec{p}_{h^-}$ correspond to the reconstructed momenta of the positively and negatively charged hadrons in the central detector.

\begin{figure}[htbp!]
        \centering
        \includegraphics[width = .8\linewidth, clip=true, trim=1pt 2pt 1pt 1pt]{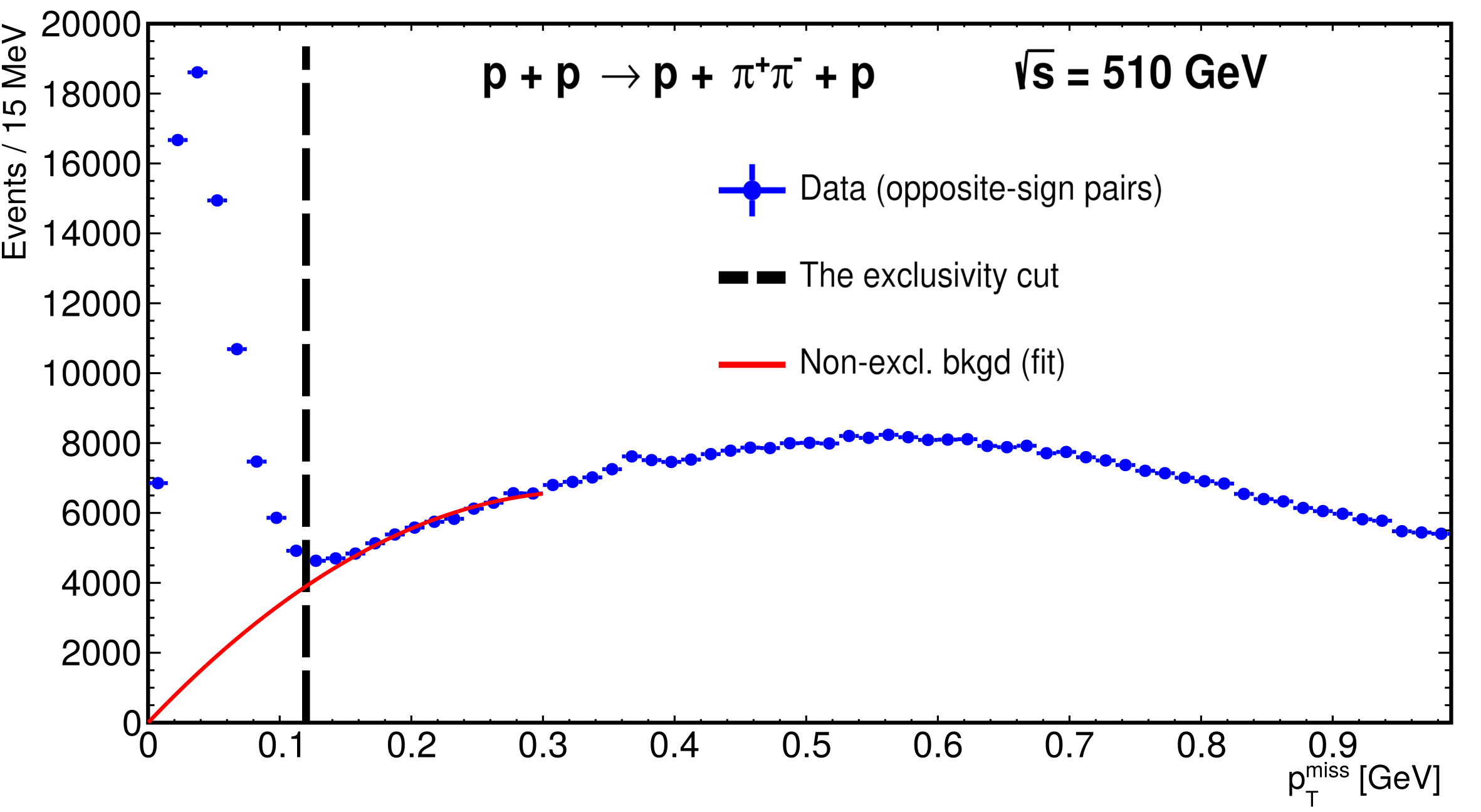}
        \caption[$\pTmiss$ for $\pi^+\pi^-$ CEP event candidates]{The distribution $\pTmiss$ for $\pi^+\pi^-$ CEP event candidates with the $\pTmiss$ cut illustrated by the black dashed line. The non-exclusive background is fitted using a second-degree polynomial with the constant term set to zero. The fit is depicted by the red line and projected into the signal region.}
        \label{06pTMiss}
    \end{figure}

In the CEP processes, the $\pTmiss$ should be equal to zero due to momentum conservation. Figure~\ref{06pTMiss} shows the $\pTmiss$ distribution of CEP event candidates before the exclusivity requirement. The peak at low $\pTmiss$ is due to the exclusive events. In each coordinate in the ($p_x^{\text{miss}}, p_y^{\text{miss}}$) space, the peaks are centered at zero. Due to the finite resolution in both $p_x^{\text{miss}}$ and $p_y^{\text{miss}}$, the distribution of $\pTmiss$ is shifted away from zero and peaks at a non-zero value. Its width is predominantly determined by the beam angular divergence of 90~$\mu$rad%
\footnote{
For a 255 GeV beam, a beam angular divergence of 90~$\mu$rad corresponds to a smearing of approximately 
$\sigma_{p_{x,y}}^{\text{divergence}} \approx 23$~MeV for each transverse component of the forward proton momentum. 
The intrinsic RP momentum resolution contributes at the level of $\sim$10~MeV and is therefore subdominant. 
The resulting forward-proton transverse-momentum resolution is $\sigma_{p_{x,y}}^{\text{proton}} \approx 25$~MeV, leading to $\sigma(\pTmiss) \approx 35$~MeV. 
The contribution from the momentum resolution of the central system is significantly smaller and is therefore neglected.}.
Events are selected if they have $\pTmiss < 120$~MeV as illustrated by the black dashed line in~figure~\ref{06pTMiss}. The red line is the fit and its extrapolation into the signal region.

\subsection{Particle identification} 
\label{sec:particle_identification}
The PID is performed using information from the TPC, particle's $\langle dE/dx\rangle$, and its time of flight in the TOF detector. The combined information is used to identify hadron pairs: $\pi^+\pi^-, K^+K^-, \text{ or } p\bar{p}$. 

\begin{figure}[htbp!]
        \centering
        \includegraphics[width = .49\linewidth]{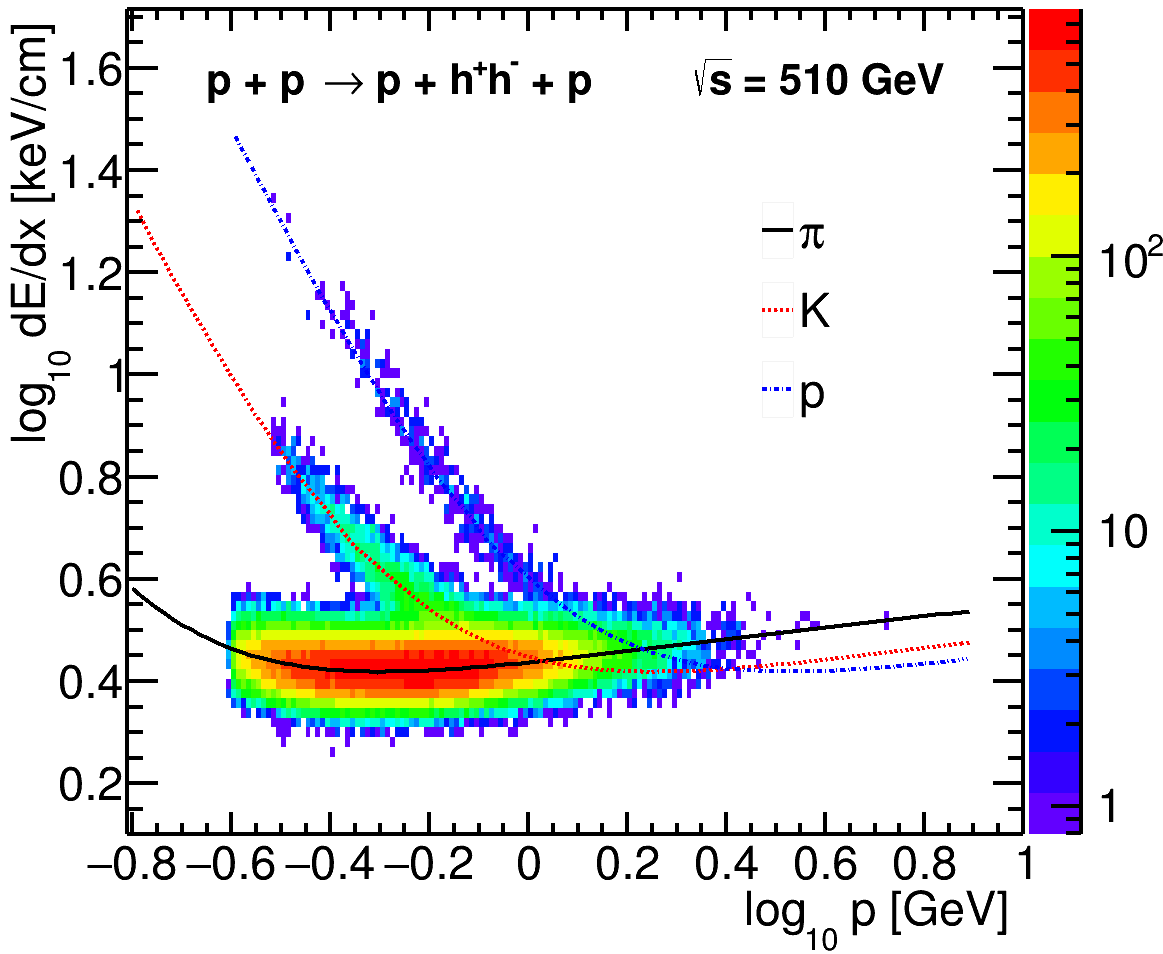}
        \hfill
        \includegraphics[width = .49\linewidth]{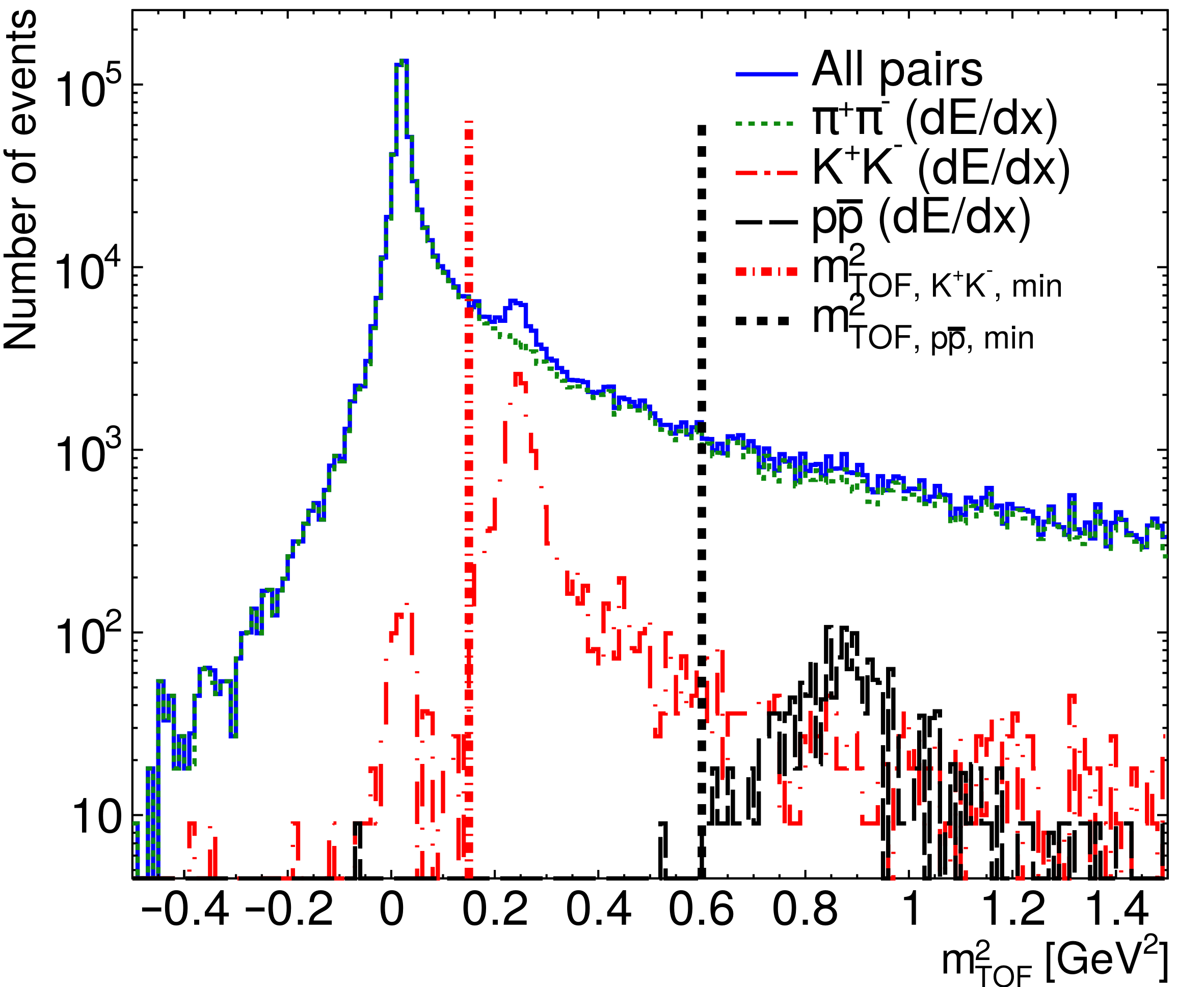} 
        \caption[$\langle dE/dx\rangle$ and $m^2_{\text{TOF}}$ of the central system]{Left: the energy loss $\langle dE/dx\rangle$ of central charged particles as a function of their momenta for CEP event candidates. Colored curves illustrate expected values for pions, kaons, and protons. Right: distributions of invariant mass squared $m^{2}_{\text{TOF}}$ of pions, kaons, and protons, based solely on their $dE/dx$ information. Applied $m^2_{\text{TOF},K^+K^-,\text{min}}$ and $m^2_{\text{TOF},p\bar{p},\text{min}}$ cuts are highlighted by dot-dashed lines.}
        \label{06dedx}
    \end{figure}

The $dE/dx$ component of the PID is quantified by defining a $\chi^2$ for a given hadron-pair hypothesis, defined as:
    \begin{equation}\label{eq:dEdxChisq}
        {\chi}^{2}_{\langle dE/dx\rangle}(h^+h^-) = \left(n \sigma_{h^+} \right)^2 + \left(n \sigma_{h^-} \right)^2,
    \end{equation} 
where $n \sigma_h$ is the number of standard deviations between the measured and the expected energy loss~\cite{Bichsel:2006cs} for a hadron, $h$. Figure~\ref{06dedx} (left) shows the log of the $\langle dE/dx\rangle$ of charged particles as a function of the log of their momentum. Pions, kaons, and protons are clearly distinguishable, with colored curves indicating their expected values.

To extend the PID capabilities of the $\langle dE/dx\rangle$ method, the TOF information is used. However, the standard method based on the inverse particle velocity ($1/\beta$) cannot be used for CEP events, which is due to lack of the VPD start time of the collision. Instead, the difference in the time of flight of each of the two hadrons can be used for identifying particle pairs without an explicitly measured starting time. The method assumes that both tracks originate from the same vertex and have the same mass. Hence, they have the same squared mass, $m^2_{\text{TOF}}$, that can be derived from the measured TOF time difference between particles:
    \begin{equation} \label{eq:msquared}
        t_2 - t_1 = l_2 \sqrt{1 + \frac{m^2_{\text{TOF}}}{p_2^2}} - l_1 \sqrt{1 + \frac{m^2_{\text{TOF}}}{p_1^2}},  
    \end{equation}
where $t_{1,2}$ are the time of the tracks' detection in TOF, while $l_{1,2}$ and $p_{1,2}$ denote the track lengths and momenta of the particles, respectively, as determined from their trajectories measured by the TPC.

The $m^2_{\text{TOF}}$ distributions of pions, kaons, and protons, identified solely based on $\langle dE/dx\rangle$ information, can be seen in~figure~\ref{06dedx} (right). The $m^2_{\text{TOF},p\bar{p},\text{min}} = 0.6$ GeV$^2$ and $m^2_{\text{TOF},K^+K^-,\text{min}} = 0.15$ GeV$^2$ are the minimum invariant mass squared values required for $p\bar{p}$ and $K^+K^-$ pairs in the selection. Peaks close to each particles’ invariant mass squared values are seen. Due to the resolution effects, negative, nonphysical values can be seen. 

Particle pairs are identified based on the following hypothesis. First, the $p\bar{p}$ pair hypothesis is checked:
    \begin{equation}
        ({\chi}^{2}_{\langle dE/dx\rangle}(p\bar{p}) < 9) \land (m^2_{\text{TOF}} > 0.6 \text{ GeV}^2) \land ({\chi}^{2}_{\langle dE/dx\rangle}(K^+K^-) > 9) \land ({\chi}^{2}_{\langle dE/dx\rangle}(\pi^+\pi^-) > 9).
        \label{PID:proton}
    \end{equation} 

If the hypothesis is satisfied, the pair is identified as a $p\bar{p}$ pair, and if not, the $K^+K^-$ pair hypothesis is tested:
    \begin{equation}
        ({\chi}^{2}_{\langle dE/dx\rangle}(K^+K^-) < 9) \land (m^2_{\text{TOF}} > 0.15\text{ GeV}^2) \land ({\chi}^{2}_{\langle dE/dx\rangle}(p\bar{p}) > 9) \land ({\chi}^{2}_{\langle dE/dx\rangle}(\pi^+\pi^-) > 9).
        \label{PID:kaon}
    \end{equation} 

If neither $p\bar{p}$ nor $K^+K^-$ pair hypothesis is satisfied, the $\pi^+\pi^-$ pair is assumed and the $\pi^+\pi^-$ pair hypothesis is checked: 
    \begin{equation}
        {\chi}^{2}_{\langle dE/dx\rangle}(\pi^+\pi^-) < 12.
        \label{PID:pion}
    \end{equation}

This PID technique is designed to minimize misidentification of $\pi^+\pi^-$ pairs as $K^+K^-$ or $p\bar{p}$ pairs. In addition, the PID is restricted to a fiducial acceptance with high track reconstruction efficiency and high pair identification efficiency. The misidentified pairs form a negligible exclusive background ($< 1\%$), as discussed in~section~\ref{chap:Backgrounds}. The fiducial acceptance of the central hadronic state is defined as:
\begin{equation}
    \label{PID:pionRestriction}
    (\pT(\pi^+) > 0.25 \text{ GeV}) \land (\pT(\pi^-) > 0.25 \text{ GeV}),
\end{equation}
\begin{equation}
    \label{PID:kaonRestriction}
    (\pT(K^+) > 0.3 \text{ GeV})  \land (\pT(K^-) > 0.3 \text{ GeV})  \land  (\text{min}(\pT(K^+), \pT(K^-)) < 0.7 \text{ GeV}),
\end{equation}
\begin{equation}
    \label{PID:protonRestriction}
    (\pT(p) > 0.4 \text{ GeV})  \land (\pT(\bar{p}) > 0.4 \text{ GeV})  \land  (\text{min}(\pT(p), \pT(\bar{p})) < 1.1 \text{ GeV}).
\end{equation}

After all the above selection cuts, there are 86008 $\pi^+\pi^-$, 2454 $K^+K^-$, and 225 $p\bar{p}$ CEP events.

\section{Background estimation}
\label{chap:Backgrounds}
Both exclusive and non-exclusive backgrounds to the signal CEP-events are considered.

The exclusive background comes primarily from incorrectly identified particles species. This occurs when the event passes all the selection criteria but is misidentified due to the incorrectly identified particle species. The dominant contribution to this background comes from $\pi^+\pi^-$ pairs misidentified as $K^+K^-$ or $p\bar{p}$ pairs. The phase space of the measurement is limited to preserve high pair identification efficiency and low probability of misidentification. This background is estimated using the PID efficiency evaluated for hadron pairs as a function of $(\pT^{\text{max}}, \pT^{\text{min}})$, defined as the larger and smaller transverse momenta of the two hadrons. The true yield is obtained by correcting the measured transverse-momentum distributions for the corresponding PID probabilities. The resulting contribution of exclusive background is found to be $< 1\%$, and therefore negligible. Photoproduction can also be an exclusive background. In most cases, the transverse momentum of the protons originating from
the photon vertex is too small to be detected within the fiducial acceptance of the RP detectors, but there is a non-negligible tail to higher values. In addition, events with proton dissociation may contribute~\cite{LangPDF}. These sources of background have not been precisely evaluated.

The non-exclusive background is expected to be the dominant source of background. It arises from non-exclusive events that imitate the $h^+h^-$ topology of CEP: two forward protons, two opposite sign central tracks and large rapidity gaps. The most common sources of the non-exclusive background are: 

\begin{enumerate}
    \item Accidental coincidences (pile-up): there can be accidental coincidences of the forward and central system. For example, the overlap of elastic scattering with another inelastic interaction in the central system can imitate a CEP event. Or, it can also be an overlap of a single diffraction event with a proton from the beam halo.
    \item Central Diffraction: for example, a CEP event with a higher number of produced particles, where only two are detected. 
\end{enumerate}
    
The non-exclusive background subtraction is carried out using missing transverse momentum. The method described here is described for the invariant mass of $\pi^+\pi^-$, $m(\pi^+\pi^-)$. The same procedure is used for all other observables studied in this analysis.

\begin{enumerate}
    \item The amount of non-exclusive background is estimated in two steps. First, the $\pTmiss$ distribution (see~figure~\ref{06pTMiss}) is fitted in the sideband region with a second-degree polynomial with the constant term fixed to zero. The sideband region is [140, 300]~\text{MeV} for $\pi^+\pi^-$ pairs and [140, 350]~\text{MeV} for $K^+K^-$ and $p\bar{p}$ pairs. Second, the fit is extrapolated into the signal region to obtain the amount of background. 
    \item The background template is taken from the $m(\pi^{+}\pi^{-})$ distribution of events in the sideband region defined by $\pTmiss \in [160, 220]~\text{MeV}$, i.e. just outside the signal window. Events in this interval are assumed to have the same background mass shape as those in the signal region, and are therefore used to model the background contribution.
    \item The obtained background distribution is normalized to the expected background amount determined in the first step.
    \item Finally, the normalized background distribution is subtracted separately from the $m(\pi^+\pi^-)$ distribution for each bin of the distribution.
\end{enumerate}

Using this method, the amount of non-exclusive background is found to be $19.4\pm0.4$\% for $\pi^+\pi^-$ pairs, $12.1\pm0.8$\% for $K^+K^-$ pairs, and $43\pm5$\% for $p\bar{p}$ pairs.
\section{Corrections}
\label{chap:Corrections}
The differential fiducial cross section in every bin of the measured quantity of interest, $q$, is obtained using the formula~\eqref{eq:corr.1}:

\begin{equation}
\frac{d\sigma}{dq}(q) = \frac{1}{\Delta q} \times \frac{N^{\text{w}}(q) - N^{\text{w}}_{\text{bkgd}}(q)}{L_{\text{int}}^{\text{eff}}},
\label{eq:corr.1}
\end{equation}
where \(\Delta q\) is the width of the bin, \(N^{\text{w}}(q)\) and \(N^{\text{w}}_{\text{bkgd}}(q)\) are the weighted numbers of observed and background events in the given bin, respectively. The weight ($w_i$) is calculated as:
\begin{equation}
w_i 
= \frac{1}{\prod_k \varepsilon_{k,i}},
\end{equation}
where $\varepsilon_{k,i}$ are individual efficiencies for the given $i^{\text{th}}$ event. The minimal event weight is 0.024 for $\pi^+\pi^-$ pairs, 0.011 for $K^+K^-$ pairs, and 0.044 for $p\bar{p}$ pairs. Hence, the weighted numbers of observed events is calculated as:
\begin{equation}
N^{\text{w}}(q) 
= \sum_{i\in\text{Events}}w_i.
\end{equation}
The effective integrated luminosity \(L_{\text{int}}^{\text{eff}}\) is  defined as:

\begin{equation}
L_{\text{int}}^{\text{eff}} = \sum_{\text{run}} L_{\text{int}}^{\text{run}} \times P^{\text{CEP}}_{\text{retain}}(\langle L_{\text{inst}}^{\text{run}} \rangle),
\label{eq:corr.2}
\end{equation}
where $L_{\text{int}}^{\text{run}}$ is the integrated luminosity for the given run and $P^{\text{CEP}}_{\text{retain}}(\langle L_{\text{inst}}^{\text{run}} \rangle)$ is the probability of retaining the CEP event as a function of the average instantaneous luminosity ($\langle L_{\text{inst}}^{\text{run}} \rangle$).

\subsection{Trigger and luminosity corrections}
\label{sec:luminosity}
The integrated luminosity has to be corrected to account for the probability of retaining CEP events, which depends on the instantaneous luminosity. There is a high probability that an exclusive event overlaps with another process and is therefore rejected. This occurs because the CEP trigger includes vetoes ensuring the characteristic double-gap rapidity topology of CEP events. 

The probability of retaining CEP events is calculated using zero-bias data, taking into account both online and offline vetoes. It quantifies how often a zero-bias event is not rejected by the veto. The online vetoes were part of the trigger definition as described in~section~\ref{sec:triggers}. The offline veto comes from the requirement of one reconstructed primary TOF vertex. The probability is studied separately for each run\footnote{A run at STAR refers to a dataset collected during a specific time interval when data was recorded; this typically lies between 30 minutes and 2 hours.} as a function of the instantaneous luminosity. The obtained distribution is fitted with an exponential function, that reflects the expected Poisson-like behavior of the probability that no interaction occurs in a bunch crossing. The probability of retaining the CEP event is found to be independent of the choice of the branch of the RP combination. The values of $P^{\text{CEP}}_{\text{retain}}(\langle L_{\text{inst}}^{\text{run}} \rangle)$ vary between $15-40$\%, which leads to the effective integrated luminosity of $L_{\text{int}}^{\text{eff}} = 25.6\pm1.6$ pb$^{-1}$.

\subsection{TPC and TOF corrections}
To determine the TPC and TOF detector efficiencies, an embedding technique is used to simulate the data taking environment. First, Monte Carlo (MC) generated particles are processed using a GEANT3~\cite{Brun:1119728} detector simulation implementing STAR geometry. Second, the output, which mimics raw data from collisions, is superimposed on zero-bias data representing an underlying event. 

The TPC track acceptance, reconstruction, and selection efficiencies are studied together in order to determine an overall TPC efficiency. This efficiency is calculated by tracking how many MC particles are properly reconstructed from the embedded sample. The corrections are calculated as function of ($\pT, \eta, \zvtx$). The typical TPC efficiencies for single hadron are 83\% for $\pi^\pm$, 63\% for $K^\pm$, and 78\% for $p$ and $\overline{p}$. The typical statistical uncertainties on those values are $<1$\%. The efficiencies for kaons are lower due to weak decays of kaons in front of, or inside, the TPC. 

The TOF acceptance, hit reconstruction and TPC track matching to a TOF hit efficiencies are studied together as a TOF efficiency. The efficiency is determined from embedding and compared with a data-driven tag-and-probe method based on decays of $K^0_S$ to $\pi^+\pi^-$. This $K^0_S$ decay channel is selected as it is well established and provides a narrow invariant mass peak clearly separated from the background. The $K^0_S$ is reconstructed using TPC tracking alone. One of the pions is tagged by requiring both TPC and TOF information, while the other pion is probed after being measured in the TPC and tested for its TOF response. The TOF efficiency for the probed track \(\epsilon_{\mathrm{TOF}}^{\mathrm{probe}}\) is calculated as the ratio of two different yields of $\pi^+\pi^-$ pairs satisfying criteria for $K^0_S$ candidates:
\begin{equation}
    \epsilon_{\text{TOF}}^{\text{probe}} = \frac{N_{\text{TOF}}^{\text{both}}}{N_{\text{TOF}}^{\text{tag}}},
\end{equation}
where \(N_{\text{TOF}}^{\text{both}}\) and \(N_{\text{TOF}}^{\text{tag}}\) are the yields of \(K^0_S\) candidates where both pion tracks have TOF hits and where only the tagged pion track is matched to the TOF, respectively. The typical TOF efficiencies are 63\% for $\pi^\pm$, 60\% for $K^\pm$, and 59\% for $p$ and $\overline{p}$.

\subsection{PID corrections}
The PID efficiency is defined as the probability that the hadron ($h^+h^-$) pair passes the corresponding pair selection criteria discussed in~section~\ref{sec:particle_identification}. The same embedded samples that are used for TPC and TOF detector efficiencies are used to determine the PID efficiencies. These are studied as a function of the maximum and minimum true transverse momenta of particles in the pair ($\pT^{\text{max}}, \pT^{\text{min}}$). The efficiency is applied as a function of the ($\pT^{\text{max}}, \pT^{\text{min}}$).
The average PID efficiency for $\pi^+ \pi^-$ pairs is higher than 99\%. At higher values of $\pT^{\text{min}}$, the efficiencies for $K^+ K^-$ and $p\overline{p}$ drop significantly due to the requirement of $\chi^2_{\langle dE/dx\rangle} (\pi^+ \pi^-) > 9$, which is used to limit misidentification of $\pi^+ \pi^-$ pairs in the $K^+ K^-$ and $p\overline{p}$ samples. The typical efficiencies for $K^+ K^-$ and $p\overline{p}$ pairs are $93$\% and $95$\%, respectively.

\subsection{Vertex corrections}
\label{sec:vertexCorr}
The criterion on the TPC $z$-vertex position, discussed in~section~\ref{sec:zCut}, reduces the accepted luminosity relative to the luminosity delivered by the collider.
In order to account for this loss, $z$-positions of primary vertices for each RHIC fill are independently fitted by a normal distribution. Then, the efficiency is estimated based on the obtained values of the mean and standard deviation. Another method is also used. This involves calculating the fraction of lost luminosity directly from the $\zvtx$ distribution. The difference between the two methods is about 2\% and is independent of the fill number. Therefore, the average of the two efficiency corrections is used to correct the data. A typical vertex-cut efficiency is 90\%.

In order to determine the TPC vertex reconstruction efficiency, a data-driven method is used. A data set, which contains tracks that are not required to be associated with the primary vertex is used. Two such tracks that form a CEP-like topology are examined to determine if they form a good primary vertex. These events are required to have exactly two good-quality TPC tracks matched with TOF. The same good-quality track criteria as described in~section~\ref{sec:tpc_tracks} are applied. The efficiency is determined as follows:
\begin{equation}
    \epsilon_{\text{vertex}} = \frac{N_{\text{vertices}}^{\text{reco}}}{N_{\text{vertices}}^{\text{examined}}},
\end{equation}
where \(N_{\text{vertices}}^{\text{reco}}\) and \(N_{\text{vertices}}^{\text{examined}}\) are the number of successfully reconstructed and examined vertices, respectively. The average TPC vertex reconstruction efficiency is $89.6$\%.

\subsection{Exclusivity cut corrections}
\label{sec:ptMissEff}
A small fraction of CEP events is rejected by the exclusivity requirement,
$\pTmiss < 120$~MeV, as defined in Section~\ref{sec:excl_cut}. The fraction of CEP events that satisfy the $\pTmiss$ cut and is evaluated using both data-driven and MC methods.

In the MC-based approach, the missing transverse momentum components,
$p_x^{\mathrm{miss}}$ and $p_y^{\mathrm{miss}}$, are modeled using a
data-driven parameterization. The $p_x^{\mathrm{miss}}$ and
$p_y^{\mathrm{miss}}$ distributions measured in data are each fitted with a
function composed of a second-degree polynomial and a Gaussian term. The
polynomial component describes the non-exclusive background contribution,
while the Gaussian term represents the exclusive CEP signal. The parameters
of the Gaussian component obtained from the fits are used to generate
$p_x^{\mathrm{miss}}$ and $p_y^{\mathrm{miss}}$ in the MC, from which the
$\pTmiss$ distribution is calculated and the efficiency is
determined as a fraction of CEP events that satisfy the $\pTmiss$ cut.

An independent, fully data-driven estimate of the exclusivity efficiency is
also performed. This method relies on the assumption that the signal
contribution is negligible for $\pTmiss > 200$~MeV. The $\pTmiss$ distribution
in data is fitted with a second-degree polynomial in the range up to
200~MeV, which is taken to represent the non-exclusive background. The total signal yield is obtained by integrating the $\pTmiss$ distribution
in data in the range $0 < \pTmiss < 200$~MeV and subtracting the background contribution estimated from a second-degree polynomial fit. The same procedure is repeated in the interval $0 < \pTmiss < 120$~MeV, corresponding to the exclusivity requirement. The exclusivity efficiency is finally calculated as the ratio of the signal yield
passing the $\pTmiss < 120$~MeV cut to the total signal yield obtained in the
range $0 < \pTmiss < 200$~MeV.

The final exclusivity efficiency is taken as the average of the efficiencies
obtained from the data-driven and MC-based methods and is found to be above 99\%.

\subsection{RP corrections}
In order to calculate the RP acceptance and track reconstruction and selection efficiency, a dedicated tool (\textit{ppSim}) based on GEANT4~\cite{AGOSTINELLI2003250} is developed to simulate the response of the RP detectors, as described in Refs.~\cite{Wlodek, STAR:2023vdw}. It includes full implementation of the beamline elements. It also accounts for the background contribution from scattered protons interacting with the material in front of the RPs, and inoperable SVX readout chips~\cite{STAR:2023vdw}. In order to fully reproduce the collision environment, a single proton is generated with total energy of 254.9 GeV. The proton momentum is smeared by the beam angular divergence and propagated through the beamline to the RPs. The \textit{ppSim} output is embedded in the zero-bias events. 

There are two parts to the RP efficiency: the detector efficiency and the proton selection efficiency. The detector efficiency for a given branch of the RP describes the probability that a proton is measured and reconstructed in that branch of the RP. It is calculated as the probability that a single good-quality RP track is reconstructed in the branch of the RP. This track must be matched with a true-level primary forward proton. The typical RP detector efficiency is about 98\%. That efficiency is compared with a data-driven method, which is based on reconstruction of elastic events. The efficiencies agree within $2\sigma$ of statistical uncertainties. The proton selection efficiency reflects the probability that the reconstructed proton will be selected by the reconstruction algorithm. Any additional background from pile-up or noise will lower the proton selection efficiency. The typical value of the RP efficiency for a given branch of the RP is about 88\%.

Efficiencies quoted above have statistical uncertainties at a fraction of a percent level. Hence, they are negligible compared to the statistical uncertainties of the results. Therefore, they are neither quoted nor propagated.
\section{Systematic uncertainties}
\label{chap:SysUncertainties}

In this section, we describe how systematic uncertainties are obtained. We find that only the background subtraction systematic uncertainty is bin dependent for all presented results. All other systematic uncertainties are not bin dependent. The following systematic uncertainties are evaluated for each presented distribution separately:
\begin{enumerate}
    \item Background subtraction: the systematic uncertainty related to the non-exclusive background subtraction is discussed in~section~\ref{chap:Backgrounds}. It is studied by varying the range of the projection: from 170 to 210 MeV (smaller range) and from 140 to 250 MeV (wider range). The subtraction is performed for the ranges described previously. The obtained fiducial differential cross sections for CEP of $h^+h^-$ pairs as a function $m(h^+h^-)$ are compared with the nominal one. The uncertainty is taken as the average of absolute deviations for each bin separately. There is another source of the systematic uncertainty, which is due to the statistical uncertainty of the size of the non-exclusive background sample. This uncertainty is common for all the bins. The weighted mean of uncertainties of the background subtraction in the invariant mass distribution of $\pi^+\pi^-$, $K^+ K^-$, and $p \overline{p}$ pairs are 0.4\%, 0.8\%, and 3.3\%, respectively. Only this systematic uncertainty is bin dependent for all presented results.

    \item RP efficiency correction: the systematic uncertainty associated with the RP efficiency corrections for a single proton is studied from run-by-run variations in each branch of the RP separately. An average variation is approximately 1.5\% for a single proton, resulting in the total RP systematic uncertainty on the fiducial differential cross sections being 2.4\%.
    \item TPC efficiency correction: the uncertainty is studied by varying the TPC track selection criteria ($\nFit$ and $\nDedx$) and applying the TPC efficiency corrections corresponding to the given set of selection criteria. The obtained fiducial differential cross sections are compared with the cross section calculated with corrections as described in~section~\ref{chap:Corrections}. The typical  uncertainties on the fiducial differential cross sections are $^{+7}_{-4}$\%, $^{+7}_{-4}$\%, and $^{+4}_{-4}$\% for $\pi^+ \pi^-$, $K^+ K^-$, and $p\overline{p}$ pairs, respectively. 
    \item TOF efficiency correction: the uncertainty is studied by comparing the corrections obtained from the embedding and the data-driven tag-and-probe method based on $K_S^0 \rightarrow \pi^+\pi^-$ decays. The difference between the embedding and the tag-and-probe method is found to be about 1.0\% per single track. In addition, the corrections from the embedding are calculated as a function of ($\pT, \eta, \zvtx$) with a different binning applied. The obtained fiducial differential cross sections are compared with the cross section calculated with corrections as described in~section~\ref{chap:Corrections}. This method of applying the TOF matching efficiency corrections allows to estimate the related source of uncertainty. The typical total TOF uncertainty on the fiducial differential cross sections is $^{+1}_{-2}$\% for $\pi^+ \pi^-$, $K^+ K^-$, and $p\overline{p}$ pairs.  
    \item TPC vertex reconstruction: the systematic uncertainty is determined by varying the good primary vertex selection criteria in $\mathrm{DCA}_{xy}$ and $\mathrm{DCA}z$, and by applying the corresponding efficiency corrections to the given set of vertex criteria. The nominal selections are $\mathrm{DCA}{xy} < 1.5$ cm and $-1.0 < \mathrm{DCA}z < 1.0$ cm. These are varied to $\mathrm{DCA}{xy} < 2.0$ cm ($1.25$ cm) and $-1.5 < \mathrm{DCA}_z < 1.5$ cm ($-0.8 < \mathrm{DCA}z < 0.8$ cm) to assess the systematic effect. The resulting uncertainty on the fiducial differential cross section is $^{+2.6}_{-1.0}\%$.
    \item TPC $z$-vertex criterion: the uncertainty is calculated as half of the difference between the two methods used as discussed in~section~\ref{sec:vertexCorr}. The systematic uncertainty is determined to be independent of the fill and is 1.1\%. 
    \item Exclusivity $\pTmiss$ cut: the systematic uncertainty is $0.4\%$ obtained as the difference between efficiencies obtained using two different methods discussed in section~\ref{sec:ptMissEff}.
    \item Particle identification: the uncertainty is studied by varying the pair identification criteria. A typical change of $<1\%$, $^{+4}_{-5}$\% and $^{+4}_{-5}$\% in the fiducial differential cross section is observed for $\pi^+ \pi^-$, $K^+ K^-$, and $p\overline{p}$ pairs, respectively. This change results from applying looser and tighter identification criteria and correcting for the corresponding PID efficiency. 
    The looser (tighter) condition replace the value of 9 by 12 (7) in~\eqref{PID:proton} and in~\eqref{PID:kaon} and the value of 12 by 15 (9) in~\eqref{PID:pion}. The values for the cuts $m^2_{\text{TOF}}$ are 0.1 (0.2) and 0.55 (0.7) GeV$^2$ for the kaon and proton hypotheses, respectively.
\item The luminosity uncertainty. There are two parts of the luminosity uncertainty: luminosity calibration based on van der Meer scans~\cite{vanderMeer:1968zz} and the probability of retaining a CEP event, dependent on instantaneous luminosity. The first of these, contributes to the integrated luminosity uncertainty and is found to be $5\%$~\cite{ref:Angelika}. The second contribution is evaluated by propagating the uncertainties of the fitted exponential function described in section~\ref{sec:luminosity} and amounts to 4\%. Thus, the total uncertainty on effective integrated luminosity~\eqref{eq:corr.2} is $6.4\%$. This systematic vertical scale uncertainty is not plotted in figures~\ref{fig:deltaPhi}--\ref{fig:tSumDiffInvMass}.
\end{enumerate}

Systematic uncertainties for the integrated fiducial cross sections for CEP of $\pi^+\pi^-$, $K^+K^-$, and $p\bar{p}$ pairs are shown in~table~\ref{Tab:intCrossSectionUnct}.

\begin{table}[htbp!]
    \centering
    \begin{tabular}{ccccc|c}
        \multicolumn{6}{c}{$\delta_{\text{syst}}/\sigma_{\text{fid}}$ [\%]} \\
        Particle species & TPC & TOF & RP & Other & Total \\ \hline
        \rule{0pt}{14pt} $\pi^+ \pi^-$ & $^{+7}_{-4}$ & $^{+1}_{-2}$ & 2.4 & $^{+3}_{-2}$ & $^{+8}_{-5}$ \\ [0.3em]
        $K^+ K^-$ & $^{+7}_{-4}$ & $^{+1}_{-2}$ & 2.4 & $^{+5}_{-4}$ & $^{+9}_{-7}$ \\ [0.3em]
        $p \bar{p}$ & $^{+4}_{-4}$ & $^{+1}_{-2}$ & 2.4 & $^{+5}_{-4}$ & $^{+7}_{-6}$  \\
    \end{tabular}
    \caption[Systematic uncertainties for the integrated fiducial cross section for CEP of $\pi^+\pi^-$, $K^+K^-$, and $p\bar{p}$ pairs.]{Systematic uncertainties for the integrated fiducial cross section for CEP of $\pi^+\pi^-$, $K^+K^-$, and $p\bar{p}$ pairs. The numbers represent the relative systematic uncertainty of the integrated fiducial cross section in percentage.}
    \label{Tab:intCrossSectionUnct}
\end{table}

\section{Results}
\label{chap:Results}

In this section, the results on differential fiducial cross sections and the integrated fiducial cross sections are presented. The fiducial volume of the measurement is common for all the presented results and it is defined in~\eqref{eq:fiducialVolume} for the protons in the RPs, and depending the specific central hadronic state in~\eqref{eq:etarange}, and~\eqref{PID:pionRestriction}--\eqref{PID:protonRestriction}. By definition, the reported fiducial cross sections are restricted to the experimentally accessible phase space of the STAR detector and do not represent full phase space cross sections.

\subsection{Differential fiducial cross sections}
In this section, differential fiducial cross sections are presented and compared with predictions from GRANIITTI~\cite{Mieskolainen}, an MC event generator designed for high energy diffraction with focus on the CEP. It combines up-to-date phenomenological models and approaches~\cite{Ewerz2014, HarlandLang20140626, Lebiedowicz2016}. GRANIITTI is the only model that includes a full parametric resonant spectrum, and continuum production with significant interference effects between them. In this analysis, the newest version of GRANIITTI v.~1.090 is compared with our results. That version was tuned to the latest STAR CEP results at $\sqrt{s} = 200$ GeV~\cite{Rafal20,Graniitti2023} and is in good agreement with those published results.

GRANIITTI predictions are calculated including both continuum (Cont.) and resonance (Res.) contributions, and the data are compared to predictions obtained using the pure continuum alone as well as the combined resonance-plus-continuum model (Res.+Cont.). The resonances used in the model are summarized in~table~\ref{Tab:graniitti}. This is a minimum set of resonance giving the best description of the data without additional tuning. Subsequently, the predictions are scaled to match the integrated cross section. If the scaling factor is within 10\% of unity, then no scaling is applied, and factor $\times1.0$ is used in the legend. The scaling is done only for plotting purposes and has no physical interpretation. 
Since GRANIITTI is in good agreement with results at $\sqrt{s} = 200$ GeV~\cite{Rafal20,Graniitti2023}, there is a need for additional tuning of GRANIITTI using results presented here. This may be needed because the two data sets are at different energies and $t$-ranges, hence two different fiducial regions of the phase space. Moreover, the pure continuum is shown without scaling applied for $\pi^+\pi^-$ and $K^+K^-$ pairs and with scaling for $p\Bar{p}$ pairs. The hadron pairs are smeared based on the $\pT$ resolution and the same fiducial criteria for forward protons and central hadrons as described in~section~\ref{chap:EventSelection} are applied.

\begin{table}[htbp!]
    \centering
    \begin{tabular}{c|c}
        Particle species & Resonances  \\
        \hline
        $\pi^+ \pi^-$ & $f_0(980)$, $f_2(1270)$, and $f_0(1710)$ \\
        $K^+ K^-$ & $f_0(980)$, $f_0(1500)$, $f_2(1525)$, and $f_0(1710)$  \\
        $p \bar{p}$ & only continuum  \\ 
    \end{tabular}
    \caption[The summary of resonances used in the calculation of GRANIITTI predictions.]{The summary of resonances used in the calculation of GRANIITTI~\cite{Mieskolainen} predictions for CEP of $\pi^+ \pi^-$, $K^+ K^-$, and $p \bar{p}$ pairs.}
    \label{Tab:graniitti}
\end{table}

\begin{figure}[htbp!]
\centering
\includegraphics[width = .325\linewidth]{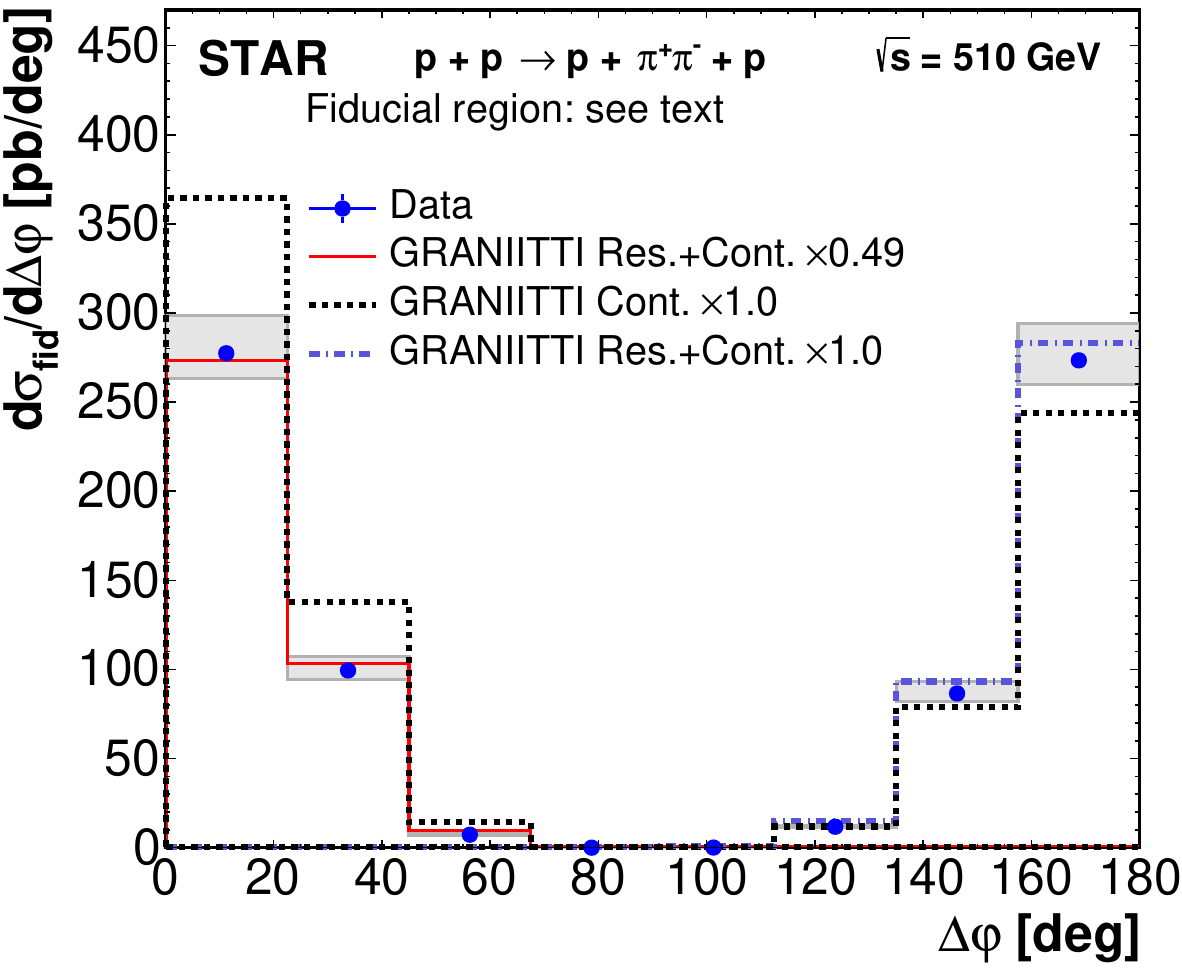}
\hfill
\includegraphics[width = .325\linewidth]{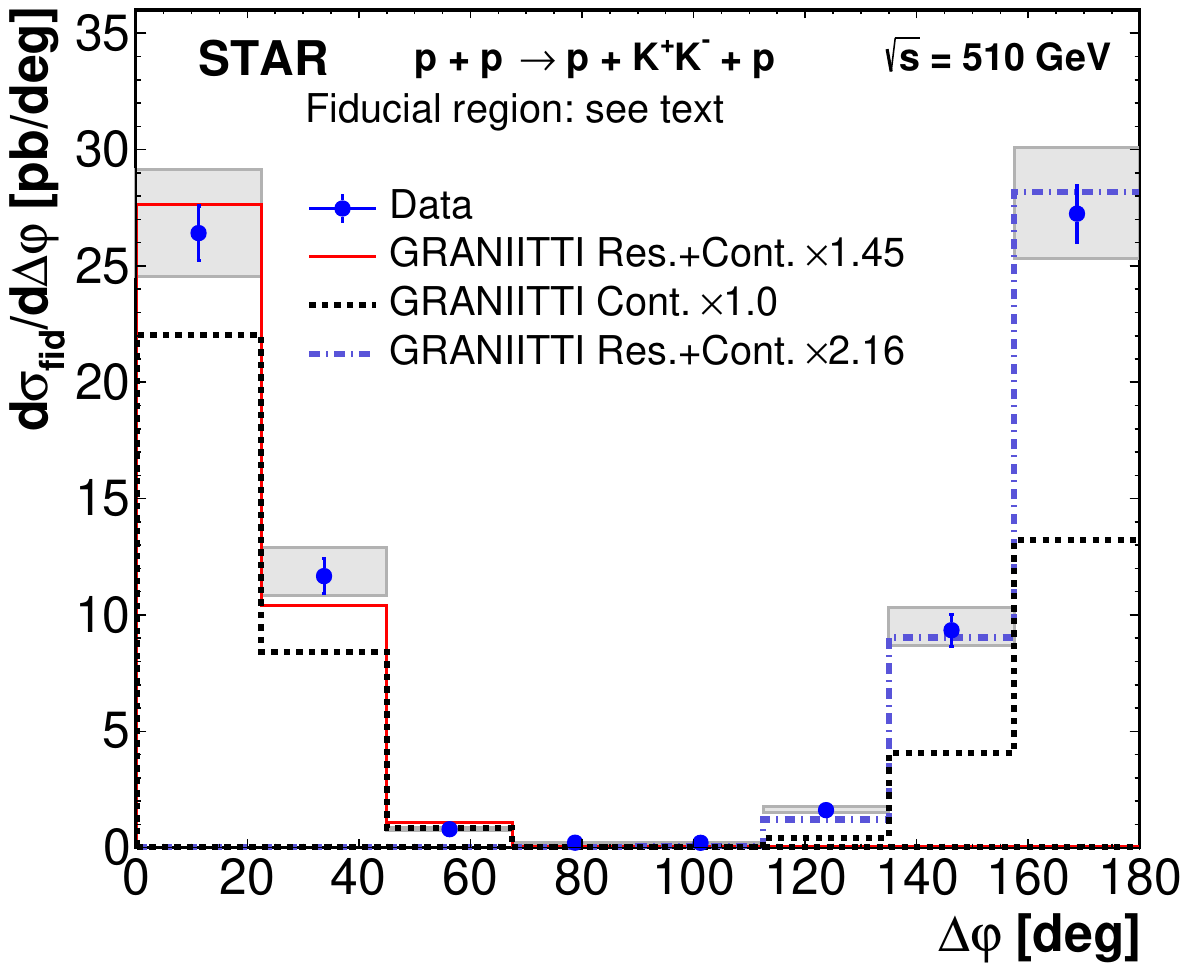}
\hfill
\includegraphics[width = .325\linewidth]{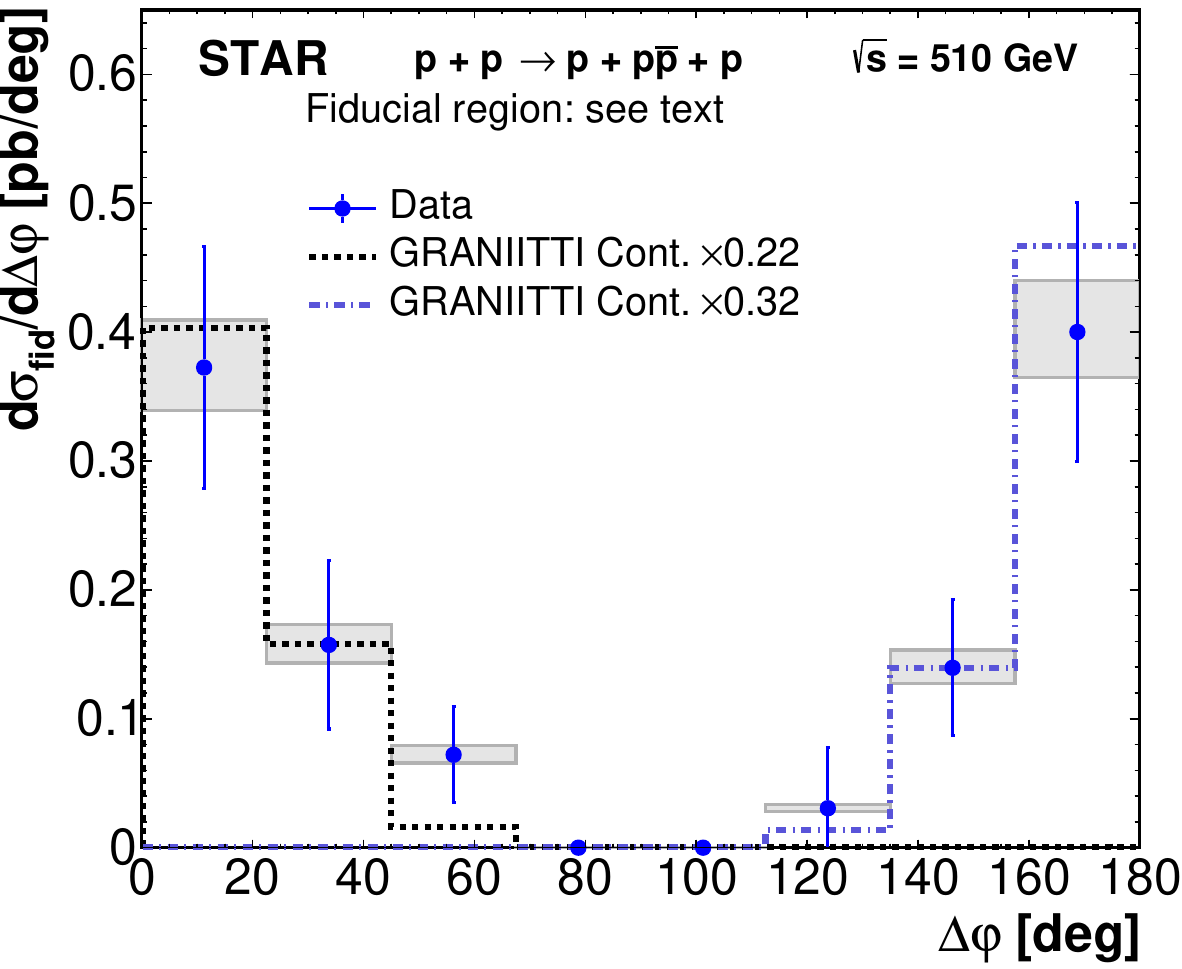}
\caption[Differential fiducial cross sections of hadron pairs as a function of $\Delta \upvarphi$]{Differential fiducial cross sections of $\pi^+\pi^-$ (left), $K^+K^-$ (middle) and $p\Bar{p}$ (right) pairs as a function of $\Delta \upvarphi$ of the forward scattered protons measured in the fiducial volume explained in the text. Data are shown as solid blue points with error bars representing the statistical uncertainties. The systematic uncertainties are shown as gray boxes. The scale uncertainty on the vertical axis due to the effective integrated luminosity is 6.4\% and is not shown. Predictions from MC model GRANIITTI~\cite{Mieskolainen} are shown separately for each $\Delta \upvarphi$ region within the experimentally accessible fiducial acceptance of the STAR detector. The normalization of the model is performed independently in the two $\Delta\upvarphi$ regions and is used solely to compare the shapes of the distributions.}
\label{fig:deltaPhi}
\end{figure}

Figure~\ref{fig:deltaPhi} shows the differential fiducial cross sections for CEP of $\pi^+\pi^-$, $K^+K^-$, and $p\Bar{p}$ pairs as a function of the difference between azimuthal angles of the forward protons $\Delta \upvarphi$. In this measurement, the $\Delta \upvarphi$ is the analogy to the $\dpt$ filter proposed originally by Close and Kirk~\cite{CLOSE1997333} and observed by WA102 experiment~\cite{BARBERIS1997339}. The $q\bar{q}$ mesons and possible glueball candidates could be distinguish by using so called $\dpt$ filter, where $\dpt$ is the difference in transverse momentum between the exchanged particles. Namely, when the $\dpt$ is small, $\dpt < 0.2$~GeV, the $q\bar{q}$ states would be suppressed and the glueball candidates would survive (small $\dpt$ corresponds to small $\Delta\upvarphi$).

Due to the limited azimuthal acceptance of the RP system, the region around $\Delta\upvarphi\approx 90^\circ$ is not accessible in the present measurement, which was verified by using a simple MC simulation with a flat $\Delta \upvarphi$ distribution. Therefore, all $\Delta\upvarphi$-dependent distributions are measured and interpreted strictly within the experimentally defined fiducial volume and are not extrapolated to the full azimuthal phase space. As a result, these distributions should not be interpreted as representing the complete $\Delta\upvarphi$ spectrum.
However, the same suppression is expected as was seen in CMS and TOTEM~\cite{CMS}, where it was attributed to the additional \Pom omeron exchanges between the incoming protons. The acceptance naturally divides the phase space into two ranges of $\Delta \upvarphi$, where different \Pom omeron dynamics and absorption effects are expected. Therefore, GRANIITTI predictions were produced separately for each $\Delta \upvarphi$ region: $\Delta \upvarphi > 90^{\circ}$ and $\Delta \upvarphi < 90^{\circ}$.

\begin{figure}[htbp!]
\centering
\includegraphics[width = .49\linewidth]{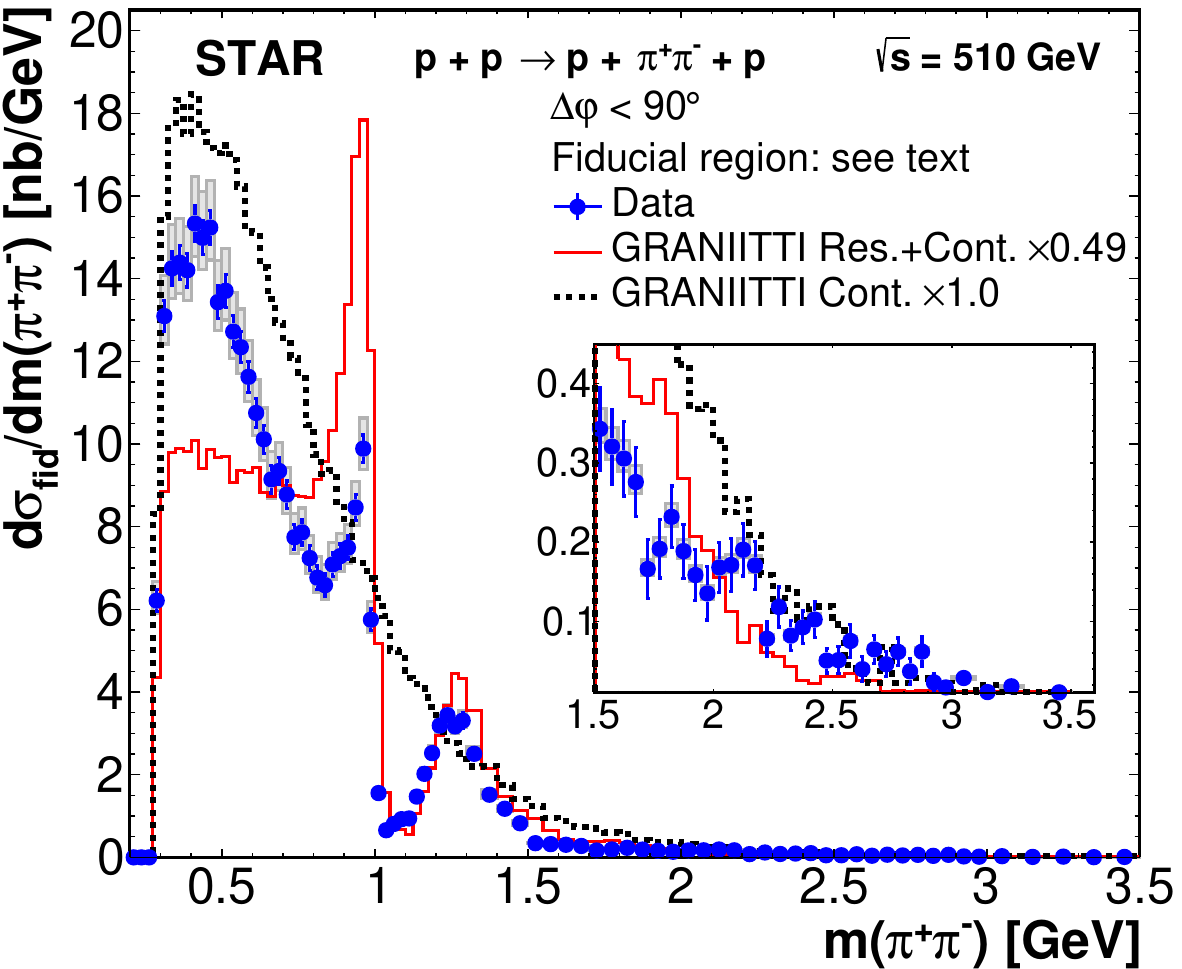}
\hfill
\includegraphics[width = .49\linewidth]{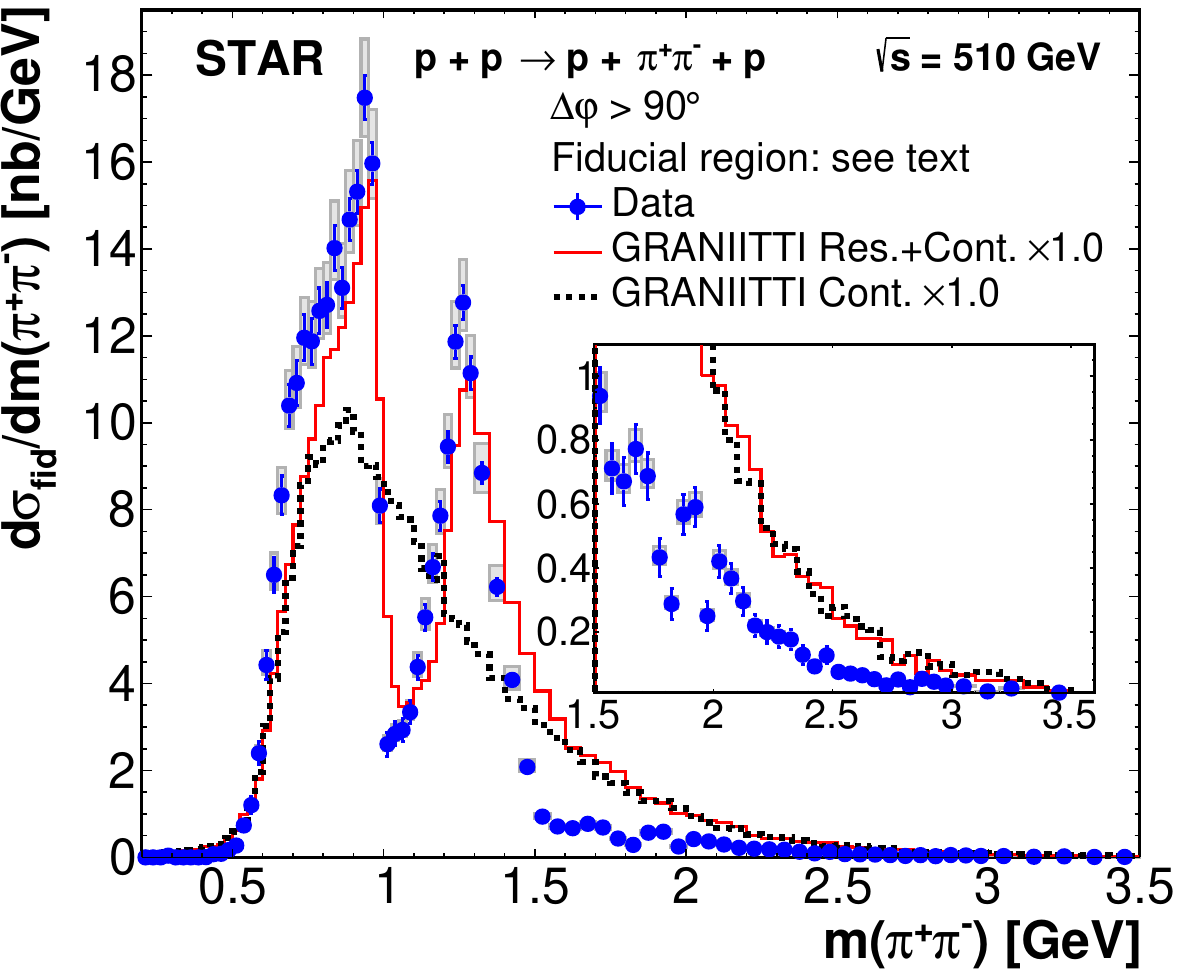}
\includegraphics[width = .49\linewidth]{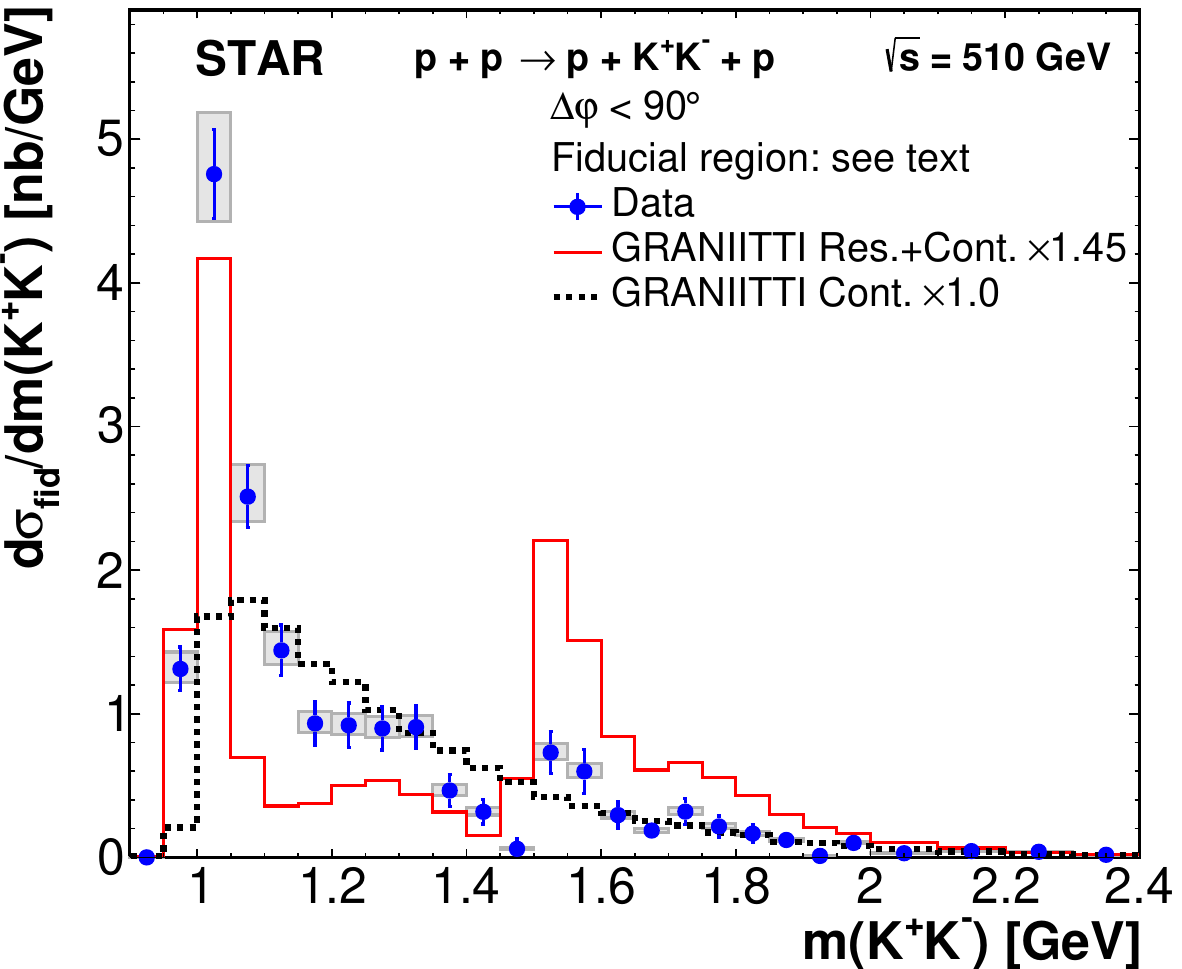}
\hfill
\includegraphics[width = .49\linewidth]{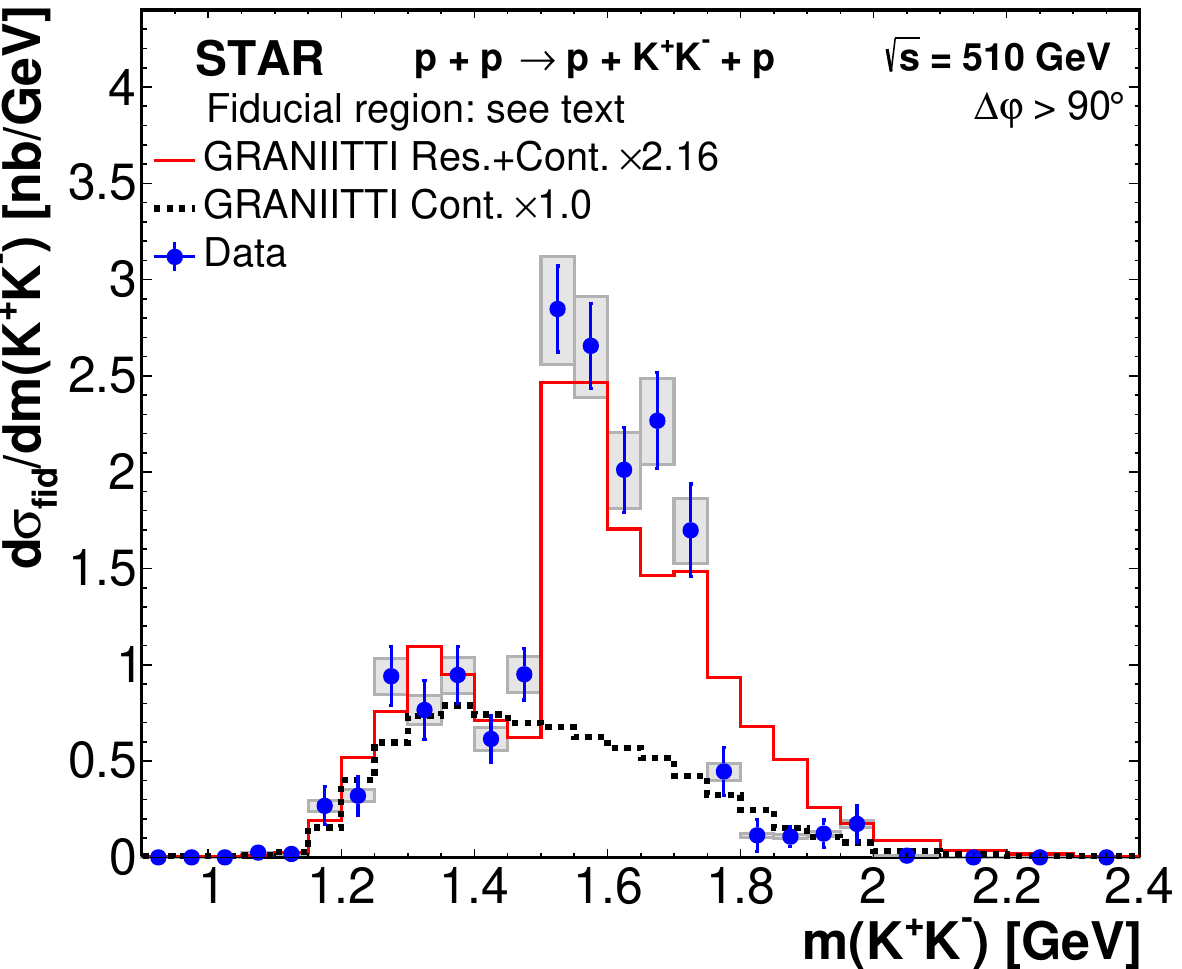}
\includegraphics[width = .49\linewidth]{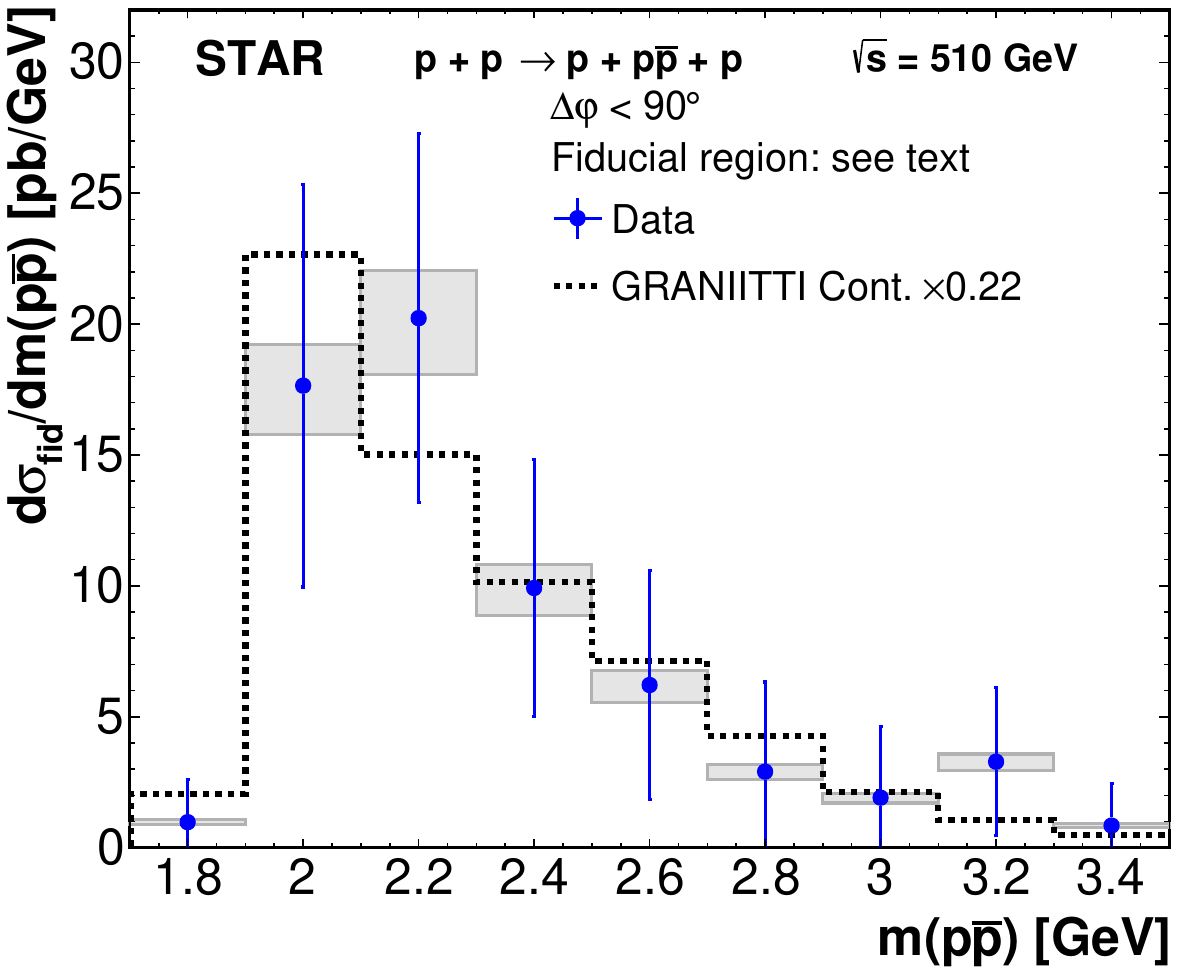}
\hfill
\includegraphics[width = .49\linewidth]{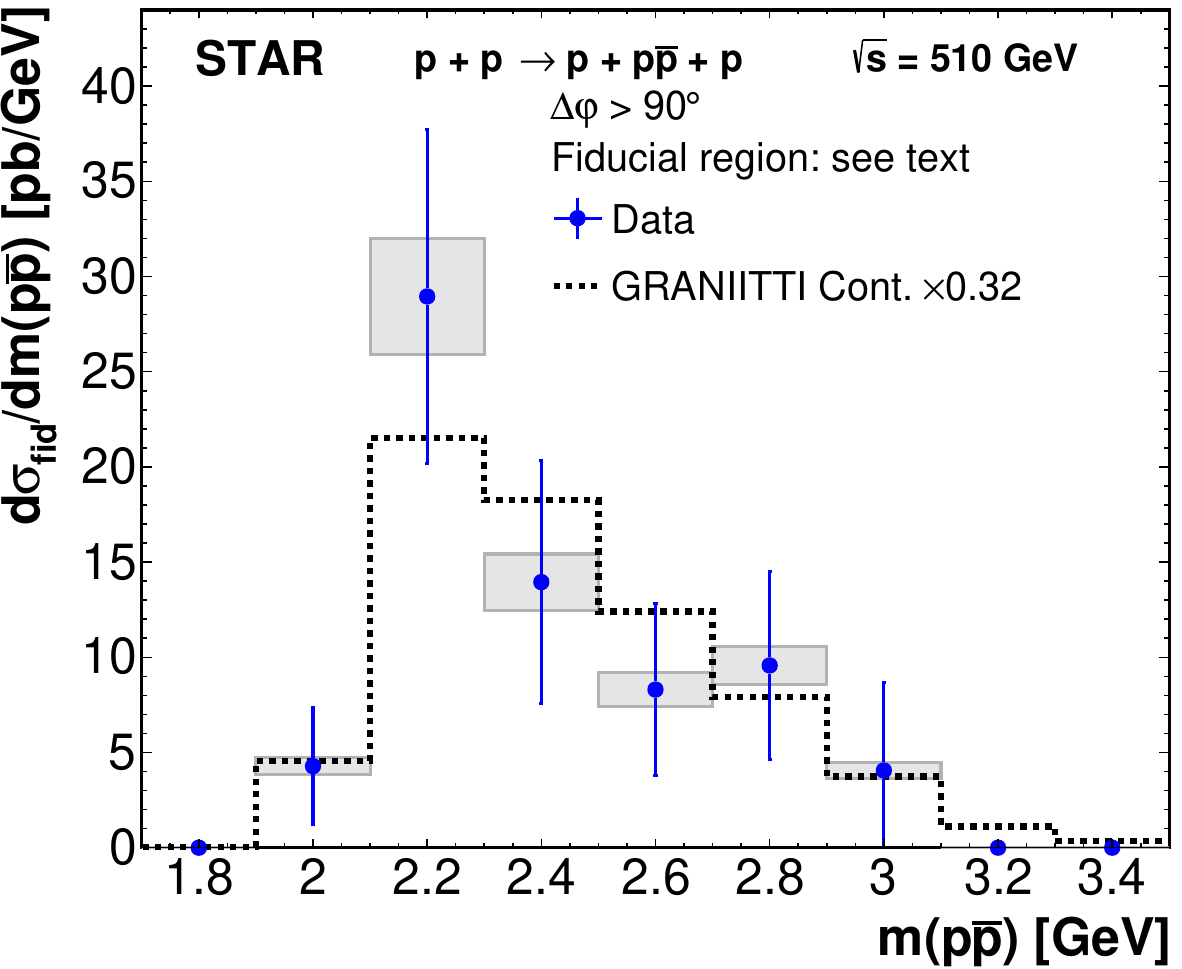}
\caption[Differential fiducial cross sections of pion, kaon, and proton pairs as a function of the invariant mass of the pair in two regions of $\Delta \upvarphi$]{Differential fiducial cross sections of $\pi^+\pi^-$ (top), $K^+K^-$ (middle), and $p\Bar{p}$ (bottom) pairs as a function of the invariant mass of the pair in two regions of the difference in azimuthal angles $\Delta \upvarphi$ of the forward scattered protons: $\Delta \upvarphi < 90^{\circ}$ (left) and $\Delta \upvarphi > 90^{\circ}$ (right), measured in the fiducial volume explained in the text. Data are shown as solid blue points with error bars representing the statistical uncertainties. The systematic uncertainties are shown as gray boxes. The scale uncertainty on the vertical axis due to the effective integrated luminosity is 6.4\% and is not shown. Predictions from MC model GRANIITTI~\cite{Mieskolainen} are shown separately for each $\Delta \upvarphi$ region within the experimentally accessible fiducial acceptance of the STAR detector. The normalization of the model is performed independently in the two $\Delta \upvarphi$ regions and is used solely to compare the shapes of the distributions.}
\label{fig:invMassDiffInAngle}
\end{figure}

Figure~\ref{fig:invMassDiffInAngle} shows differential fiducial cross sections of $\pi^+\pi^-$, $K^+K^-$, and $p\Bar{p}$ pairs as a function of the invariant mass of the pair in two $\Delta \upvarphi$ regions measured within the STAR acceptance in proton-proton collisions at $\sqrt{s} = 510$ GeV. GRANIITTI predictions are shown as well with the same scaling as in~figure~\ref{fig:deltaPhi}.

In the $\pi^+ \pi^-$ invariant mass distribution in both $\Delta \upvarphi$ regions, expected features seen in previous CEP~\cite{Rafal20, Akesson1986, CMS} and central production~\cite{PhysRevD.91.091101} measurements are observed: a sharp drop at about 1 GeV attributed to the quantum mechanical negative interference of $f_0(980)$ with the continuum contribution, and a peak consistent with the $f_2(1270)$. A clear difference between the two $\Delta \upvarphi$ regions can be seen. In $\Delta \upvarphi < 90^{\circ}$, an enhancement at low invariant mass and a suppression in the region of $f_2(1270)$ resonance are observed. The enhancement is mainly due to the kinematics of the forward protons at $\Delta \upvarphi < 90^{\circ}$ as it opens an acceptance compared to the $\Delta \upvarphi > 90^{\circ}$. Above $m(\pi^+ \pi^-) > 1.5$ GeV, there are no significant structures in the cross section, which generally decreases with increasing invariant mass. Indications of structures at 1.7 and 1.9 GeV in $\Delta \upvarphi > 90^{\circ}$ can be seen. The first one is consistent with the $f_0(1710)$ resonance that was considered as a ``pure'' glueball candidate~\cite{PhysRevD.91.052006}. Moreover, there is no peak around the mass $m(\pi^+ \pi^-) \sim 0.8$ GeV that could be attributed to the $\rho(770)$ meson. Hence, the contributions from photoproduction exchanges should be insignificant within the measured fiducial acceptance, in particular due to the requirement of detecting both forward protons in the RP detectors. The observed spectrum is consistent with \DPE.

Although the GRANIITTI predictions had to be scaled to match the data, the spectrum with its main features is described quite well. There are a few notable differences: the higher predicted cross section above $m(\pi^+ \pi^-) > 1.5$ GeV in $\Delta \upvarphi > 90^{\circ}$ and the more pronounced predicted $f_0(980)$ resonance in $\Delta \upvarphi < 90^{\circ}$. Also, the enhancement in the measured cross sections at the invariant mass about 500 MeV compared to the prediction, including both continuum and resonances, could be attributed to the $f_0(500)$ resonance that is missing in the GRANIITTI predictions. However, that contribution to the mass spectrum would have to be tuned in GRANIITTI.

In the invariant mass of $K^+ K^-$ pairs, a strong dependence of differential fiducial cross sections on the azimuthal separation between forward protons ($\Delta \upvarphi < 90^{\circ}, \Delta \upvarphi > 90^{\circ}$) is observed. 
For $\Delta \upvarphi < 90^{\circ}$, peaks at 1~GeV and 1.5~GeV are seen. Based only on a comparison with GRANIITTI, these are consistent with the presence of $f_0(980)$ and ($f_0(1500)$, $f_2(1525)$) resonances. Also, a dip is observed at $m(K^+ K^-) \lesssim 1.5$~GeV. The dip can be explained as due to the negative interference of $f_0(1500)$ with the continuum production. Also, an enhancement corresponding to $f_0(1710)$ is seen.


We observe a peak at a mass of $\sim 1$~GeV in the $K^+ K^-$ channel for $\Delta \upvarphi < 90^{\circ}$. Contributions to this peak could be
due to $f_0(980)$ production in DPE processes or to the $\upvarphi(1020)$ meson~\cite{CISEK2010168} produced via the photoproduction process.

For $\Delta \upvarphi > 90^{\circ}$, two structures are seen at about 1.3 and 1.5 GeV. In GRANIITTI, the broad structure around 1.3~GeV is due to the acceptance cut off of the $K^+ K^-$ continuum invariant mass distribution at the lower masses, while the peak at $\sim 1.5$~GeV can be explained by $f_0(1500)$ and $f_2(1525)$. There is also a possible peak at $\sim 1.7$~GeV, which could be due to the $f_0(1710)$ resonance. Also a possible enhancement at $m(K^+ K^-) \sim 1.975$ GeV is seen.

Data do not support the presence of the $f_2(1270)$ resonance and GRANIITTI reproduces the invariant mass spectrum quite well without the $f_2(1270)$.

The differential fiducial cross sections of $p \overline{p}$ pairs do not show any significant resonances or any notable $\Delta \upvarphi$ asymmetry. Hence, it is compared with the GRANIITTI predictions based solely on continuum contributions. An enhancement at the level of $1\sigma$ at $m(p \overline{p}) \sim 2.2$~GeV for $\Delta \upvarphi > 90^{\circ}$ is seen. It was also observed at the same level at $\sqrt{s} = 200$~GeV~\cite{Rafal20}.

\begin{figure}[htbp!]
\centering
\includegraphics[width = .49\linewidth]{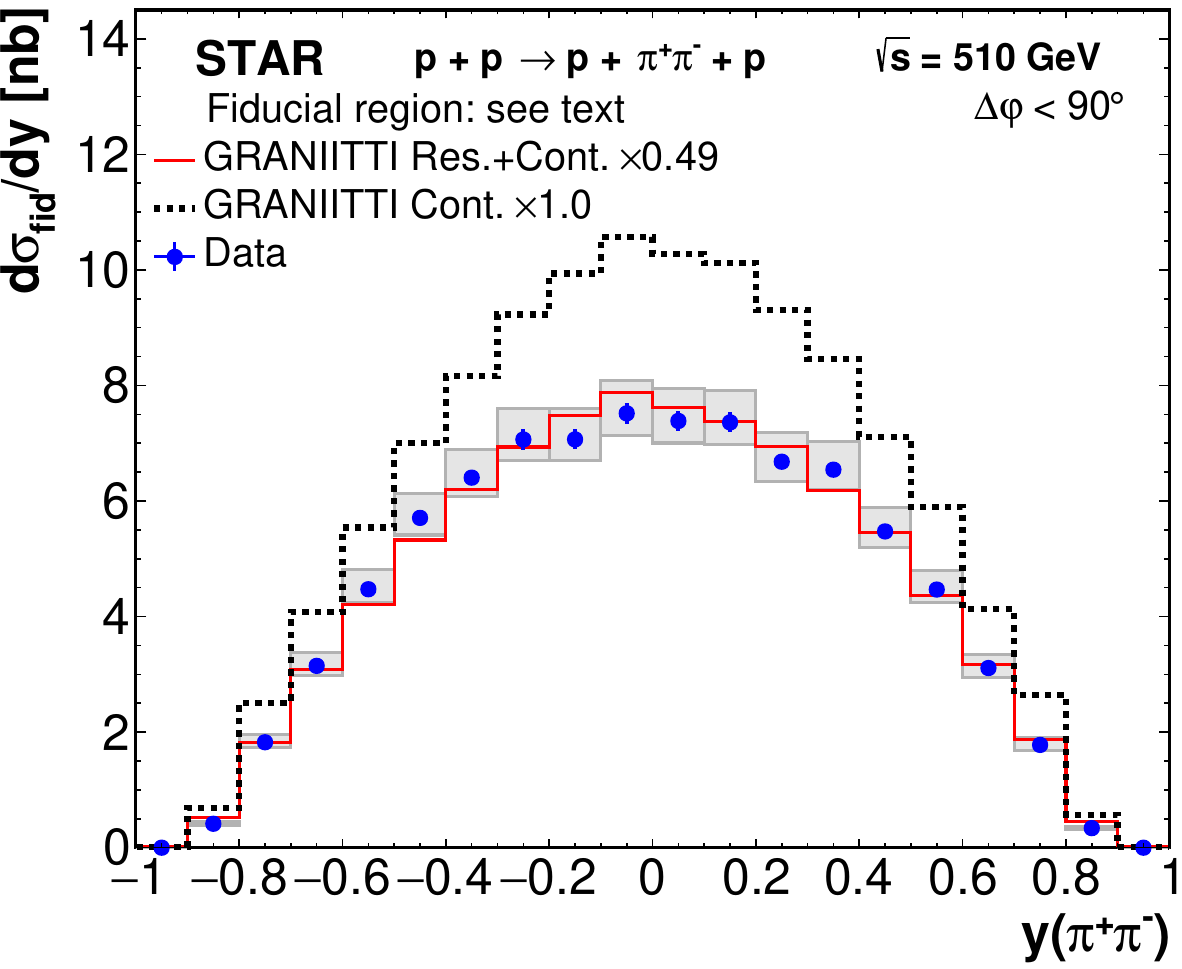}
\hfill
\includegraphics[width = .49\linewidth]{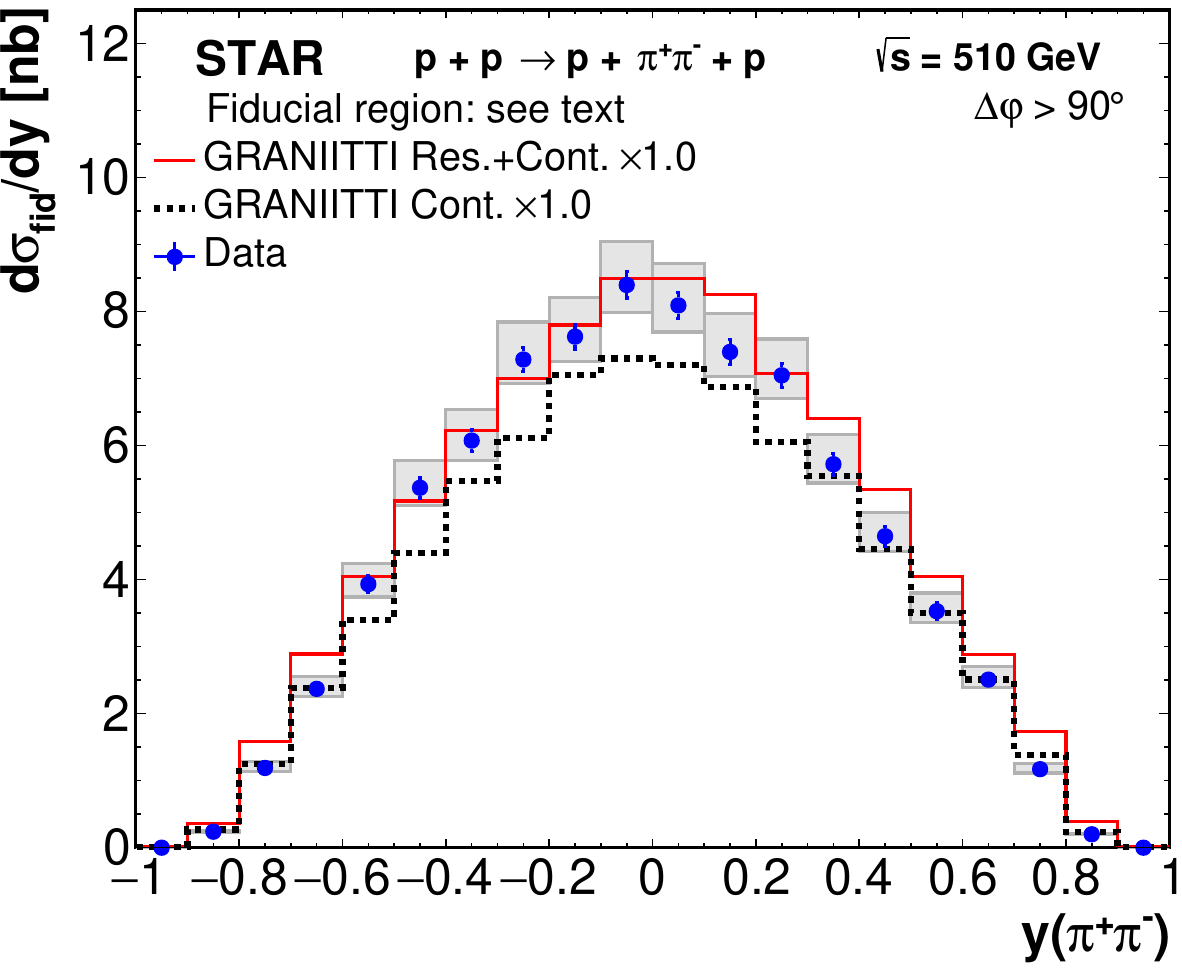}
\includegraphics[width = .49\linewidth]{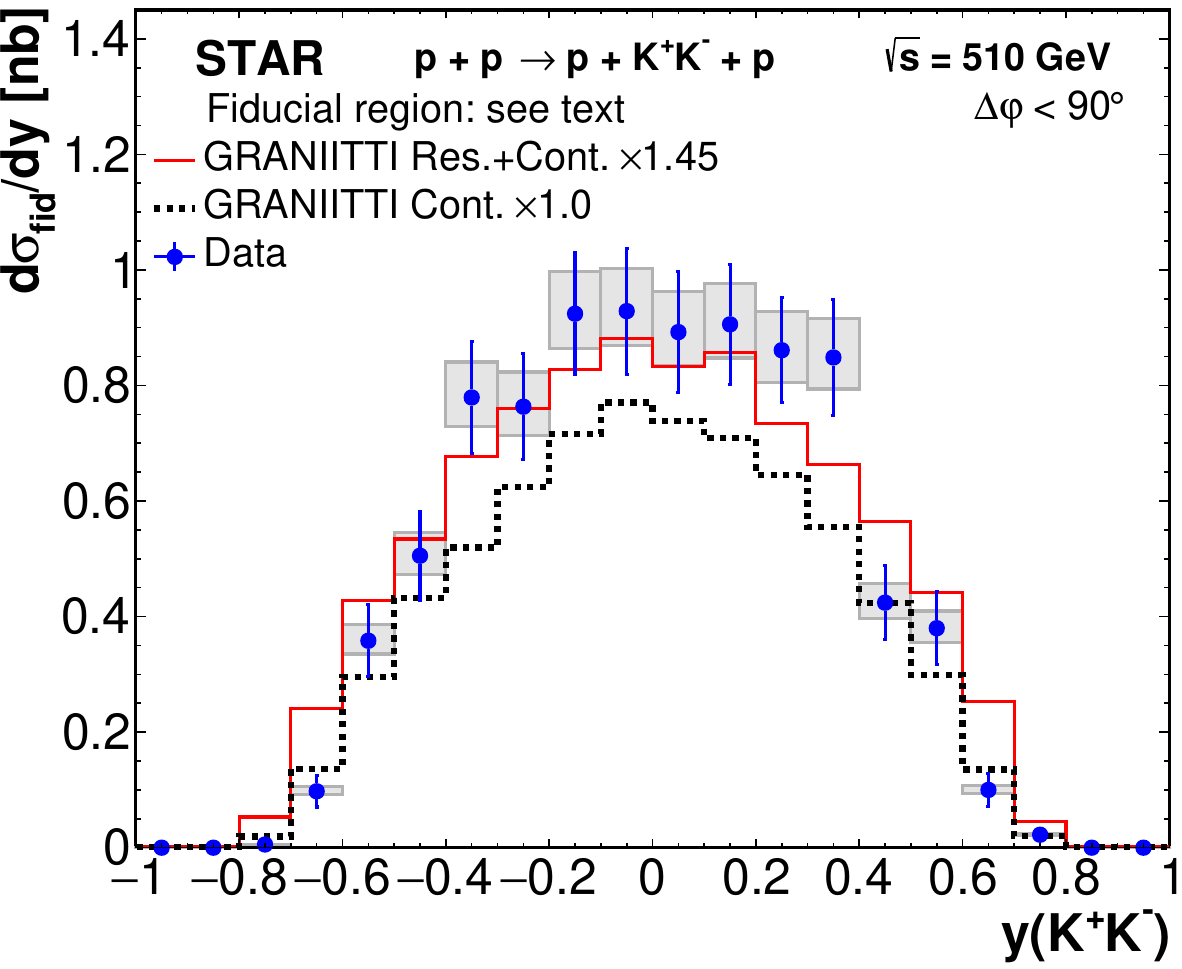}
\hfill
\includegraphics[width = .49\linewidth]{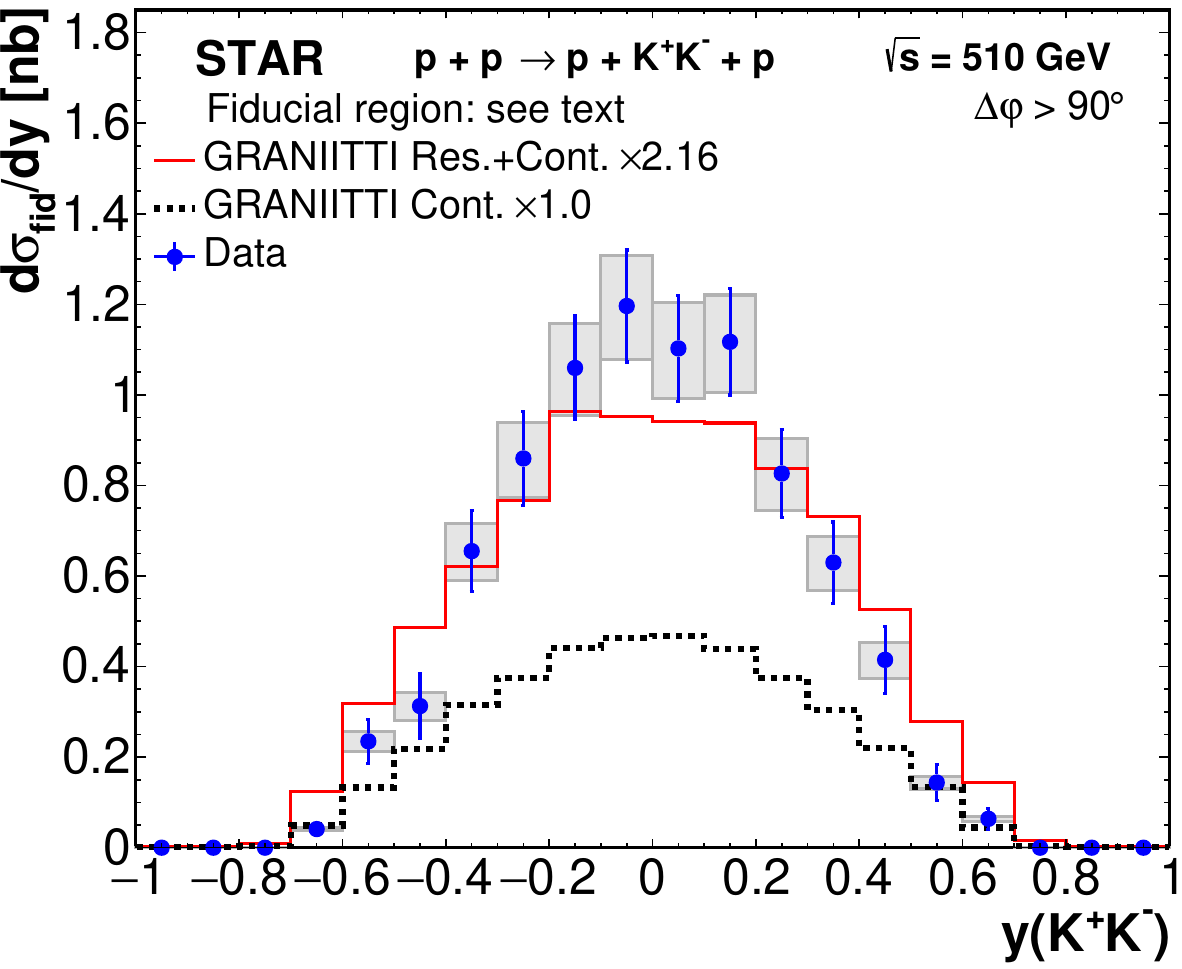}
\includegraphics[width = .49\linewidth]{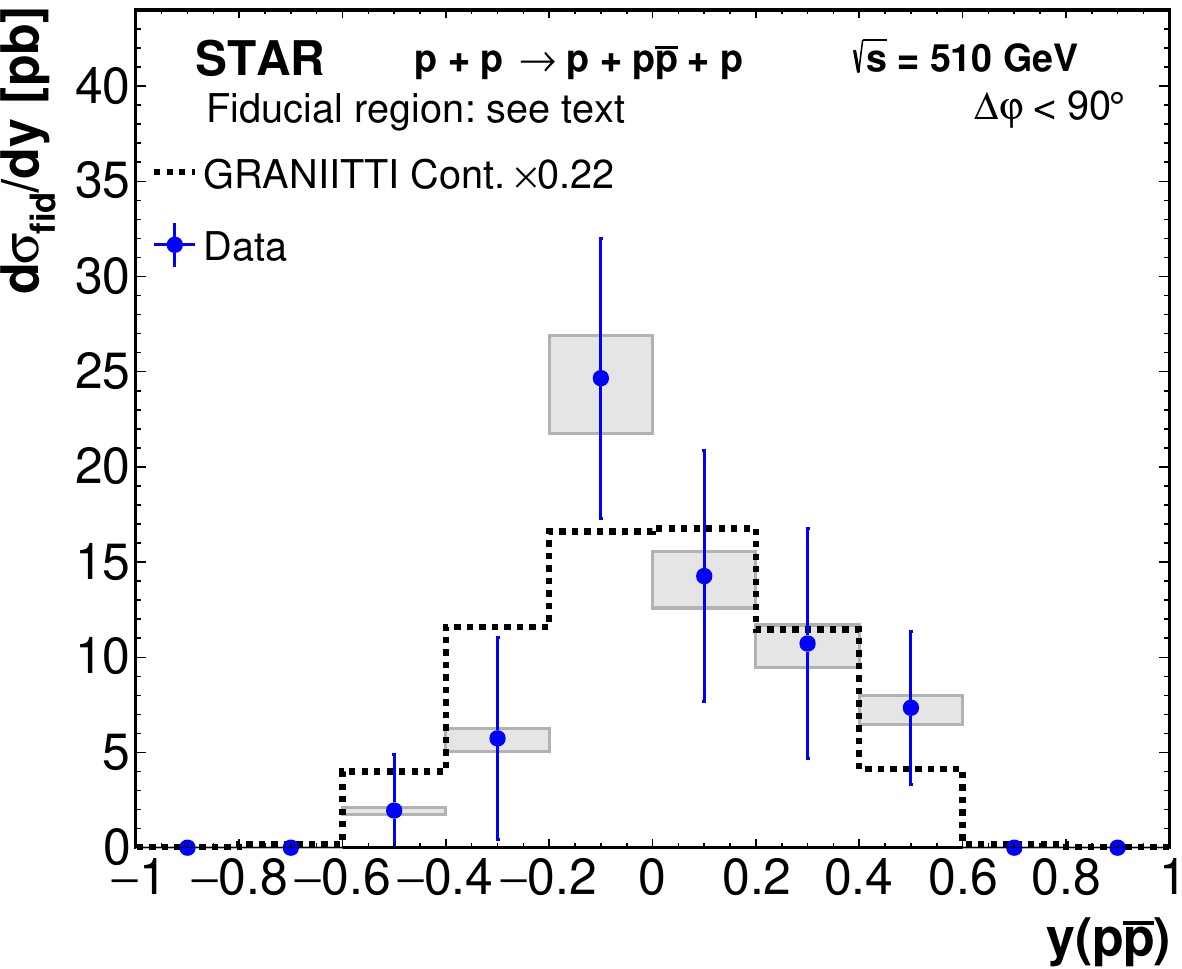}
\hfill
\includegraphics[width = .49\linewidth]{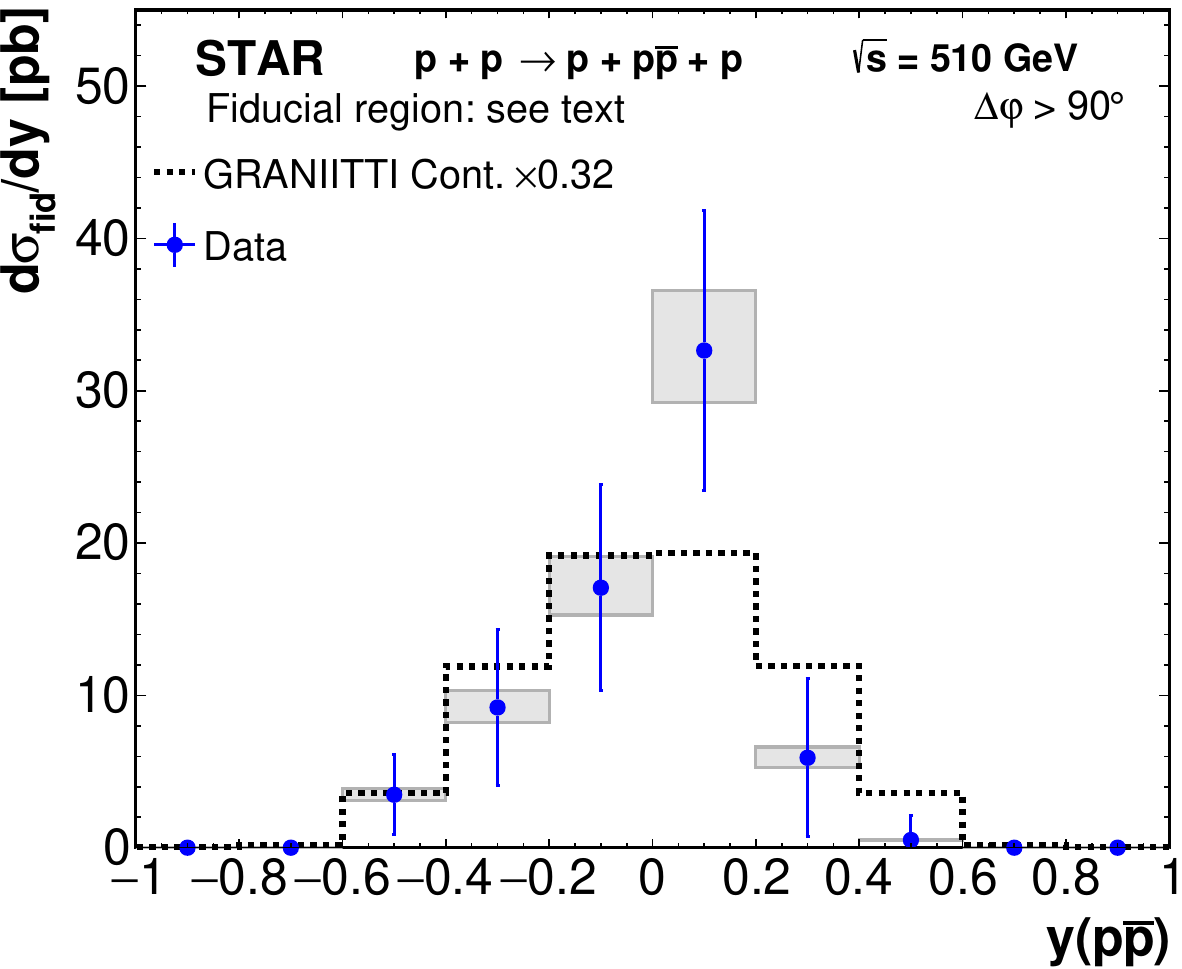}
\caption[Differential fiducial cross sections of pion, kaon, and proton pairs as a function of the rapidity of the pair in two regions of $\Delta \upvarphi$]{Differential fiducial cross sections of $\pi^+\pi^-$ (top), $K^+K^-$ (middle), and $p\Bar{p}$ (bottom) pairs as a function of the pair rapidity in two regions of the difference in azimuthal angles $\Delta \upvarphi$ of the forward scattered protons: $\Delta \upvarphi < 90^{\circ}$ (left) and $\Delta \upvarphi > 90^{\circ}$ (right), measured in the fiducial volume explained in the text. Data are shown as solid blue points with error bars representing the statistical uncertainties. The systematic uncertainties are shown as gray boxes. The scale uncertainty on the vertical axis due to the effective integrated luminosity is 6.4\% and is not shown. Predictions from MC model GRANIITTI~\cite{Mieskolainen} are shown separately for each $\Delta \upvarphi$ region within the experimentally accessible fiducial acceptance of the STAR detector. The normalization of the model is performed independently in the two $\Delta\upvarphi$ regions and is used solely to compare the shapes of the distributions.}
\label{fig:rapDiffInAngle}
\end{figure}

Figure~\ref{fig:rapDiffInAngle} shows differential fiducial cross sections of $\pi^+\pi^-$, $K^+K^-$, and $p\Bar{p}$ pairs as a function of the pair rapidity in two $\Delta \upvarphi$ regions. The GRANIITTI predictions are shown as well with the same scaling as in~figure~\ref{fig:deltaPhi}. In general, the shapes of the measured distributions are well described by the GRANIITTI predictions and show the same behavior as at $\sqrt{s} = 200$ GeV~\cite{Rafal20}.

\begin{figure}[htbp!]
\centering
\includegraphics[width = .49\linewidth]{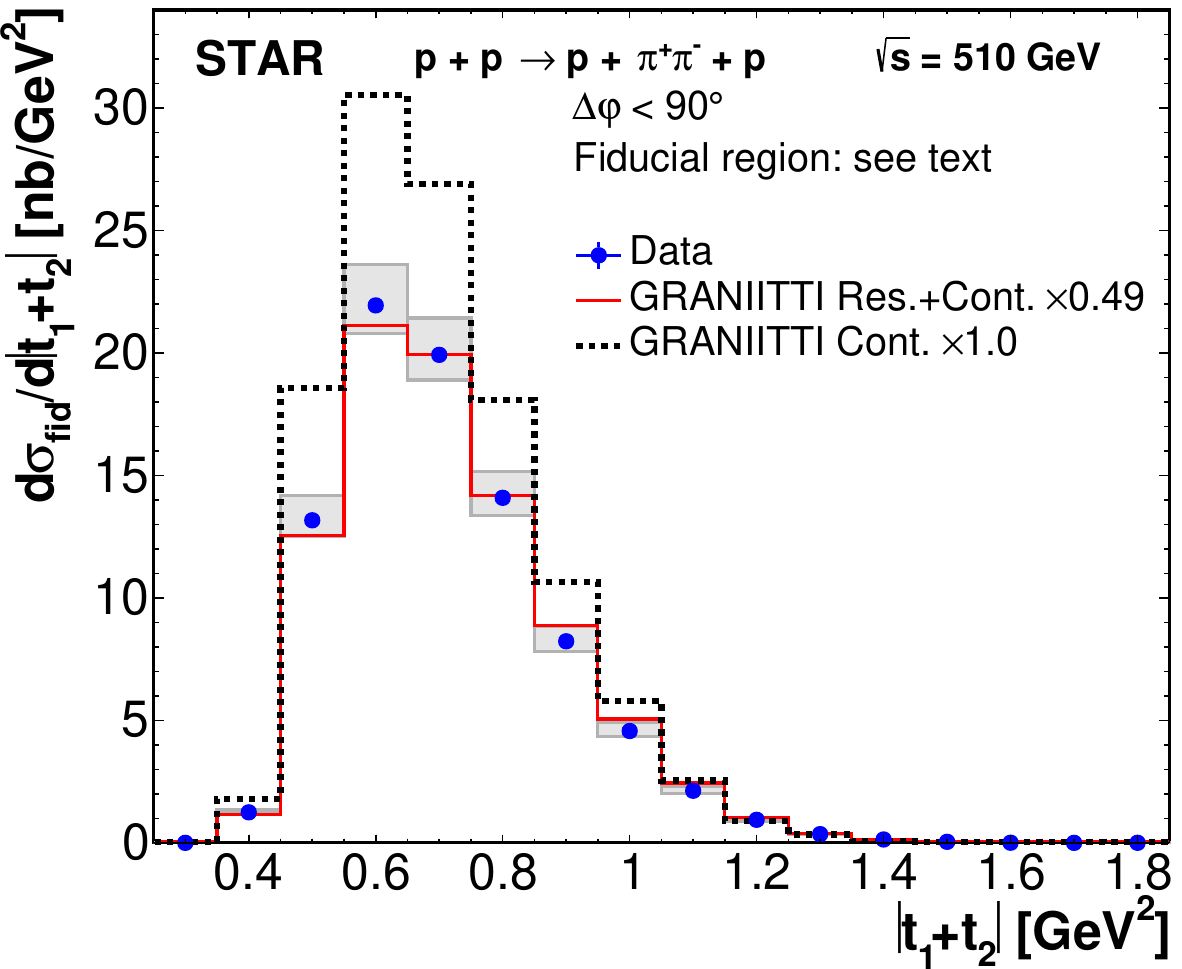}
\hfill
\includegraphics[width = .49\linewidth]{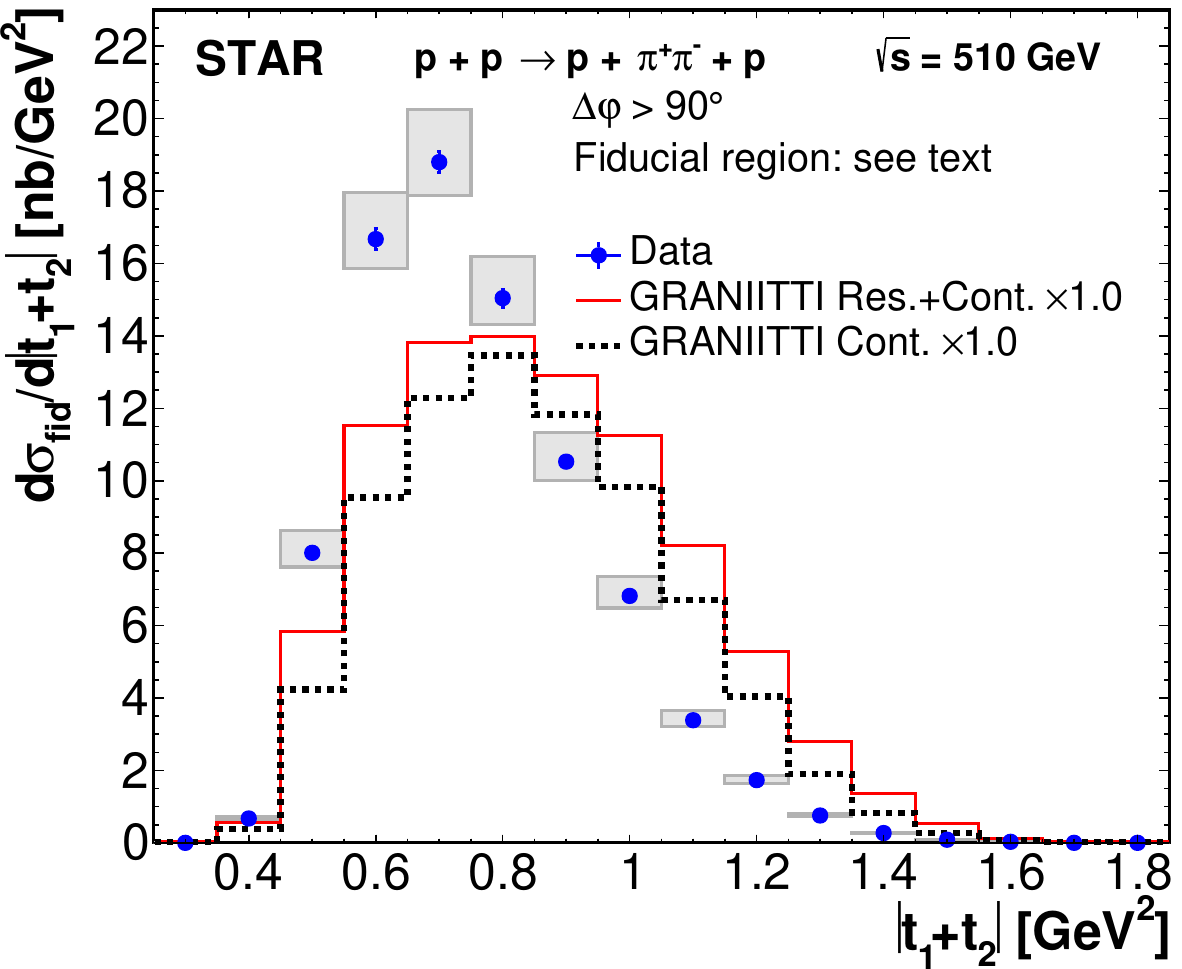}
\includegraphics[width = .49\linewidth]{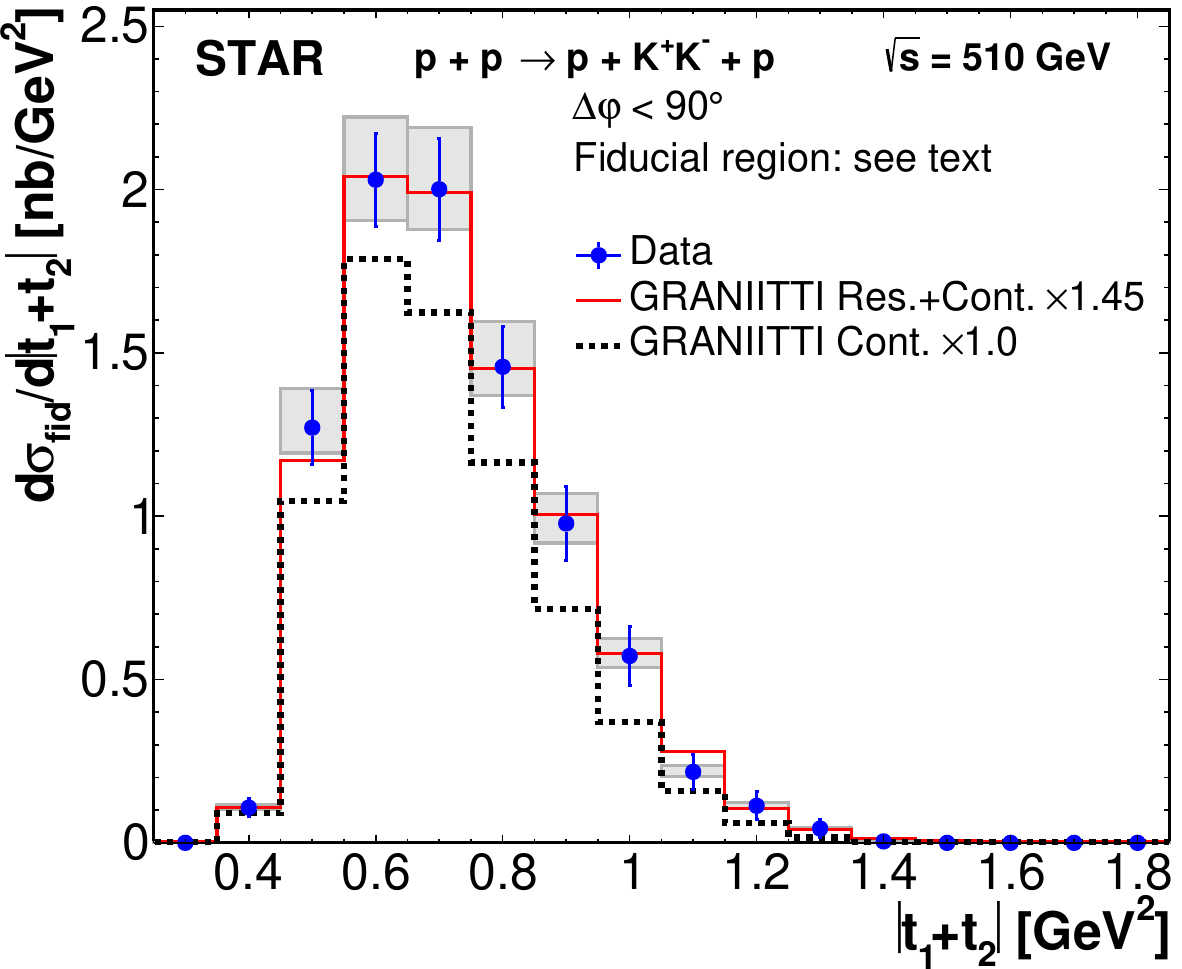}
\hfill
\includegraphics[width = .49\linewidth]{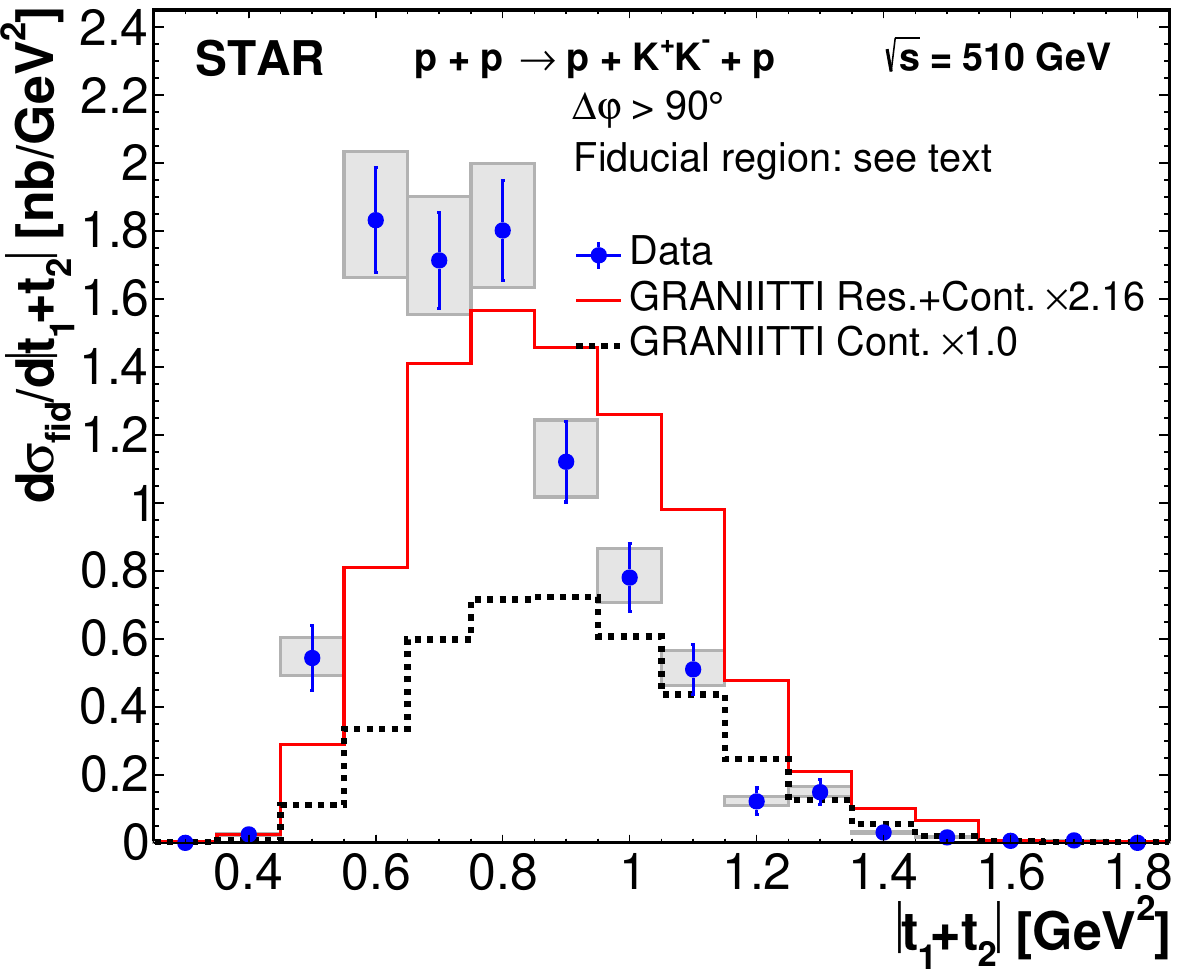}
\includegraphics[width = .49\linewidth]{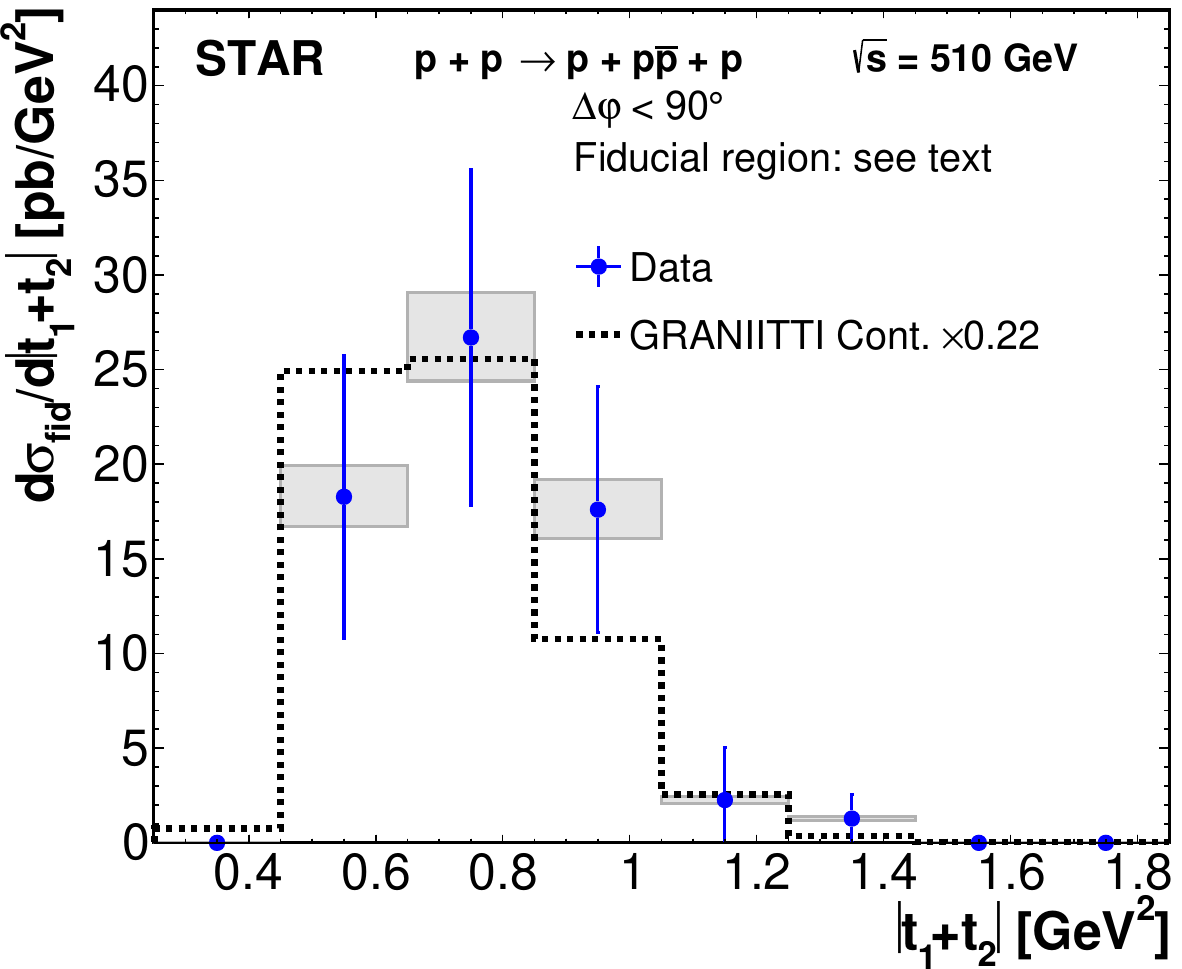}
\hfill
\includegraphics[width = .49\linewidth]{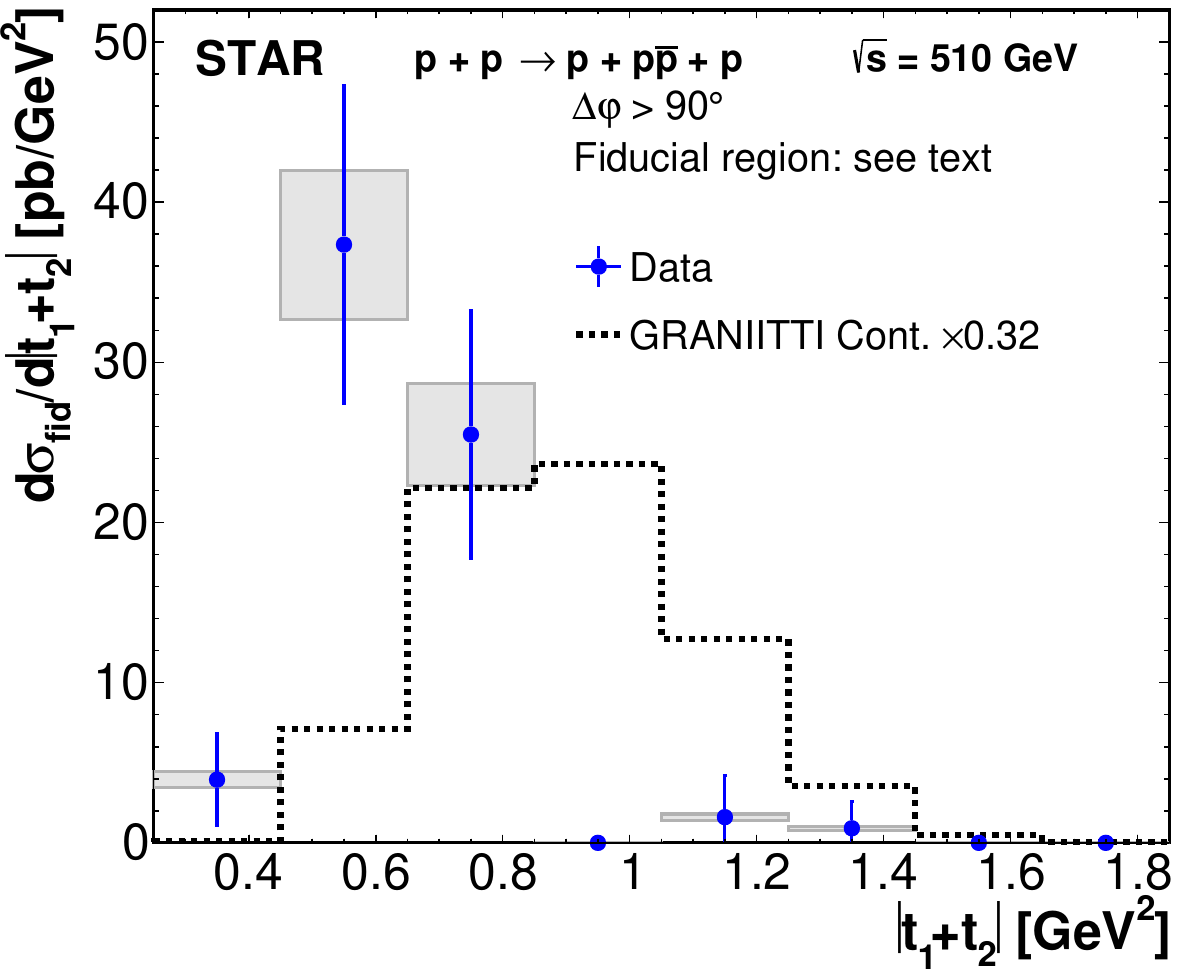}
\caption[Differential fiducial cross sections of pion, kaon, and proton pairs as a function of the $|t_1 + t_2|$ in two regions of $\Delta \upvarphi$]
{Differential fiducial cross sections for CEP of $\pi^+\pi^-$ (top), $K^+K^-$ (middle), and $p\Bar{p}$ (bottom) pairs as a function of the absolute value of the sum of the squares of the four-momentum transfer of the forward protons in two regions of the difference in azimuthal angles $\Delta \upvarphi$ of the forward scattered protons: $\Delta \upvarphi < 90^{\circ}$ (left) and $\Delta \upvarphi > 90^{\circ}$ (right), measured in the fiducial volume explained in the text. Data are shown as solid blue points with error bars representing the statistical uncertainties. The systematic uncertainties are shown as gray boxes. The scale uncertainty on the vertical axis due to the effective integrated luminosity is 6.4\% and is not shown. Predictions from MC model GRANIITTI~\cite{Mieskolainen} are shown separately for each $\Delta \upvarphi$ region within the experimentally accessible fiducial acceptance of the STAR detector. The normalization of the model is performed independently in the two $\Delta\upvarphi$ regions and is used solely to compare the shapes of the distributions.}
\label{fig:tDiffInAngle}
\end{figure}

Figure~\ref{fig:tDiffInAngle} shows differential fiducial cross sections of $\pi^+\pi^-$, $K^+K^-$, and $p\Bar{p}$ pairs as a function of the absolute value of the sum of the squares of the four-momentum transfer of the forward protons ($|t_1 + t_2|$) in two $\Delta \upvarphi$ regions. GRANIITTI predictions are shown as well. The same scaling as in the invariant mass distributions is applied. The shapes of the measured distributions are strongly affected by the fiducial cuts applied to the forward scattered protons as verified with a simple MC study. In general, the shapes are well described by the GRANIITTI predictions in $\Delta \upvarphi < 90^{\circ}$ while the distributions in $\Delta \upvarphi > 90^{\circ}$ are slightly shifted to the higher values in the model. The shapes are similar to those observed at $\sqrt{s} = 200$ GeV~\cite{Rafal20}.

The large statistics of the $\pi^+\pi^-$ sample enables a more detailed study. The differential fiducial cross sections are studied in three characteristic ranges of the invariant mass of the pair: $m(\pi^+\pi^-) < 1$ GeV, 1 GeV $ < m(\pi^+\pi^-) < 1.5$ GeV and $m(\pi^+\pi^-) > 1.5$ GeV. The first region is considered to be dominated by continuum production. The second is dominated by resonant production, namely by $f_2(1270)$. 

\begin{figure}[htbp!]
\centering
\includegraphics[width = .32\linewidth]{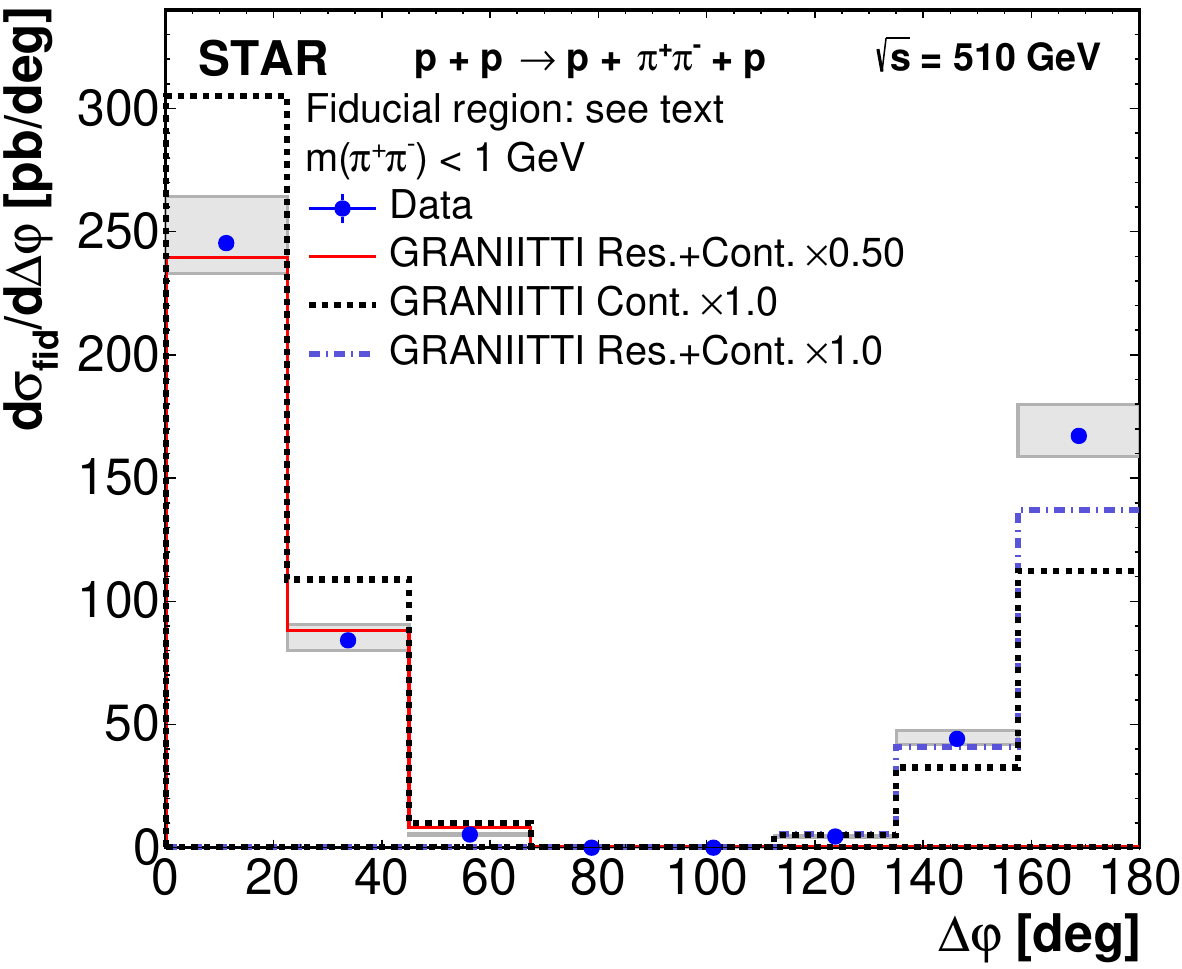}
\hfill
\includegraphics[width = .32\linewidth]{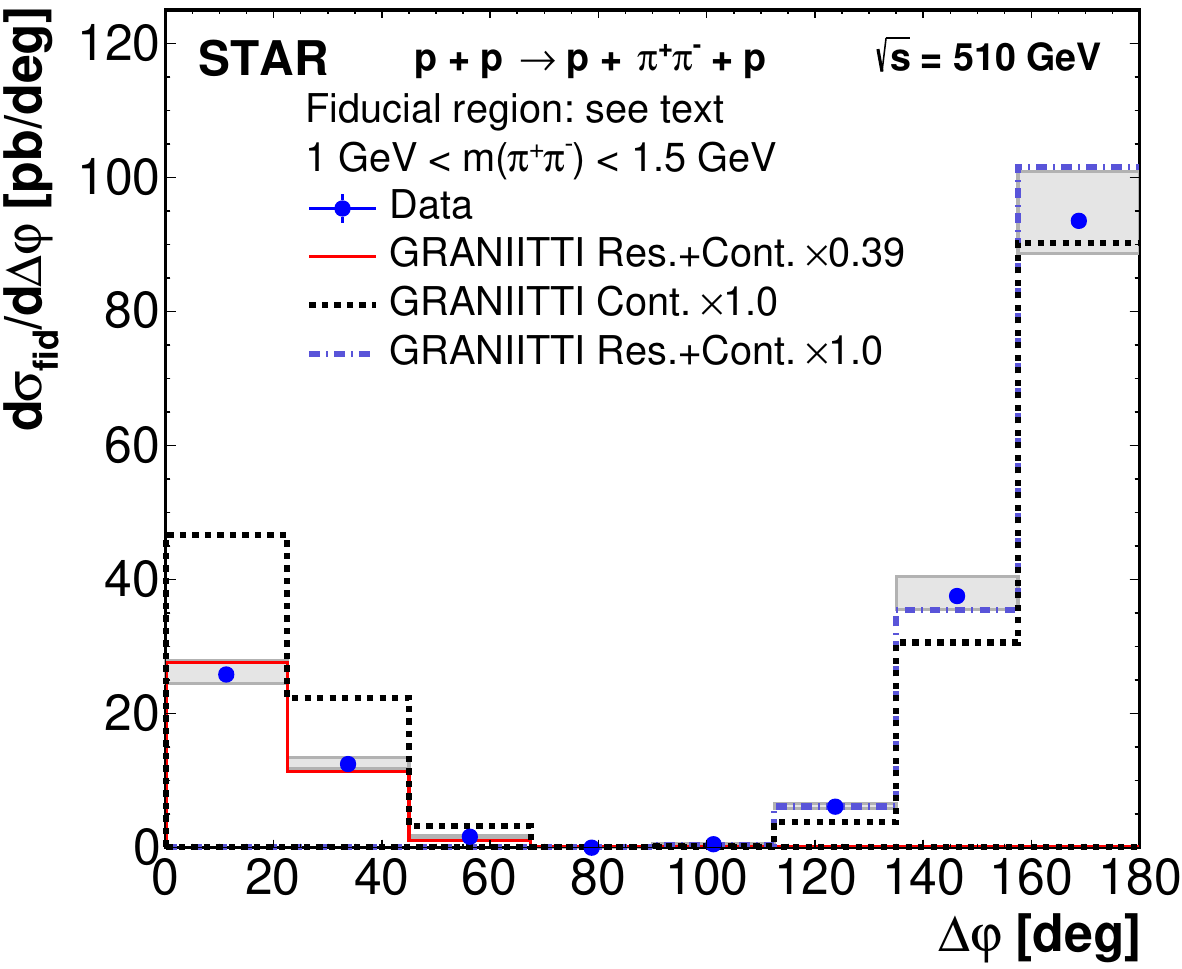}
\hfill
\includegraphics[width = .32\linewidth]{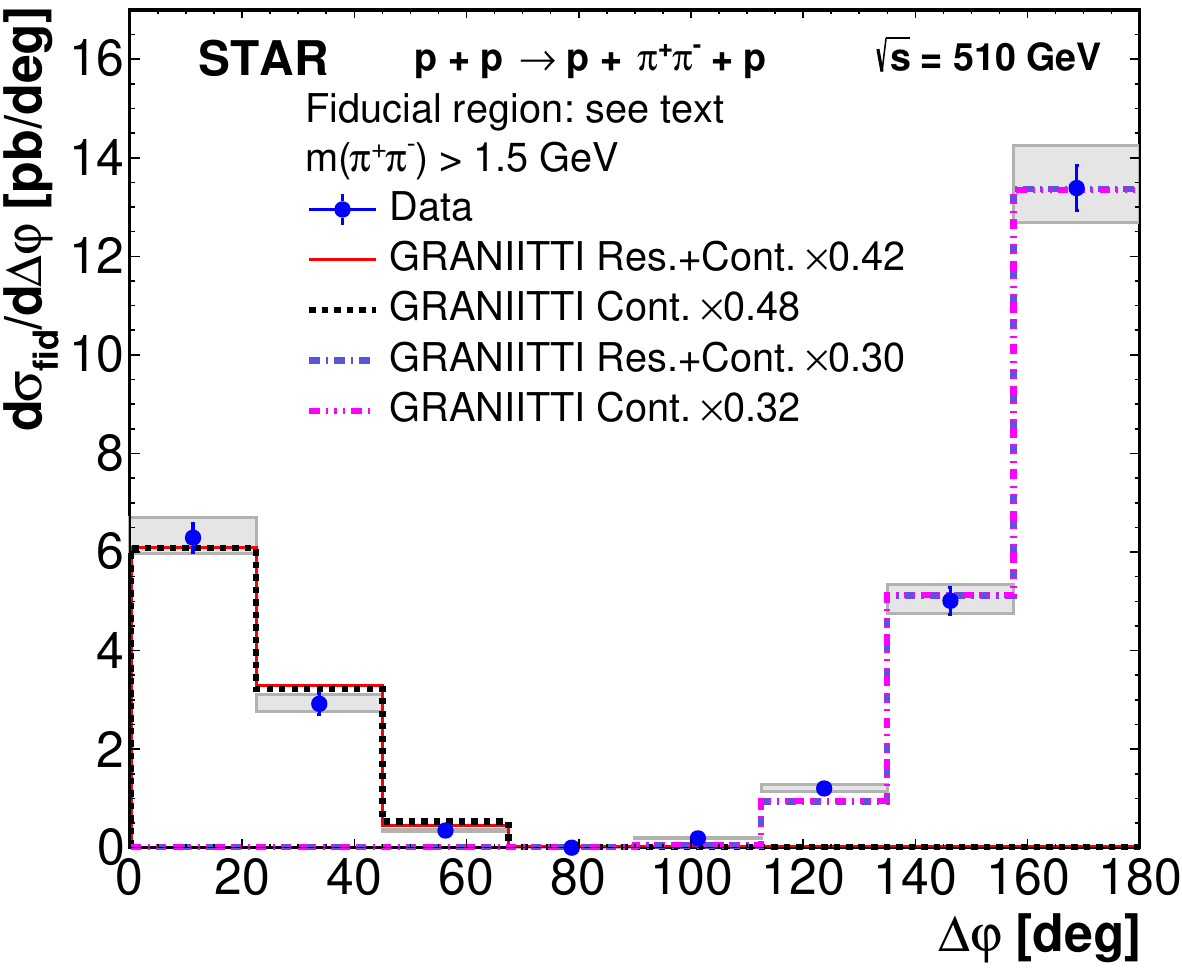}
\caption[Differential fiducial cross sections of pion pairs as a function of the $\Delta \upvarphi$ in three regions of the invariant mass of the pair]{Differential fiducial cross sections of $\pi^+\pi^-$ pairs as a function of the difference in azimuthal angles $\Delta \upvarphi$ of the forward scattered protons in three ranges of the $\pi^+\pi^-$
pair invariant mass: $m(\pi^+\pi^-) < 1$ GeV (left), 1 GeV $ < m(\pi^+\pi^-) < 1.5$ GeV (middle) and $m(\pi^+\pi^-) > 1.5$ GeV (right), measured in the fiducial volume explained in the text. Data are shown as solid blue points with error bars representing the statistical uncertainties. The systematic uncertainties are shown as gray boxes. The scale uncertainty on the vertical axis due to the effective integrated luminosity is 6.4\% and is not shown. Predictions from MC model GRANIITTI~\cite{Mieskolainen} are shown separately for each $\Delta \upvarphi$ region within the experimentally accessible fiducial acceptance of the STAR detector. The normalization of the model is performed independently in the two $\Delta\upvarphi$ regions and is used solely to compare the shapes of the distributions.}
\label{fig:DeltaPhiDiffInvMass}
\end{figure}

Figure~\ref{fig:DeltaPhiDiffInvMass} shows differential fiducial cross sections of $\pi^+\pi^-$ pairs as a function of the difference in azimuthal angles ($\Delta \upvarphi$) of the forward scattered protons in three ranges of the $\pi^+\pi^-$ pair invariant mass. GRANIITTI predictions are shown as well. The strong suppression close to $\Delta\upvarphi=90^{\circ}$ is due to the limited azimuthal acceptance in the RP detector system. The differential fiducial cross sections in $\Delta\upvarphi>90^{\circ}$ in the first range of the $\pi^+\pi^-$ pair invariant mass is suppressed due to the STAR TPC acceptance. The same suppression was observed also at $\sqrt{s} = 200$~GeV~\cite{Rafal20}. In the other ranges, suppressions of differential fiducial cross sections in $\Delta\upvarphi<90^{\circ}$ are seen. The suppression in the middle range was also seen at $\sqrt{s} = 200$~GeV~\cite{Rafal20} and is also predicted by GRANIITTI calculations assuming only the continuum contribution. In the last range, GRANIITTI predicts the same asymmetry as seen in the data.

\begin{figure}[htbp!]
\centering
\includegraphics[width = .32\linewidth]{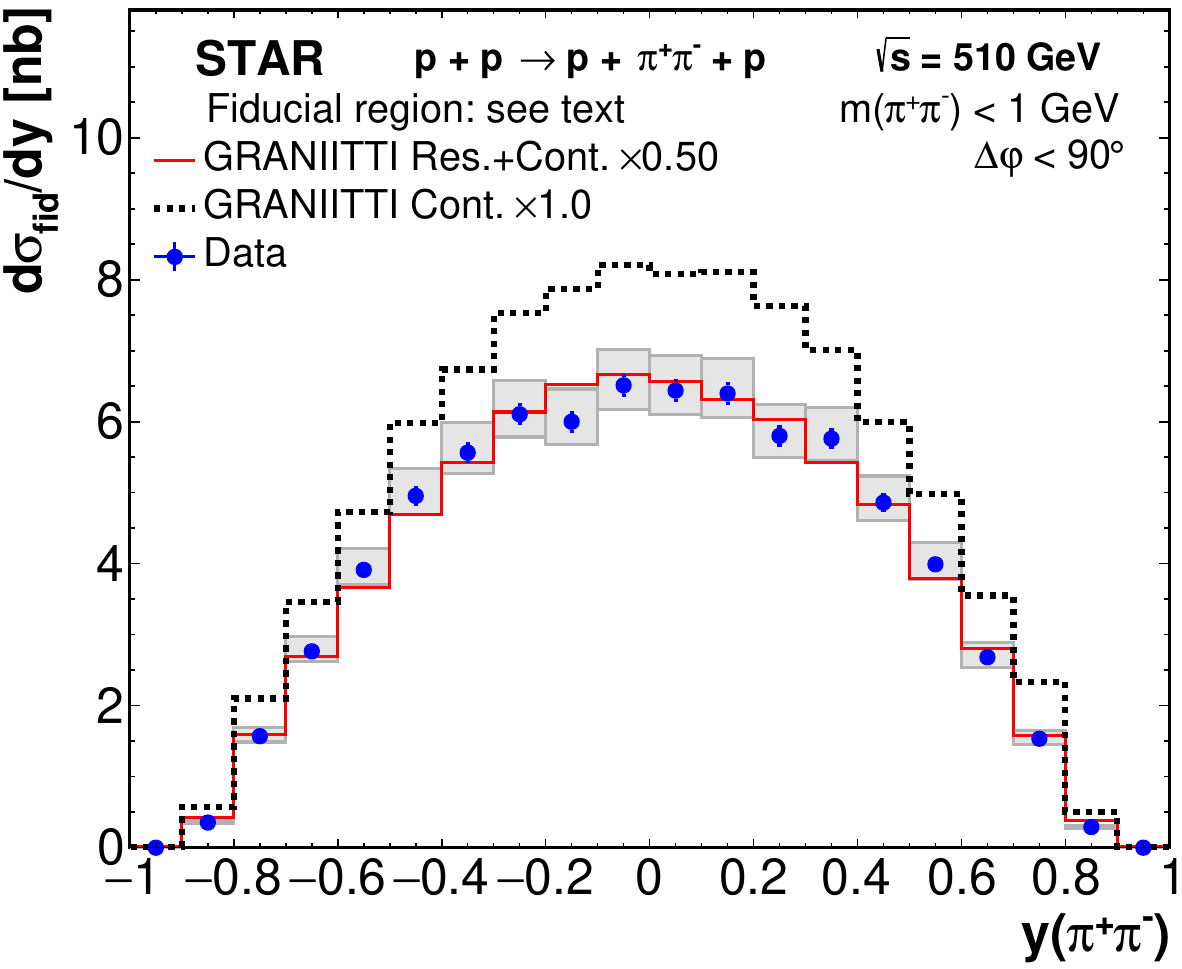}
\hfill
\includegraphics[width = .32\linewidth]{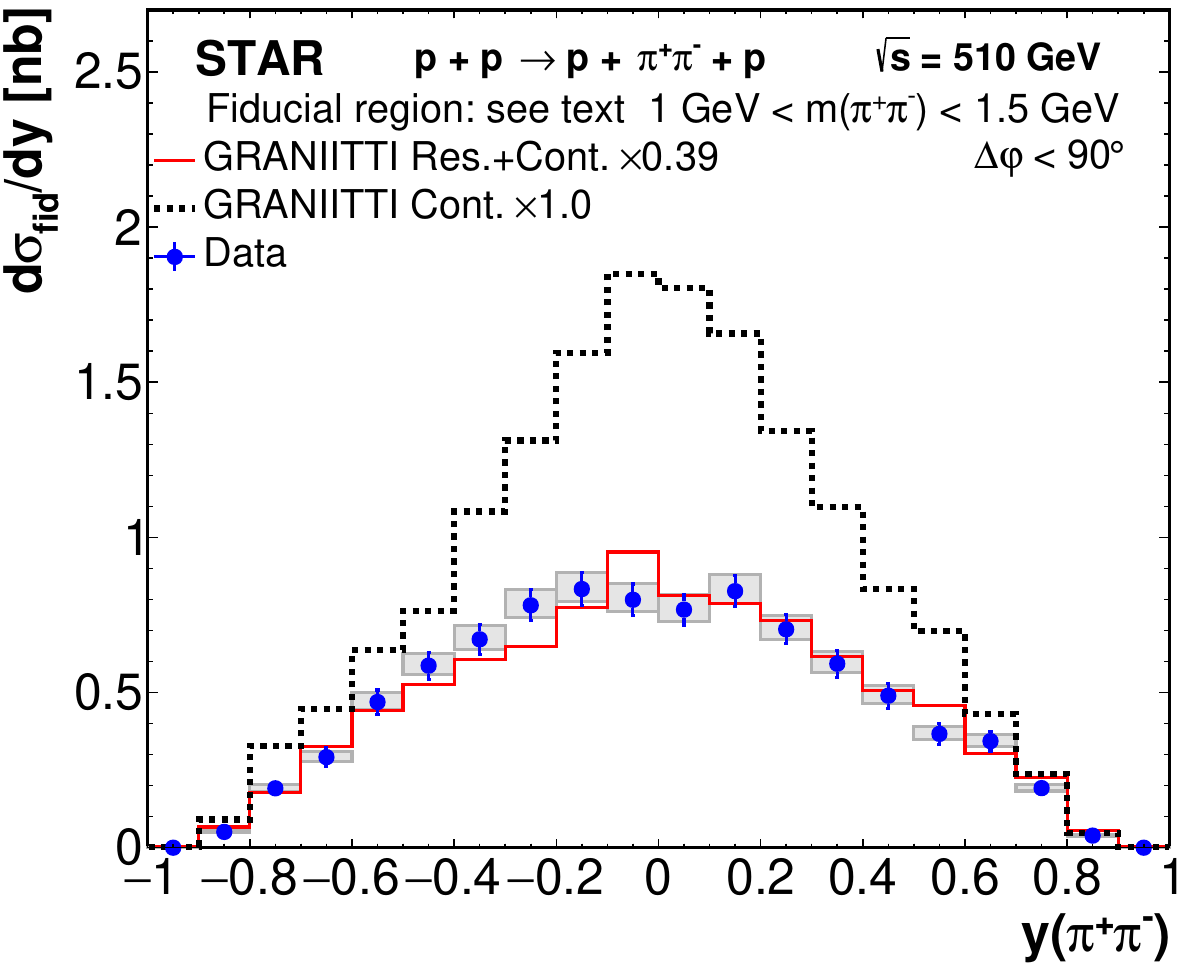}
\hfill
\includegraphics[width = .32\linewidth]{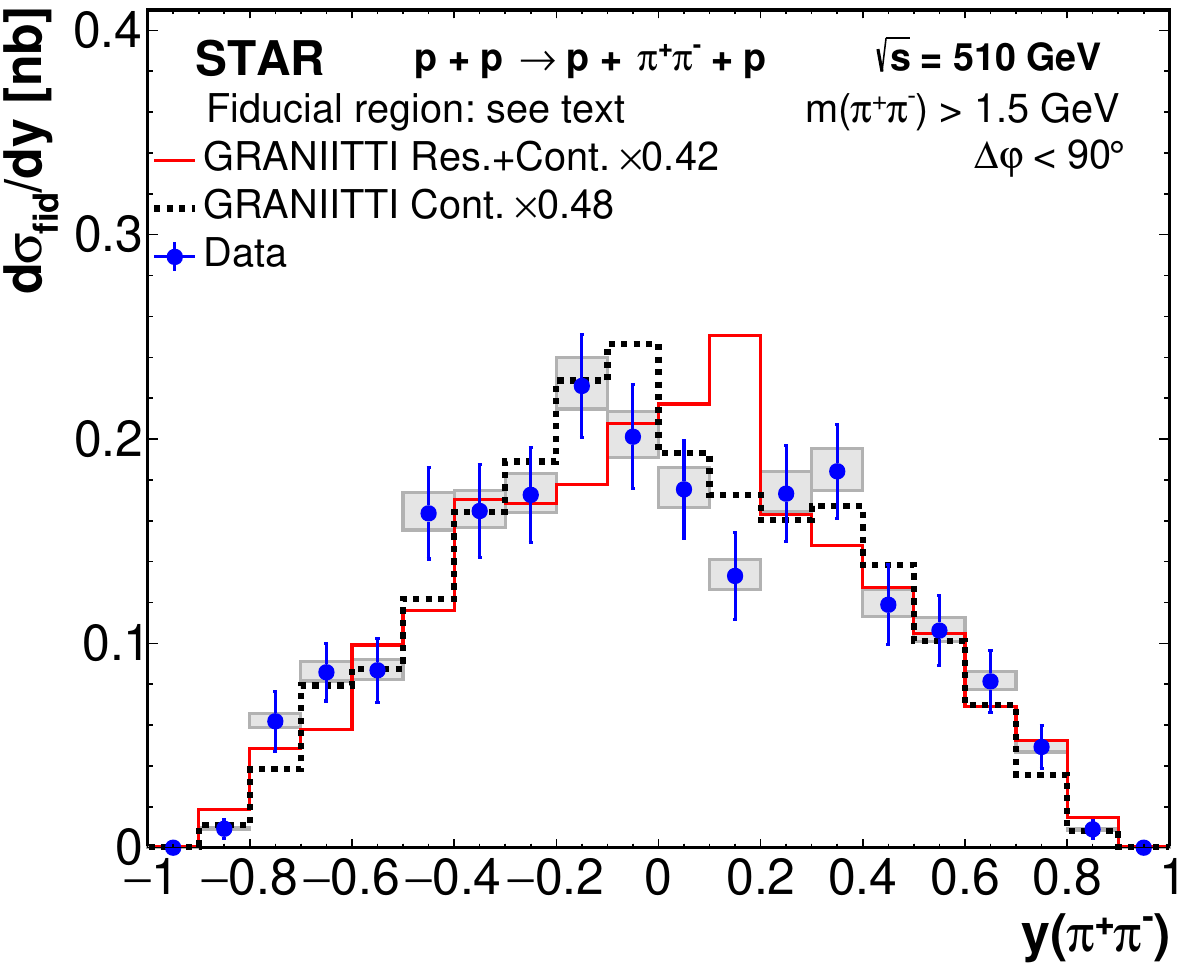}
\includegraphics[width = .32\linewidth]{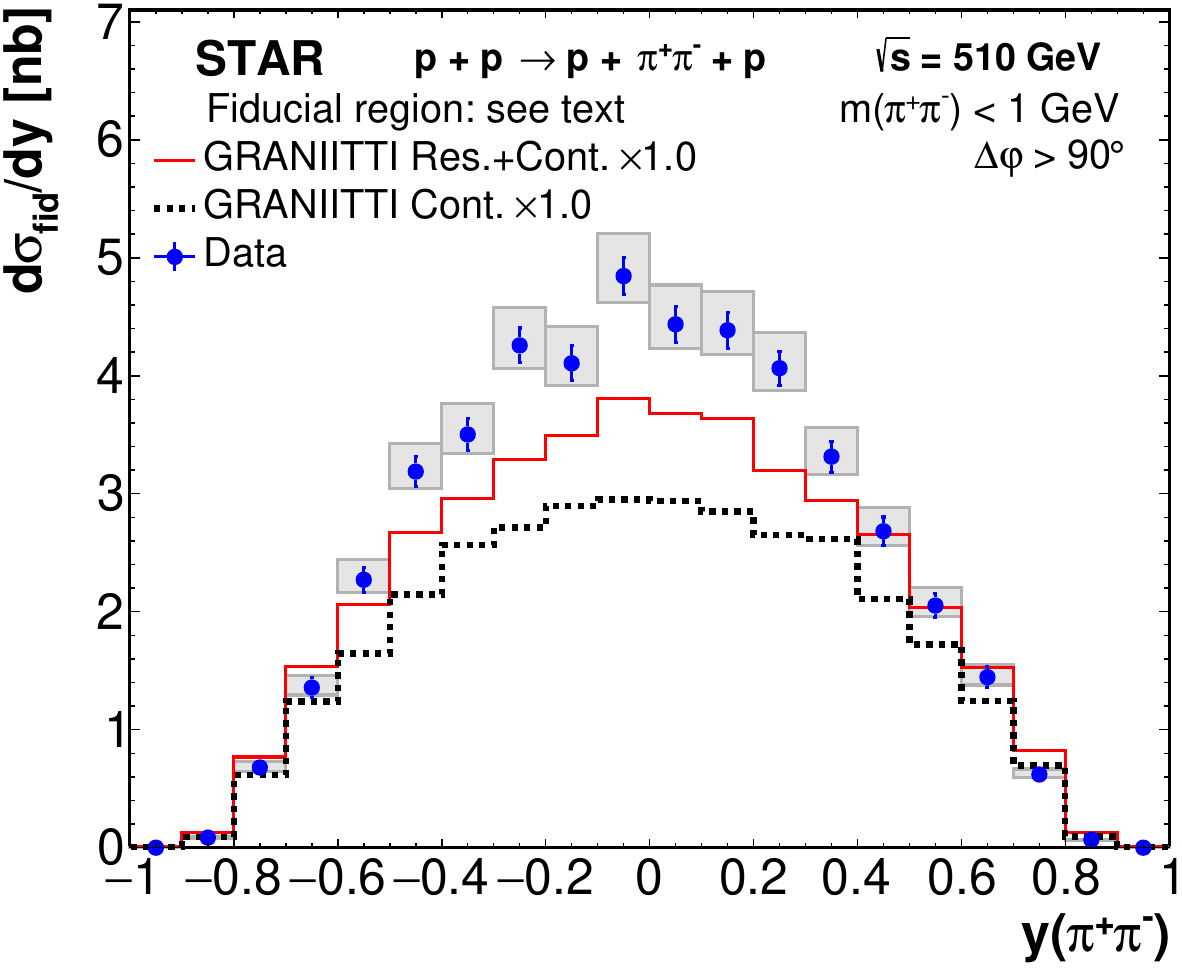}
\hfill
\includegraphics[width = .32\linewidth]{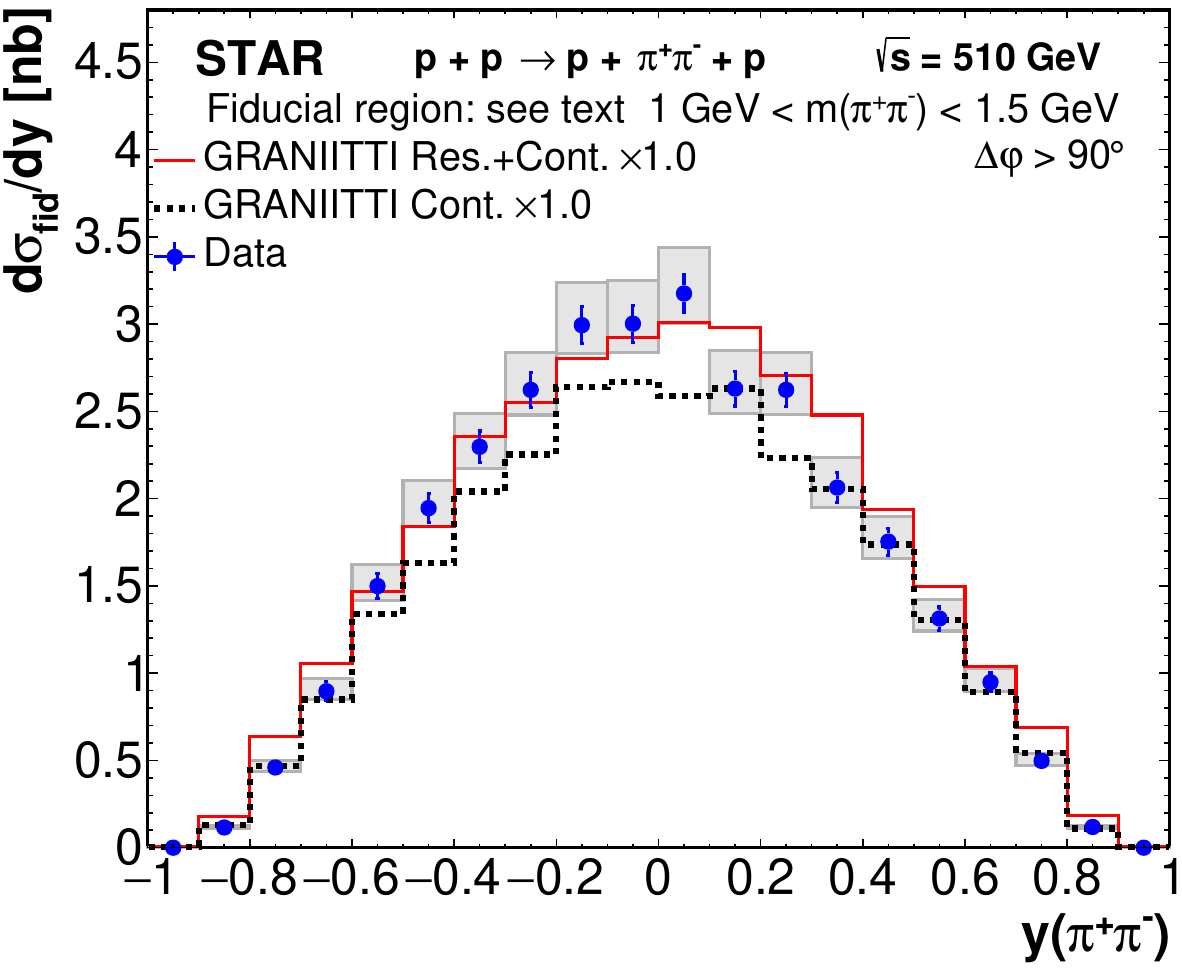}
\hfill
\includegraphics[width = .32\linewidth]{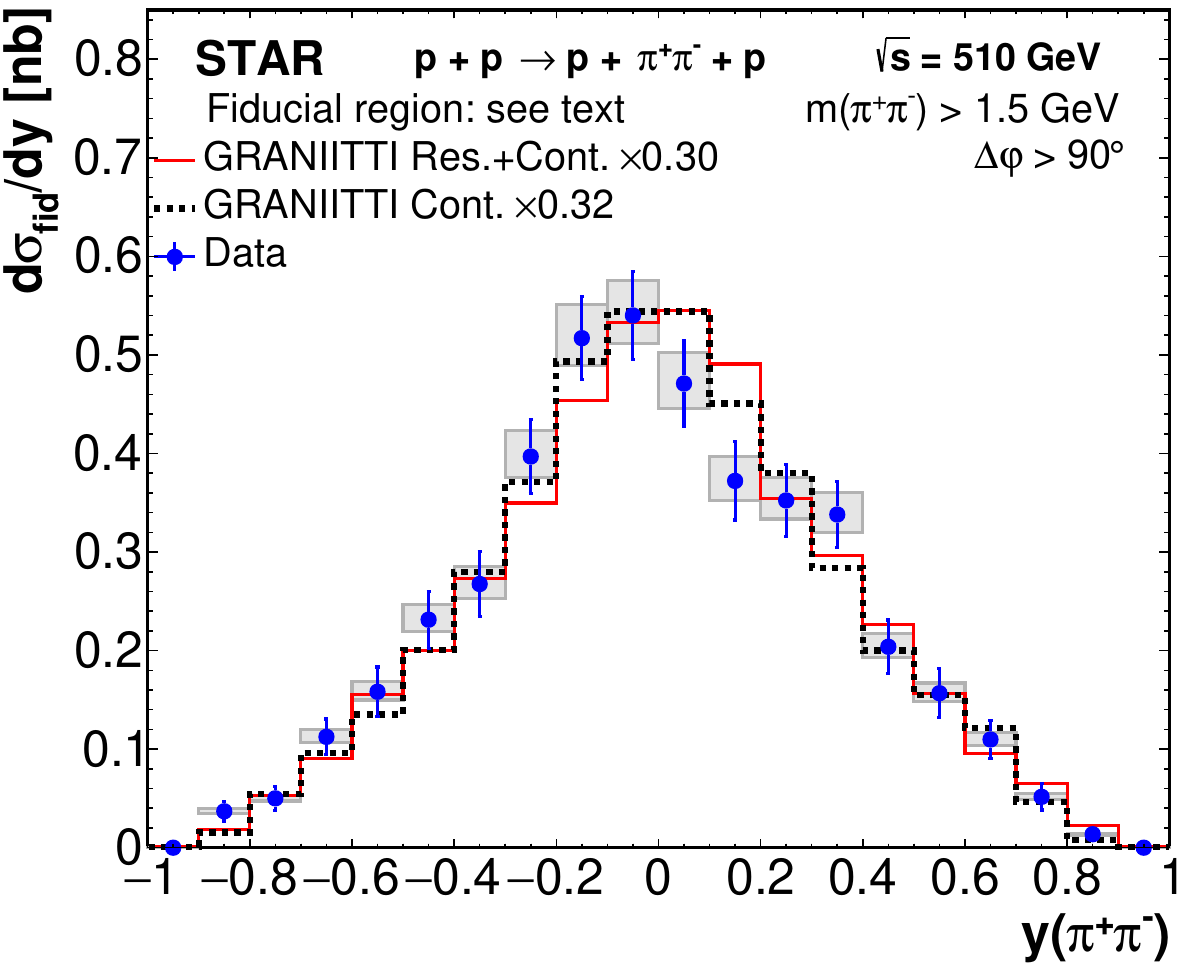}
\caption[Differential fiducial cross sections of pion pairs as a function of the $y$ in three regions of the invariant mass of the pair]
{Differential fiducial cross sections of $\pi^+\pi^-$ pairs as a function of the pair rapidity in three ranges of the $\pi^+\pi^-$ pair invariant mass: $m(\pi^+\pi^-) < 1$ GeV (left), 1 GeV $ < m(\pi^+\pi^-) < 1.5$ GeV (middle) and $m(\pi^+\pi^-) > 1.5$ GeV (right) and in to two regions of the difference in azimuthal angles $\Delta \upvarphi$ of the forward scattered protons: $\Delta \upvarphi < 90^{\circ}$ (top) and $\Delta \upvarphi > 90^{\circ}$ (bottom), measured in the fiducial volume explained in the text. Data are shown as solid blue points with error bars representing the statistical uncertainties. The systematic uncertainties are shown as gray boxes. The scale uncertainty on the vertical axis due to the effective integrated luminosity is 6.4\% and is not shown. Predictions from MC model GRANIITTI~\cite{Mieskolainen} are shown separately for each $\Delta \upvarphi$ region within the experimentally accessible fiducial acceptance of the STAR detector. The normalization of the model is performed independently in the two $\Delta\upvarphi$ regions and is used solely to compare the shapes of the distributions.}
\label{fig:rapDiffInvMass}
\end{figure}

Figure~\ref{fig:rapDiffInvMass} shows differential fiducial cross sections of hadron $\pi^+\pi^-$ pairs as a function of the pair rapidity in three ranges of the $\pi^+\pi^-$ pair invariant mass and in two $\Delta \upvarphi$ regions. GRANIITTI predictions are shown as well with the same scaling as in~figure~\ref{fig:DeltaPhiDiffInvMass}. The GRANIITTI predictions describe well the shapes of measured differential fiducial cross sections in both $\Delta \upvarphi$ regions and in all three ranges of the $\pi^+\pi^-$ pair invariant mass. The measured shapes are similar to those observed at $\sqrt{s} = 200$ GeV~\cite{Rafal20}.

\begin{figure}[htbp!]
\centering
\includegraphics[width = .32\linewidth]{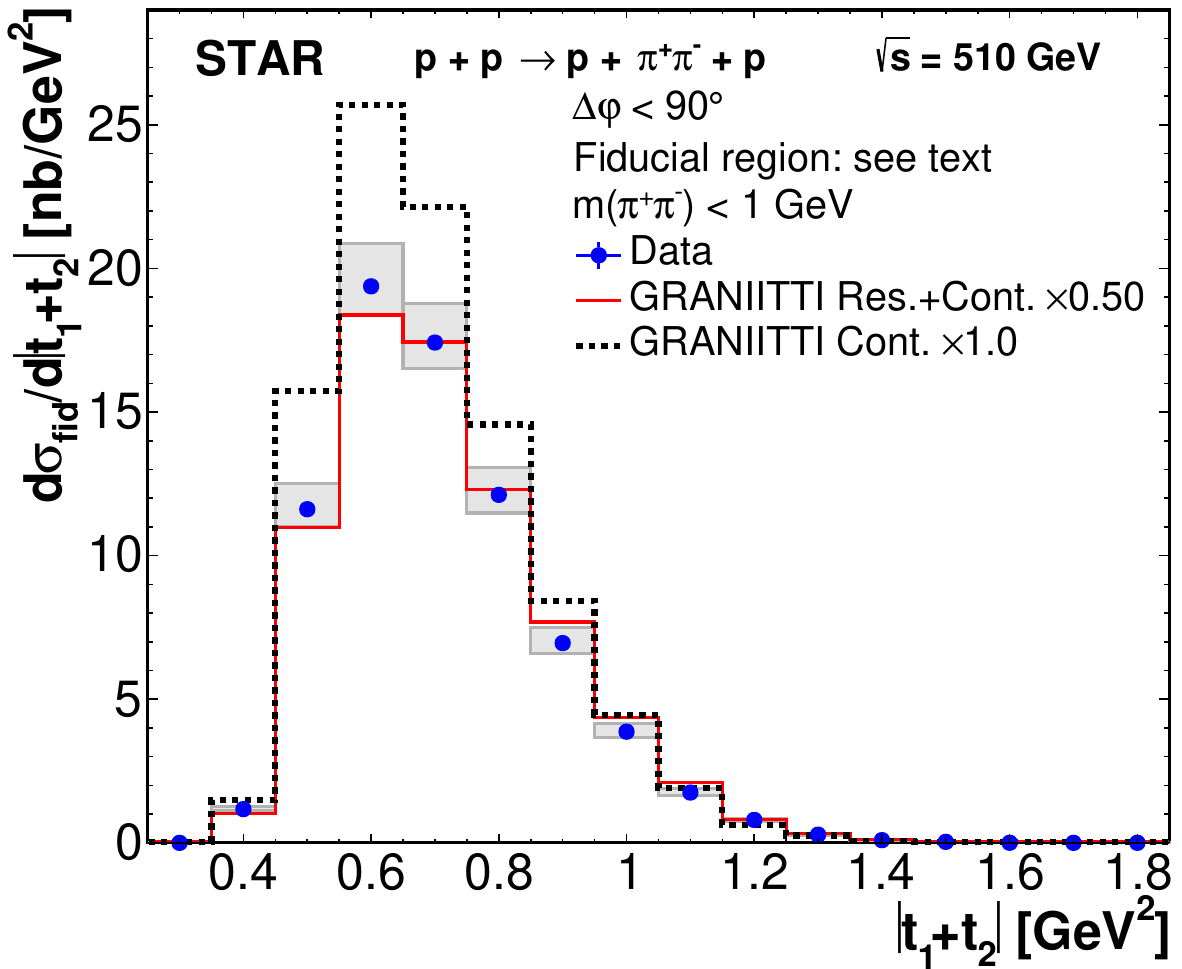}
\hfill
\includegraphics[width = .32\linewidth]{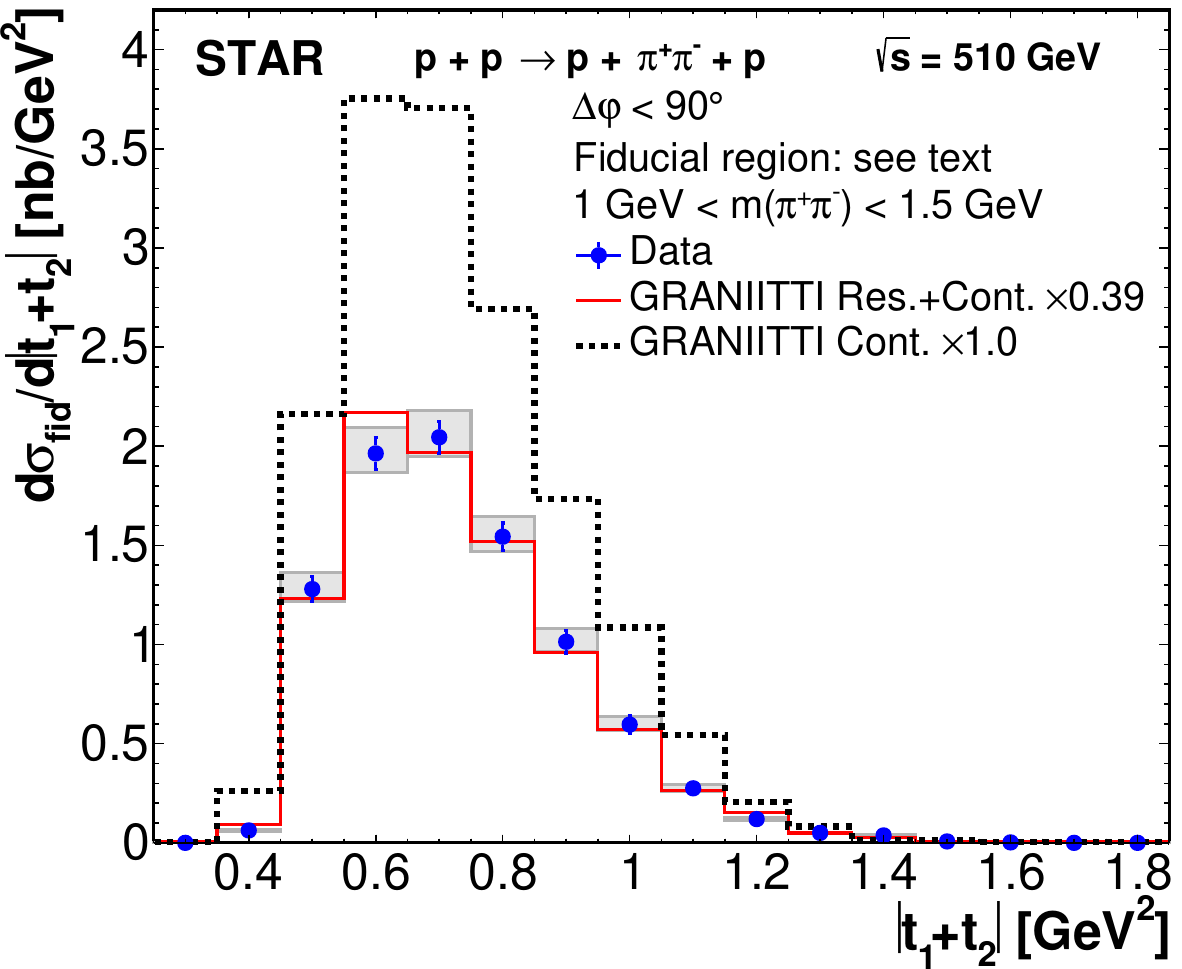}
\hfill
\includegraphics[width = .32\linewidth]{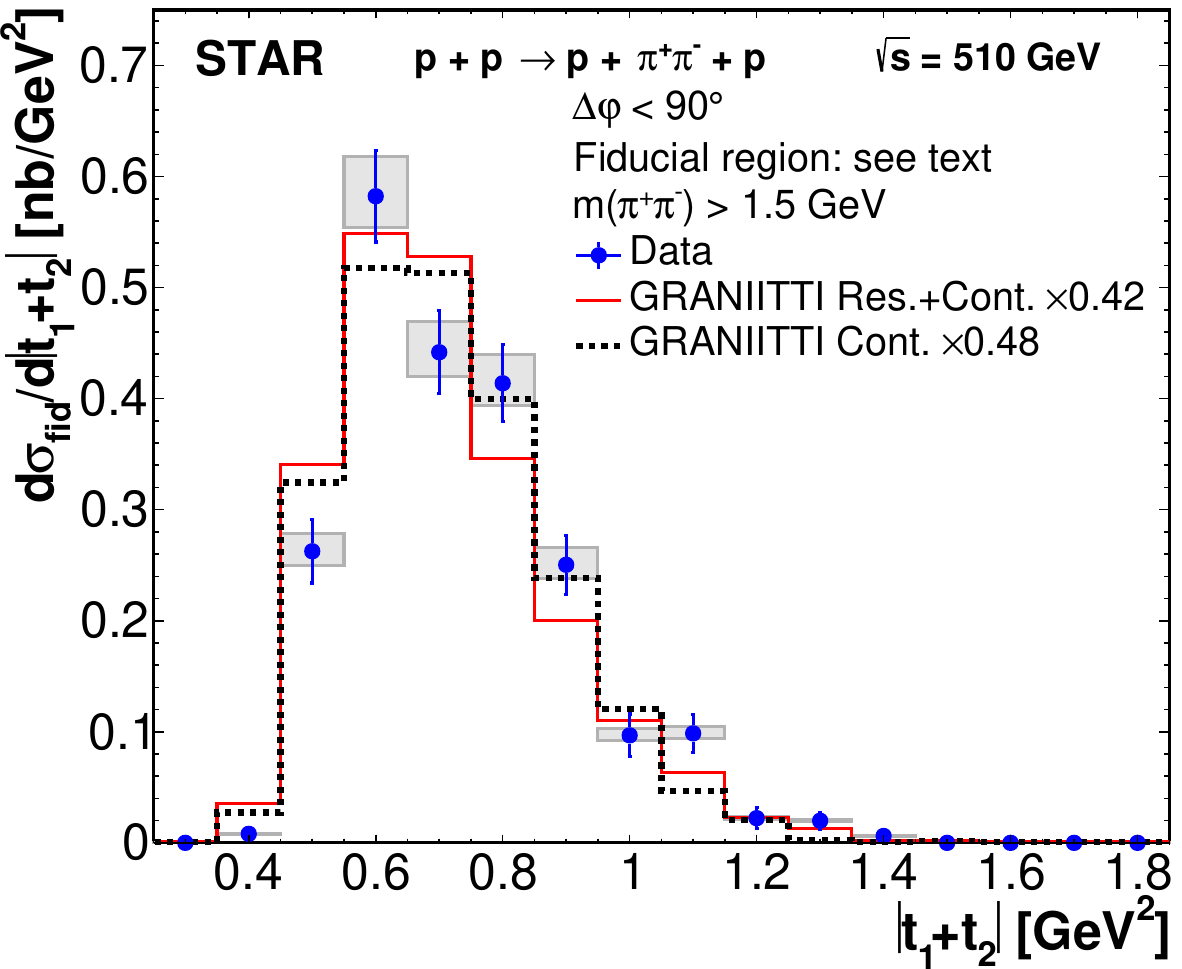}
\includegraphics[width = .32\linewidth]{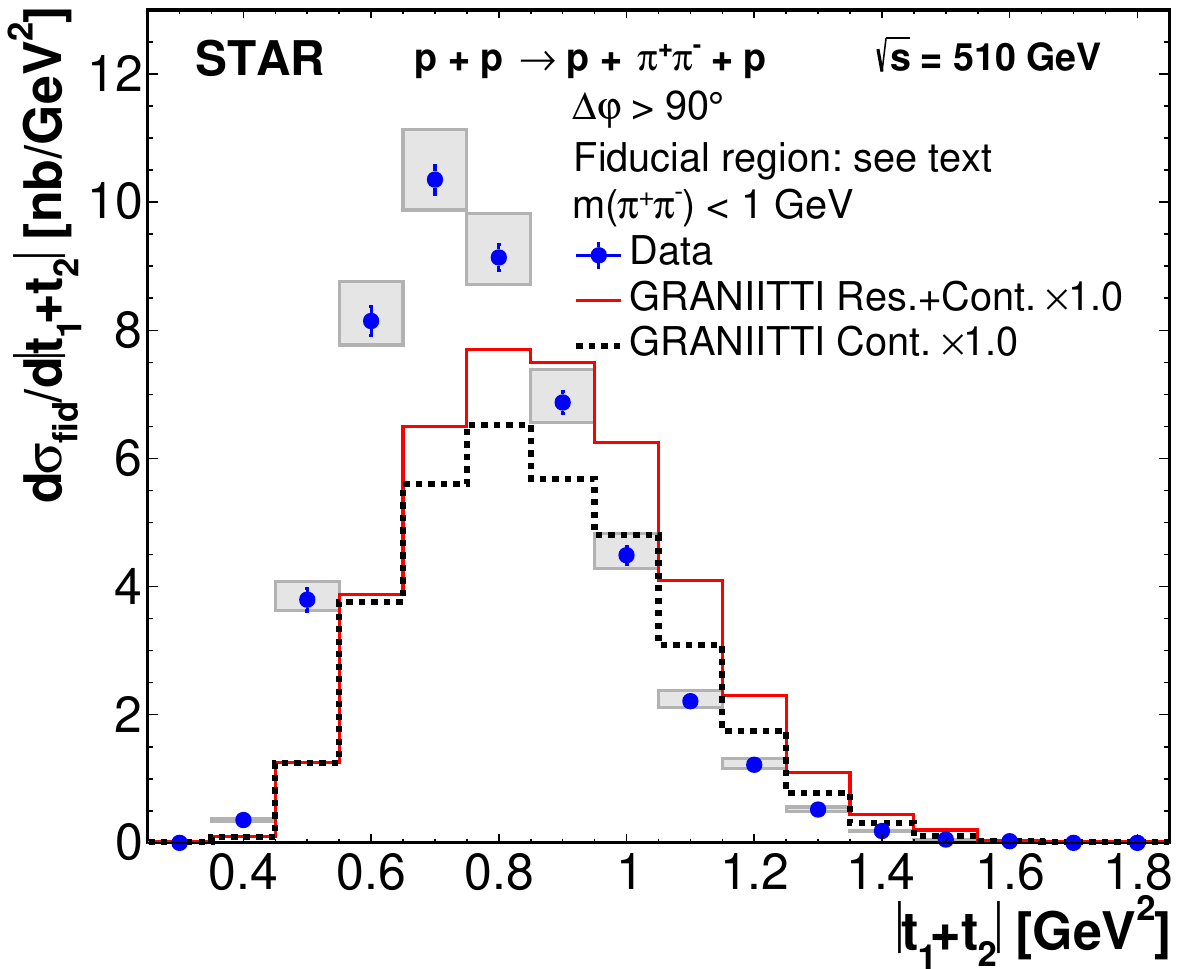}
\hfill
\includegraphics[width = .32\linewidth]{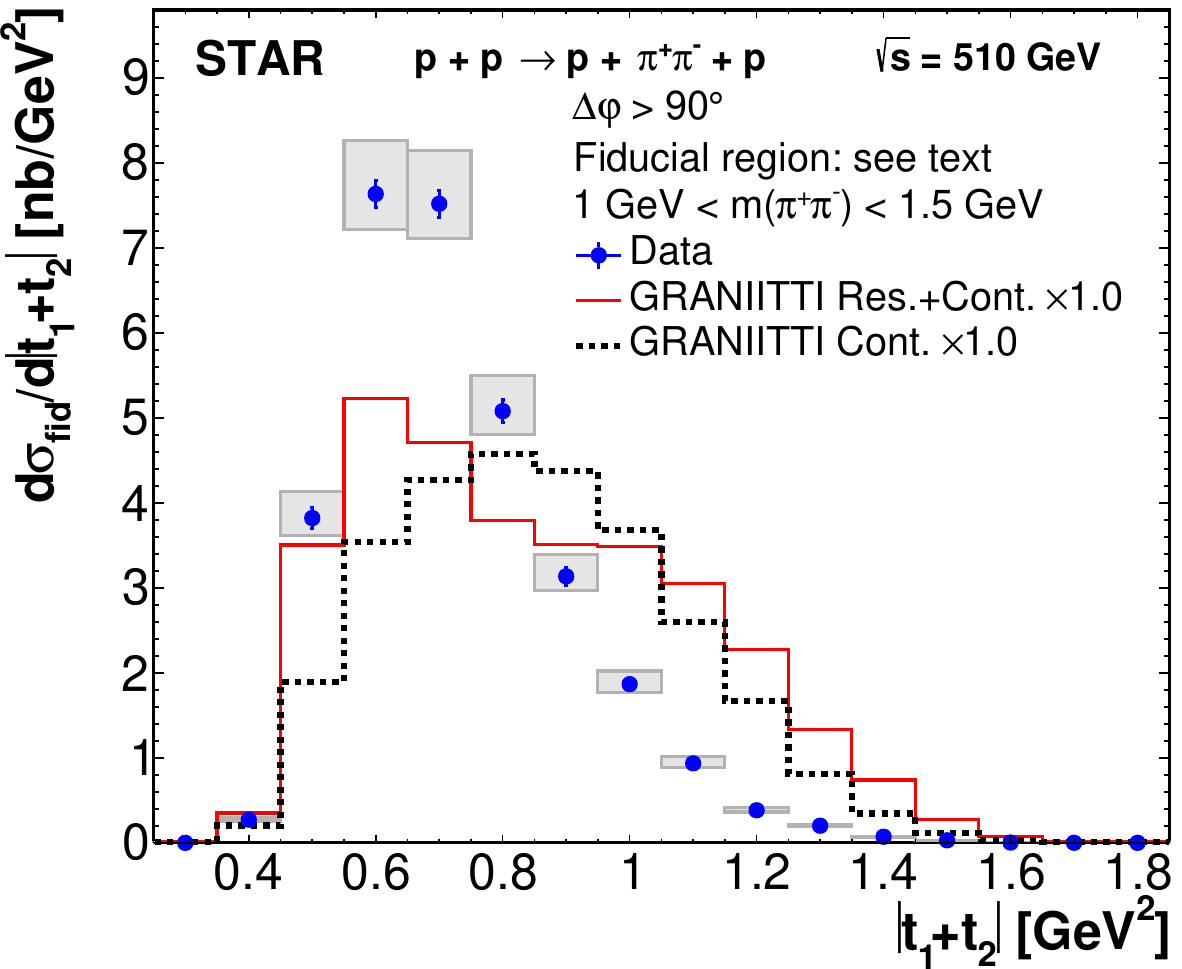}
\hfill
\includegraphics[width = .32\linewidth]{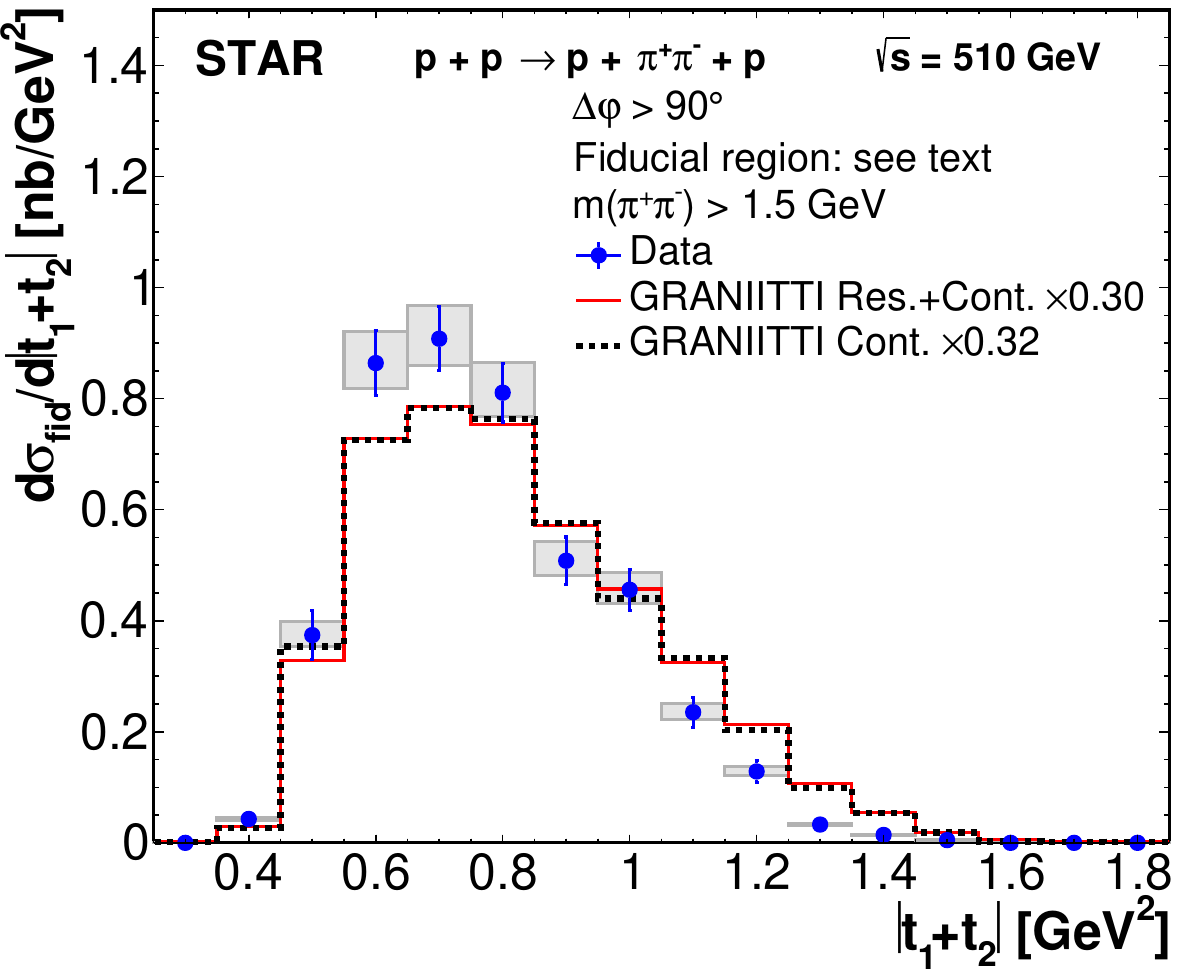}
\caption[Differential fiducial cross sections of pion pairs as a function of the $|t_1 + t_2|$ in three regions of the invariant mass of the pair]
{Differential fiducial cross sections of $\pi^+\pi^-$ pairs as a function of the absolute value of the sum of the squares of the four-momentum transfer of the forward protons in three ranges of the $\pi^+\pi^-$ pair invariant mass: $m(\pi^+\pi^-) < 1$ GeV (left), 1 GeV $ < m(\pi^+\pi^-) < 1.5$ GeV (middle) and $m(\pi^+\pi^-) > 1.5$ GeV (right) and in to two regions of the difference in azimuthal angles $\Delta \upvarphi$ of the forward scattered protons: $\Delta \upvarphi < 90^{\circ}$ (top) and $\Delta \upvarphi > 90^{\circ}$ (bottom), measured in the fiducial volume explained in the text. Data are shown as solid blue points with error bars representing the statistical uncertainties. The systematic uncertainties are shown as gray boxes. The scale uncertainty on the vertical axis due to the effective integrated luminosity is 6.4\% and is not shown. Predictions from MC model GRANIITTI~\cite{Mieskolainen} are shown separately for each $\Delta \upvarphi$ region within the experimentally accessible fiducial acceptance of the STAR detector. The normalization of the model is performed independently in the two $\Delta\upvarphi$ regions and is used solely to compare the shapes of the distributions.}
\label{fig:tSumDiffInvMass}
\end{figure}

Figure~\ref{fig:tSumDiffInvMass} shows differential fiducial cross sections of $\pi^+\pi^-$ pairs as a function of $|t_1 + t_2|$ in three ranges of the $\pi^+\pi^-$ pair invariant mass and in two $\Delta \upvarphi$ regions. GRANIITTI predictions are shown as well with the same scaling as in~figure~\ref{fig:DeltaPhiDiffInvMass}. The GRANIITTI predictions describe well the shapes of measured differential fiducial cross sections in $\Delta \upvarphi < 90^{\circ}$ in all three ranges of the $\pi^+\pi^-$ pair invariant mass. In $\Delta \upvarphi > 90^{\circ}$, the GRANIITTI prediction reproduces the shape only for the higher invariant masses while the predictions are shifted to the higher values of $|t_1 + t_2|$ for the first two ranges. The measured shapes are comparable to those observed at $\sqrt{s} = 200$ GeV~\cite{Rafal20}. The higher range of $|t_1 + t_2|$ is due to higher momentum of the forward protons compared to that at $\sqrt{s} = 200$ GeV~\cite{Rafal20}.

\subsection{Integrated fiducial cross sections}

In $m(h^+h^-)$, the integrated fiducial cross sections of $\pi^+\pi^-$, $K^+K^-$, and $p\bar{p}$ pairs, measured in the fiducial volume explained in the text above, are calculated separately for two regions of the difference in azimuthal angles $\Delta \upvarphi$ of the forward scattered protons: $\Delta \upvarphi > 90^{\circ}$ and $\Delta \upvarphi < 90^{\circ}$. The results are presented in~table~\ref{Tab:intCrossSection}.

\begin{table}[htbp!]
    \centering
    \begin{tabular}{c|c|c|c}
        \multirow{2}*{Particle species} & \multirow{2}*{Unit} & \multicolumn{2}{c}{$\sigma_{\text{fid}} \pm \delta_{\text{stat}} \pm \delta_{\text{syst}}$} \\
        &  & $\Delta \upvarphi < 90^\circ$ & $\Delta \upvarphi > 90^\circ$ \\
        \hline
        \rule{0pt}{14pt} $\pi^+ \pi^-$ & nb &  $8.68 \pm 0.04 \ ^{+0.64}_{-0.43}$  & $8.42 \pm 0.04 \ ^{+0.64}_{-0.41}$ \\ [0.3em]
        $K^+ K^-$ & pb & $879 \pm 27 \ ^{+78}_{-57}$  & $868 \pm 28 \ ^{+81}_{-84}$ \\ [0.3em]
        $p \bar{p}$ & pb & $13.9 \pm 1.4 \ ^{+0.8}_{-1.3}$  & $14.4 \pm 1.5 \ ^{+1.1}_{-1.1}$ \\
    \end{tabular}
    \caption[Integrated fiducial cross sections of $\pi^+\pi^-$, $K^+K^-$, and $p\bar{p}$ pairs.]
    {Integrated fiducial cross sections for CEP of $\pi^+\pi^-$, $K^+K^-$, and $p\bar{p}$ pairs measured in the fiducial volume explained in the text. Integrated cross sections are presented in to two regions of the difference in azimuthal angles $\Delta \upvarphi$ of the forward scattered protons: $\Delta \upvarphi > 90^{\circ}$ and $\Delta \upvarphi < 90^{\circ}$. Statistical and systematic uncertainties are provided for each cross section.}
    \label{Tab:intCrossSection}
\end{table}

\section{Summary}

We present the first results on the CEP of oppositely charged hadron pairs ($\pi^+\pi^-$, $K^+K^-$, and $p\bar{p}$) in proton-proton collisions at $\sqrt{s} = 510$ GeV using the STAR detector at RHIC. The measurement of forward-scattered protons enabled full control of the process kinematics and verification of its exclusivity. The measurements are presented within the STAR detector’s acceptance, which is determined from the transverse momenta and pseudorapidities of the centrally produced charged hadron pairs, as well as the momenta of forward-scattered protons.

The differential fiducial cross sections are presented as functions of the $\Delta\upvarphi$, $|t_1 + t_2|$, $m(h^+h^-)$, and $y(h^+h^-)$ in two regions of $\Delta\upvarphi$. Structures observed in the invariant mass spectra of $\pi^+\pi^-$ and $K^+K^-$ pairs suggest that \DPE~is the dominant production mechanism in this fiducial phase space. We observe a strong dependence of the cross section measured here, $d\sigma/dm(h^+h^-)$, on $\Delta\upvarphi$ suggesting different Pomeron interactions in each $\Delta\upvarphi$ region.

For $\pi^+\pi^-$ pairs, the invariant mass spectrum is consistent with the $f_0(980)$ and $f_2(1270)$ resonances, which have been observed in previous measurements at other collision energies~\cite{Akesson1986, PhysRevD.91.091101, Rafal20, CMS}. Also, an enhancement consistent with the $f_0(1710)$ is observed. 


For $K^+K^-$ pairs, the observed features can be explained as due to the $f_0(980)$, $\upvarphi(1020)$, $f_0(1500)$, $f_2(1525)$, and $f_0(1710)$ resonances. The limited statistics do not allow for any significant conclusions about the resonances in the $p\bar{p}$ invariant mass spectrum.

We also present the rapidity distributions of the central states and the $|t_1+t_2|$, and $\Delta\upvarphi$ distributions of the forward protons. In the $\pi^+\pi^-$ channel, these distributions are present in three mass regions.

We also present a comparison of the results with predictions from GRANIITTI v1.090 event generator, which was tuned to and is in good agreement with the results at $\sqrt{s} = 200$ GeV~\cite{Rafal20,Graniitti2023}. In general, GRANIITTI describes the shape of the distributions, but additional scaling factor in the range 0.3 to 1.8 is often needed to match the results. The new data presented here will allow for additional tuning of GRANIITTI and other models of CEP. For example, tuning the contribution from the $f_0(500)$ resonance may improve the agreement between GRANIITTI and the data. 


\acknowledgments

We thank the RHIC Operations Group and SCDF at BNL, the NERSC Center at LBNL, and the Open Science Grid consortium for providing resources and support.  This work was supported in part by the Office of Nuclear Physics within the U.S. DOE Office of Science, the U.S. National Science Foundation, National Natural Science Foundation of China, Chinese Academy of Science, the Ministry of Science and Technology of China and the Chinese Ministry of Education, NSTC Taipei, the National Research Foundation of Korea, Czech Science Foundation and Ministry of Education, Youth and Sports of the Czech Republic, Hungarian National Research, Development and Innovation Office, New National Excellency Programme of the Hungarian Ministry of Human Capacities, Department of Atomic Energy and Department of Science and Technology of the Government of India, the National Science Centre and WUT ID-UB of Poland, German Bundesministerium f\"ur Bildung, Wissenschaft, Forschung and Technologie (BMBF), Helmholtz Association, Ministry of Education, Culture, Sports, Science, and Technology (MEXT), Japan Society for the Promotion of Science (JSPS), and Agencia Nacional de Investigacion y Desarrollo de Chile (ANID), Chile.


\bibliographystyle{JHEP}

\begingroup
\interlinepenalty=10000
\bibliography{PaperCEPat510GEV.bib}
\endgroup

\appendix

\raggedright
\setcounter{secnumdepth}{0}
\section{STAR Collaboration}

\begingroup
\footnotesize
\setlength{\parindent}{0pt}
\setlength{\parskip}{0.35em}

\noindent
B.~E.~Aboona$^{59}$,
J.~Adam$^{18}$,
L.~Adamczyk$^{4}$,
I.~Aggarwal$^{46}$,
M.~M.~Aggarwal$^{46}$,
Z.~Ahammed$^{67}$,
A.~K.~Alshammri$^{34}$,
E.~C.~Aschenauer$^{8}$,
S.~Aslam$^{23}$,
J.~Atchison$^{3}$
% ...

\medskip
\begin{flushleft}
\noindent $^{1}$Academia Sinica, Nankang, 115\\
\noindent $^{2}$Panchayat College (PC), Bargarh, affiliated with Sambalpur University (SU), Odisha 768028, India\\
\noindent $^{3}$Abilene Christian University, Abilene, Texas   79699\\
\noindent $^{4}$AGH University of Krakow, FPACS, Cracow 30-059, Poland\\
\noindent $^{5}$Argonne National Laboratory, Argonne, Illinois 60439\\
\noindent $^{6}$American University in Cairo, New Cairo 11835, Egypt\\
\noindent $^{7}$Ball State University, Muncie, Indiana, 47306\\
\noindent $^{8}$Brookhaven National Laboratory, Upton, New York 11973\\
\noindent $^{9}$University of Calabria \& INFN-Cosenza, Rende 87036, Italy\\
\noindent $^{10}$University of California, Berkeley, California 94720\\
\noindent $^{11}$University of California, Davis, California 95616\\
\noindent $^{12}$University of California, Los Angeles, California 90095\\
\noindent $^{13}$University of California, Riverside, California 92521\\
\noindent $^{14}$Central China Normal University, Wuhan, Hubei 430079\\
\noindent $^{15}$University of Illinois at Chicago, Chicago, Illinois 60607\\
\noindent $^{16}$Chongqing University, Chongqing, 401331\\
\noindent $^{17}$Creighton University, Omaha, Nebraska 68178\\
\noindent $^{18}$Czech Technical University in Prague, FNSPE, Prague 115 19, Czech Republic\\
\noindent $^{19}$Technische Universit\"at Darmstadt, Darmstadt 64289, Germany\\
\noindent $^{20}$National Institute of Technology Durgapur, Durgapur - 713209, India\\
\noindent $^{21}$ELTE E\"otv\"os Lor\'and University, Budapest, Hungary H-1117\\
\noindent $^{22}$Frankfurt Institute for Advanced Studies FIAS, Frankfurt 60438, Germany\\
\noindent $^{23}$Fudan University, Shanghai, 200433\\
\noindent $^{24}$Guangxi Normal University, Guilin, 541004\\
\noindent $^{25}$University of Heidelberg, Heidelberg 69120, Germany\\
\noindent $^{26}$University of Houston, Houston, Texas 77204\\
\noindent $^{27}$Huzhou University, Huzhou, Zhejiang  313000\\
\noindent $^{28}$Indian Institute of Science Education and Research (IISER), Berhampur 760010 , India\\
\noindent $^{29}$Indian Institute of Science Education and Research (IISER) Tirupati, Tirupati 517507, India\\
\noindent $^{30}$Indian Institute Technology, Patna, Bihar 801106, India\\
\noindent $^{31}$Indiana University, Bloomington, Indiana 47408\\
\noindent $^{32}$Institute of Modern Physics, Chinese Academy of Sciences, Lanzhou, Gansu 730000\\
\noindent $^{33}$University of Jammu, Jammu 180001, India\\
\noindent $^{34}$Kent State University, Kent, Ohio 44242\\
\noindent $^{35}$University of Kentucky, Lexington, Kentucky 40506-0055\\
\noindent $^{36}$Lanzhou University, Lanzhou, 730000\\
\noindent $^{37}$Lawrence Berkeley National Laboratory, Berkeley, California 94720\\
\noindent $^{38}$Lehigh University, Bethlehem, Pennsylvania 18015\\
\noindent $^{39}$Lovely Professional University, Jalandhar - Delhi G.T. Road, Pagwara, Panjab, 144411, India\\
\noindent $^{40}$Max-Planck-Institut f\"ur Physik, Munich 80805, Germany\\
\noindent $^{41}$Michigan State University, East Lansing, Michigan 48824\\
\noindent $^{42}$National Institute of Science Education and Research, HBNI, Jatni 752050, India\\
\noindent $^{43}$National Cheng Kung University, Tainan 70101\\
\noindent $^{44}$Nuclear Physics Institute of the CAS, Rez 250 68, Czech Republic\\
\noindent $^{45}$The Ohio State University, Columbus, Ohio 43210\\
\noindent $^{46}$Panjab University, Chandigarh 160014, India\\
\noindent $^{47}$Purdue University, West Lafayette, Indiana 47907\\
\noindent $^{48}$Rice University, Houston, Texas 77251\\
\noindent $^{49}$Rutgers University, Piscataway, New Jersey 08854\\
\noindent $^{50}$University of Science and Technology of China, Hefei, Anhui 230026\\
\noindent $^{51}$South China Normal University, Guangzhou, Guangdong 510631\\
\noindent $^{52}$Sejong University, Seoul, 05006, Korea, Republic Of\\
\noindent $^{53}$Shandong University, Qingdao, Shandong 266237\\
\noindent $^{54}$Shanghai Institute of Applied Physics, Chinese Academy of Sciences, Shanghai 201800\\
\noindent $^{55}$Southern Connecticut State University, New Haven, Connecticut 06515\\
\noindent $^{56}$State University of New York, Stony Brook, New York 11794\\
\noindent $^{57}$Instituto de Alta Investigaci\'on, Universidad de Tarapac\'a, Arica 1000000, Chile\\
\noindent $^{58}$Temple University, Philadelphia, Pennsylvania 19122\\
\noindent $^{59}$Texas A\&M University, College Station, Texas 77843\\
\noindent $^{60}$Texas Southern University, Houston, Texas, 77004\\
\noindent $^{61}$University of Texas, Austin, Texas 78712\\
\noindent $^{62}$Tsinghua University, Beijing 100084\\
\noindent $^{63}$University of Tsukuba, Tsukuba, Ibaraki 305-8571, Japan\\
\noindent $^{64}$University of Chinese Academy of Sciences, Beijing, 101408\\
\noindent $^{65}$United States Naval Academy, Annapolis, Maryland 21402\\
\noindent $^{66}$Valparaiso University, Valparaiso, Indiana 46383\\
\noindent $^{67}$Variable Energy Cyclotron Centre, Kolkata 700064, India\\
\noindent $^{68}$Warsaw University of Technology, Warsaw 00-661, Poland\\
\noindent $^{69}$Wayne State University, Detroit, Michigan 48201\\
\noindent $^{70}$Wuhan University of Science and Technology, Wuhan, Hubei 430065\\
\noindent $^{71}$Yale University, New Haven, Connecticut 06520\\
\end{flushleft}
\endgroup

\end{document}